\documentclass[NETN,manuscript]{stjour-new}

\usepackage{parskip}
\articletype{Review}
\usepackage{changepage}

\graphicspath{{figs/}}

\begin{document}
\title[Theoretical foundations of studying criticality in the brain]{Theoretical foundations of studying criticality in the brain}



\author[Yang Tian \and Zeren Tan \and Hedong Hou \and Guoqi Li \and Aohua Cheng \and Yike Qiu \and Kangyu Weng \and Chun Chen \and Pei Sun]
{Yang Tian\affil{1,2} \and Zeren Tan\affil{3} \and Hedong Hou\affil{4} \and Guoqi Li\affil{5,6} \and Aohua Cheng\affil{7} \and Yike Qiu\affil{7} \and Kangyu Weng\affil{7} \and Chun Chen\affil{1} \and Pei Sun\affil{1}}

\affiliation{1}{Department of Psychology \& Tsinghua Laboratory of Brain and Intelligence, Tsinghua University, Beijing, 100084, China.}
\affiliation{2}{Laboratory of Advanced Computing and Storage, Central Research Institute, 2012 Laboratories, Huawei Technologies Co. Ltd., Beijing, 100084, China.}
\affiliation{3}{Institute for Interdisciplinary Information Science, Tsinghua University, Beijing, 100084, China.}
\affiliation{4}{UFR de Math\'{e}matiques, Universit\'{e} de Paris, Paris, 75013, France.}
\affiliation{5}{Institute of Automation, Chinese Academy of Science, Beijing, 100190, China.}
\affiliation{6}{University of Chinese Academy of Science, Beijing, 100049, China.}
\affiliation{7}{Tsien Excellence in Engineering
Program, School of Aerospace Engineering, Tsinghua University, Beijing, 100084, China.}


\correspondingauthor{Pei Sun}{peisun@tsinghua.edu.cn}


\begin{abstract}
     Criticality is hypothesized as a physical mechanism underlying efficient transitions between cortical states and remarkable information processing capacities in the brain. While considerable evidence generally supports this hypothesis, non-negligible controversies persist regarding the ubiquity of criticality in neural dynamics and its role in information processing. Validity issues frequently arise during identifying potential brain criticality from empirical data. Moreover, the functional benefits implied by brain criticality are frequently misconceived or unduly generalized. These problems stem from the non-triviality and immaturity of the physical theories that analytically derive brain criticality and the statistic techniques that estimate brain criticality from empirical data. To help solve these problems, we present a systematic review and reformulate the foundations of studying brain criticality, i.e., ordinary criticality (OC), quasi-criticality (qC), self-organized criticality (SOC), and self-organized quasi-criticality (SOqC), using the terminology of neuroscience. We offer accessible explanations of the physical theories and statistic techniques of brain criticality, providing step-by-step derivations to characterize neural dynamics as a physical system with avalanches. We summarize error-prone details and existing limitations in brain criticality analysis and suggest possible solutions. Moreover, we present a forward-looking perspective on how optimizing the foundations of studying brain criticality can deepen our understanding of various neuroscience questions. 
\end{abstract}

\begin{authorsummary}
Brain criticality hypothesis is one of the most focused and controversial topics in neuroscience and biophysics. This research develops a unified framework to reformulate the physics theories of four basic types of brain criticality, i.e., ordinary criticality (OC), quasi-criticality (qC), self-organized criticality (SOC), and self-organized quasi-criticality (SOqC), into more accessible and neuroscience-related forms. For the statistic techniques used to validate brain criticality hypothesis, we also present comprehensive explanations of them, summarize their error-prone details, and suggest possible solutions. This framework may help resolve potential controversies in studying brain criticality hypothesis, especially those arising from the misconceptions about the theoretical foundations of brain criticality.
\end{authorsummary}

\section{Introduction}
Neuroscience is dawning upon revealing physics foundations of the brain \cite{abbott2008theoretical}. Ever since the 1970s, the term neurophysics has been suggested as a term to indicate the essential role of physics in understanding the brain \cite{scott1977neurophysics}. More recently, substantial progress has been accomplished in studying brain connectivity and brain functions with statistical physics theories \cite{lynn2019physics}.

For brain connectivity, physics provides insights for its emergence, organization, and evolution. Random graphs \cite{betzel2017generative,betzel2016generative}, percolation \cite{guo2021percolation,breskin2006percolation}, and other physics theories of correlated systems \cite{haimovici2013brain,wolf2005symmetry} are applied to reveal the underlying mechanisms accounting for the origins of brain network properties. Complex network theories act as the foundation of characterizing brain connectivity organizational features (e.g., community \cite{betzel2017multi,betzel2018diversity,khambhati2018modeling}, hub \cite{gong2009mapping,deco2015rethinking}, and small-world \cite{deco2015rethinking,bullmore2012economy} structures) and embedding attributes into physical space \cite{bassett2010efficient,kaiser2006nonoptimal}. Network evolution driven by neural plasticity helps to explain the dynamics of brain connectivity structures during information processing \cite{del2021unconsciousness,montague1996framework,song2000competitive,robert2021stochastic,galvan2010neural}. For brain functions, physics presents possible explanations for the origin of information processing capacities from collective neural activities. From single neuron dynamics models \cite{gerstner2014neuronal}, stochastic network models of neural populations and circuits \cite{tian2021characteristics,tian2021bridging}, mean-field neural mass models of brain regions \cite{touboul2011neural,david2003neural}, eventually to models of entire brain networks \cite{schneidman2006weak,hopfield1982neural}, important efforts have been devoted to characterize information-processing-related neural dynamics across different scales. Networks with memory capacities (e.g., Hopfield networks \cite{tyulmankov2021biological}), which are equivalent to Ising models under specific conditions \cite{lynn2019physics}, have been applied to study neural information storage and recall \cite{krotov2020large,haldeman2005critical}, adaptation to environment changes \cite{shew2015adaptation}, information transmission optimization \cite{beggs2003neuronal}, dynamic range maximization \cite{kinouchi2006optimal,shew2009neuronal}, and neural computation power \cite{bertschinger2004real}. These models are further related to maximum entropy models (e.g., specific fine-tuned Ising models) that predict long-range correlations observed among neurons \cite{schneidman2006weak,ganmor2011sparse}. Moreover, general theories of free-energy principle \cite{friston2010free,friston2009free,guevara2021synchronization} and information thermodynamics \cite{tian2022information,capolupo2013dissipation,collell2015brain,sartori2014thermodynamic} are suggested as the unified foundations of perception, action, and learning in the brain.

If one needs to specify one of the most focused and controversial topics among all the works mentioned above, brain criticality may be a potential candidate \cite{beggs2012being}. The hypothesis of the critical brain has received increasing attention in recent decades, serving as a possible mechanism underlying various intriguing but elusive phenomena in the brain. In light of our limited understanding of the complex nature of collective neural dynamics, these phenomena include, to name a few, efficient transitions between cortical states \cite{fontenele2019criticality}, maximal dynamic ranges of neural responses \cite{kinouchi2006optimal,shew2009neuronal,antonopoulos2016dynamic,gautam2015maximizing}, optimized information transmission and representation \cite{shew2011information,li2012neuronal,antonopoulos2016dynamic}, and numerous other issues concerning brain functions that we have mentioned above. One can see \cite{shew2013functional,chialvo2010emergent,beggs2007build,hesse2014self,cocchi2017criticality} for systematic reviews of the diverse function advantages implied by brain criticality and their experimental demonstrations. From a Darwinian perspective, one potential reason for the brain to feature criticality lay in that the most informative parts of external world principally occur at a borderline between purely ordered and purely disordered states (information would be trivial in a purely ordered world while it would be incomprehensible in a purely disordered world). Becoming critical may be a potential way for the brain to adapt to the complex world, where non-trivial information has a finite opportunity to occur \cite{bak2013nature,chialvo2010emergent}. To date, generic features of a critical brain with the characteristics discussed above, such as divergent correlation length, neuronal avalanches with power-law behaviours, and long-range correlations on the microscopic scale (e.g., neural populations), have been extensively observed in mathematical models in conjunction with experimental data (e.g., \cite{fosque2021evidence,shew2009neuronal,beggs2003neuronal,dalla2019modeling,poil2012critical,hardstone2014neuronal,petermann2009spontaneous,shriki2013neuronal,tagliazucchi2012criticality,poil2008avalanche,gireesh2008neuronal,tkavcik2015thermodynamics,scott2014voltage,ponce2018whole}).

Our work does not aim at repeatedly reviewing experimental advances concerning brain criticality and its biological significance, given that they have been comprehensively summarized by existing reviews \cite{shew2013functional,chialvo2010emergent,beggs2007build,hesse2014self,cocchi2017criticality,munoz2018colloquium}. On the contrary, our motivation is to present a systematic and accessible review of the theoretical methods applied to achieve these advances, which have not received necessary attention yet. These theoretical foundations are initially thought to be incomprehensible and irrelevant to neuroscience. However, practice suggests that omitting these physical and mathematical backgrounds does not significantly improve the accessibility of studies on brain criticality. Instead, the lack of detailed explanations of theoretical foundations has frequently misled neuroscientists, leading to diverse confusions about the precise meaning, identification criteria, and biological corollaries of brain criticality. As a result, criticality, an analytic statistical physics theory with solid foundations, unnecessarily becomes an elusive black box for neuroscientists. To address this issue, we use the terminology of neuroscience to present a self-contained framework of brain criticality, reviewing and reformulating (1) physical theories that analytically derive brain criticality and (2) statistic techniques that computationally estimate brain criticality from empirical data. Given the frequent misunderstanding of neural avalanches, our discussions primarily focus on brain criticality analysis on the microscopic scale of the brain. The objectives guiding through our review are tripartite: (1) explaining why brain criticality matters in the brain, (2) understanding what is brain criticality and what it conveys about the brain, and (3) confirming how to identify potential brain criticality and ensure the validity of analyses.

\section{Brain criticality: general concepts}
\subsection{Overview of brain criticality}
Brain criticality frequently confuses neuroscientists since too many distinct phenomena are studied under this name without being properly classified. In this review, brain criticality refers to a family of critical processes in neural dynamics where erratic fluctuations appear to reduce dynamic stability. To present a systematic classification framework, we discuss three fundamental perspectives concerning brain criticality. \textbf{Table 1} provides all the necessary glossaries in comprehensible forms.   

\subsubsection{Being non-equilibrium}
 First, the brain, similar to other biological systems, generally exhibits temporal evolution from initial states that are far away from equilibrium \cite{lynn2021broken,gnesotto2018broken}. These departures from equilibrium arise due to diverse endogenous causes \cite{perl2021nonequilibrium,gnesotto2018broken} to break the detailed balance to support consciousness, sensing, and adaptation \cite{perl2021nonequilibrium,lynn2021broken}. Therefore, potential critical phenomena underlying neural dynamics, at least in most neural dynamics models and empirical data sets, are basically non-equilibrium and can not be characterized by equilibrium statistic mechanics. In \textbf{Fig. 1A}, we illustrate the difference between equilibrium and non-equilibrium dynamics.
 
  \begin{table}[!t]
 \begin{adjustwidth}{-0cm}{}
\caption{Key concepts in describing brain criticality}
\label{T1}
\begin{tabular}{@{}ll@{}}
\hline
Concept     & Meaning                                                                  \\ \hline
Equilibrium & \begin{tabular}[c]{@{}l@{}}A case where the system maximizes entropy and conserves energy simultaneously. The stationary \\ probability distribution $\mathcal{P}_{eq}\left(\cdot\right)$ of system states of a system at equilibrium is the Boltzmann distribution. \\ At equilibrium, the transition dynamics between system states $c$ and $c^{\prime}$ satisfies the detailed balance \\  condition $\mathcal{P}_{eq}\left(c\right)\mathcal{W}\left(c\rightarrow c^{\prime}\right)=\mathcal{P}_{eq}\left(c^{\prime}\right)\mathcal{W}\left(c^{\prime}\rightarrow c\right)$, where $\mathcal{W}\left(\cdot\rightarrow\cdot\right)$ denotes the transition probability. \end{tabular} \\\hline  Non-equilibrium & \begin{tabular}[c]{@{}l@{}}A case where the system is out of equilibrium because the transition dynamics between system states \\ breaks the detailed balance condition. In other words, the transition dynamics between states becomes \\ directional rather than symmetric. \end{tabular} \\\hline Self-organization & \begin{tabular}[c]{@{}l@{}}A process where the internal complexity of a system increases without being tuned by any external \\ mechanism. All potentially emergent properties are created by endogenous feedback processes or other \\ internal factors inside the system. \end{tabular} \\\hline Criticality & \begin{tabular}[c]{@{}l@{}}A kind of phenomena where the systems is generally close to specific critical points separating between \\ multiple system states. Small disturbances are sufficient to make the system experience dramatic and \\ sharp transitions between system states. \end{tabular}\\\hline Quasi-criticality & \begin{tabular}[c]{@{}l@{}}A kind of phenomena where all statistical physics relations required by criticality are principally adhered \\ by the system but slight and inconstant deviations from perfect criticality can be seen on the actual values\\  of characteristic variables. These deviations robustly exist and are generally independent of data noises. \end{tabular}\\\hline
Sub-criticality & \begin{tabular}[c]{@{}l@{}}A kind of system states below criticality. They occur when the order parameter (i.e., the macroscopic \\ observable used to describe system states) remains at zero even with the addition of derives, corresponding \\ to disordered system dynamics. \end{tabular}\\\hline
Super-criticality & \begin{tabular}[c]{@{}l@{}}A kind of system states above criticality. They occur when the order parameter is positive, corresponding \\ to ordered system dynamics. \end{tabular}\\\hline
\end{tabular}
 \end{adjustwidth}
\end{table}
 
 \subsubsection{Fine tuning versus self-organization}
 Second, there exist two types of general mechanisms underlying the existence of brain criticality. One type of mechanisms either arise from the external manipulations outside the brain (e.g., researchers manipulate the tonic dopamine D1-receptor stimulation \cite{stewart2006inverted,stewart2008homeostasis} or adjust network topology \cite{kaiser2010optimal,wang2012hierarchical,rubinov2011neurobiologically}) or belong to the top-down biological processes that globally function on neural dynamics inside the brain (e.g., anesthesia effects \cite{fontenele2019criticality,ribeiro2010spike,hahn2017spontaneous} as well as sleep restoration effects \cite{meisel2013fading}). Neural dynamics is passively fine tuned towards or away from ordinary criticality (OC) by these exogenous mechanisms, similar to ordinary critical phenomena that require the fine tuning of order parameters.
 
 Another type of mechanisms includes all endogenous factors of neural dynamics (e.g., neural plasticity mechanisms such as spike-timing dependent synaptic plasticity \cite{shin2006self,meisel2009adaptive,effenberger2015self}, short-term synaptic plasticity \cite{levina2007dynamical,levina2009phase}, retro-synaptic signals \cite{hernandez2017self} and Hebbian rules \cite{de2006self,de2010learning}), which locally function on neural dynamics as drive and dissipation components. The interactions between these components naturally form feedback control loops to support the self-organization of neural dynamics towards the critical point \cite{beggs2007build,chialvo2010emergent}. This spontaneously emerged brain criticality, distinct from ordinary critical phenomena, is conjectured as a kind of self-organized criticality (SOC) \cite{chialvo2010emergent}. In \textbf{Fig. 1B}, we present conceptual illustrations of ordinary criticality and self-organized criticality in the brain.
 
 \subsubsection{Standard versus non-standard}
 Third, brain criticality frequently occurs in non-standard forms due to stimulus derives or endogenous factors. On the one hand, slight and inconstant deviations from perfect brain criticality can be seen on the actual values of characteristic variables, differentiating the characterized phenomena from the standard criticality \cite{williams2014quasicritical,fosque2021evidence}. On the other hand, all statistical physics relations required by perfect brain criticality are still adhered by these actual characteristic variables, distinguishing the brain from being non-critical \cite{williams2014quasicritical,fosque2021evidence}. 
 
 \begin{figure}
\begin{adjustwidth}{-0cm}{}
\centering
\includegraphics[width=1\hsize]{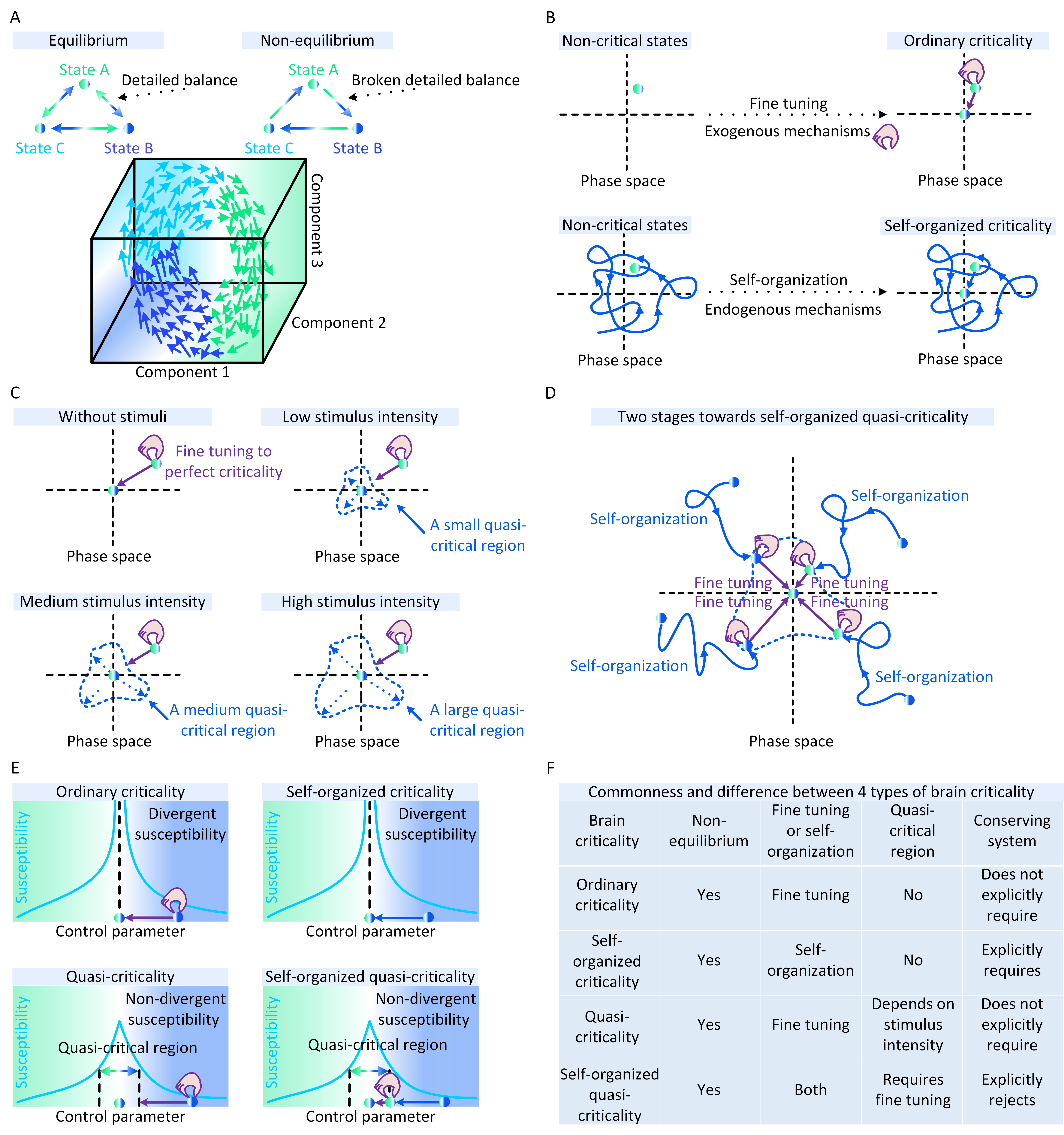}
\caption{Please see the caption on the next page.}
 \end{adjustwidth}
\end{figure}
\newpage
\addtocounter{figure}{-1}
\begin{figure}[t!]
\begin{adjustwidth}{-0cm}{}
    \caption[]{Conceptual illustrations of brain criticality. \textbf{A}, Difference between equilibrium and non-equilibrium dynamics in a three-state brain (upper parallel). Brain states are characterized by three system components. We illustrate an instance of non-equilibrium dynamics between these states (bottom parallel). \textbf{B}, Fine tuning with exogenous mechanisms (represented by animated hands) makes the brain evolve from a non-critical state (upper left) to the critical state (upper right). Endogenous mechanisms enable the brain to self-organize from a non-critical state (bottom left) to the critical state (bottom right). \textbf{C} Increasing stimulus intensity enlarges the quasi-critical region around the perfect critical point in a quasi-critical system. \textbf{D} The approaching process to a critical point in a self-organized quasi-critical system consists of two stages. In the first stage, the brain self-organizes from a non-critical state to a quasi-critical region based on certain endogenous mechanisms. In the second stage, additional exogenous mechanisms are necessary to fine tune the brain to the critical point. Otherwise, the brain just hovers within the quasi-critical region. \textbf{E} The difference between four types of brain criticality from the perspective of susceptibility. For standard brain criticality (e.g., ordinary criticality and self-organized criticality), susceptibility becomes divergent (i.e., infinite) at the critical point. For non-standard brain criticality (e.g., quasi-criticality and self-organized quasi-criticality), susceptibility is always non-divergent (i.e., finite). The quasi-critical region is defined as a set of all control parameters where susceptibility values are no less than a specific threshold (e.g., half-maximum value). \textbf{F} The commonness and difference between four types of brain criticality.
    }
     \end{adjustwidth}
\end{figure}
 
 For ordinary criticality, its non-standard form is referred to as quasi-criticality (qC) \cite{williams2014quasicritical,fosque2021evidence}. Diverse mechanisms can force the brain to depart from perfect ordinary criticality and exhibit quasi-critical neural dynamics, among which, stimulus derive may be the most common one \cite{williams2014quasicritical,fosque2021evidence}. In general, sufficiently strong stimulus drives can capture or even govern neural dynamics. Similar to the situation where external inputs suppress irregular neural dynamics \cite{molgedey1992suppressing}, the stimuli that are too strong may evoke intense but less changeable neural dynamics to make the brain depart from the perfect critical point \cite{williams2014quasicritical,fosque2021evidence}. Let us take the qC phenomenon introduced by \cite{williams2014quasicritical,fosque2021evidence} as an instance. Under specific conditions, the actual brain state may be close to a Widom line in the three-dimensional space defined by the stimulus intensity $\upsilon$, refractory period length $\tau$, and branching ratio $\kappa$ (i.e., the time-dependent average number of subsequent neural activities caused by a single neuron activation event \cite{haldeman2005critical}). The Widom line is a line of all the combinations of $\left(\upsilon,\tau,\kappa\right)$ where the susceptibility of neural dynamics is maximized \cite{williams2014quasicritical,fosque2021evidence}. The susceptibility is defined by $\lim_{x\rightarrow 0}\frac{\partial y}{\partial x}$, where $y$ is the neural dynamics state and $x$ denotes a factor that affects $y$. In general, one can understand susceptibility as the degree to which fluctuations in the state of each neuron can propagate to neighbored neurons \cite{williams2014quasicritical}. Being close to the Widom line suggests the existence of quasi-criticality in the brain. Moving along the Widom line as the stimulus intensity increases, the susceptibility of neural dynamics decreases, and the branching ratio at maximal susceptibility will decrease as well \cite{williams2014quasicritical,fosque2021evidence}. Significant deviations from the Widom line suggest non-criticality (i.e., the sub-criticality where neural dynamics is disordered and the super-criticality where neural dynamics is ordered \cite{williams2014quasicritical}). In \textbf{Fig. 1C}, we conceptually illustrate how stimuli imply qC in the brain. In \textbf{Fig. 2D}, the qC phenomenon in \cite{williams2014quasicritical,fosque2021evidence} is shown in details. 
 
 As for self-organized criticality (SOC), its non-standard form is defined according to statistical physics criteria. Perfect self-organized criticality only exists in conserved neural dynamics (e.g., see integrate-and-fire neurons analyzed by \cite{levina2007dynamical}), where system energy (i.e., neural activities) either conserves within the system and only dissipates at the system boundary, or dissipates inside the system (i.e., bulk dissipation) with a dissipation rate vanishing in the system size limit \cite{malcai2006dissipative}. Under more general conditions where neural dynamics is not conserved (e.g., see leaky integrate-and-fire neurons analyzed by \cite{millman2010self,levina2007dynamical,rubinov2011neurobiologically,stepp2015synaptic}, where neural dynamics dissipates within the system due to voltage leak), perfect self-organized criticality can be broken by any rate of bulk dissipation \cite{bonachela2009self,bonachela2010self,buendia2020self,de2015can}. Stronger bulk dissipation implies larger deviations from perfect self-organized criticality \cite{de2006self}. Consequently, the self-organization process of non-conserved neural dynamics only make the brain hover around the critical point. Any further closeness towards the critical point requires the fine tuning of order parameter by additional exogenous mechanisms, which is different from pure self-organized criticality \cite{bonachela2009self,bonachela2010self,buendia2020self,de2015can}. This non-conserved self-organization process is termed as self-organized quasi-criticality (SOqC) \cite{bonachela2009self}. Similar to SOC in conserved dynamics, neural plasticity mechanisms, such as spike-timing-dependent synaptic plasticity \cite{rubinov2011neurobiologically}, Hebbian rules \cite{de2006self}, short-term synaptic depression in conjunction with spike-dependent threshold increase \cite{girardi2021unified}, and inhibitory plasticity in conjunction with network topology \cite{ma2019cortical}, can serve as underlying self-organization mechanisms of SOqC. Because purely conserved neural dynamics is relatively rare in empirical data (e.g., neural dynamics is conserved for integrate-and-fire neurons \cite{levina2007dynamical} and leaky integrate-and-fire neurons whose pre-synaptic inputs are exactly equal to the sum of voltage leak and potential costs during neural spiking \cite{bonachela2010self}), we suggest that SOqC may be more common in the brain than SOC. In \textbf{Fig. 1D}, we present conceptual instances of the two-stage approaching process towards the critical point in the brain with SOqC.

  \subsubsection{Classification of brain criticality} The above discussion has presented a classification framework of brain criticality, i.e., ordinary criticality (OC), quasi-criticality (qC), self-organized criticality (SOC), and self-organized quasi-criticality (SOqC). In \textbf{Fig. 1E}, we compare between these four types of brain criticality in term of susceptibility. In general, susceptibility diverges at the critical point in a brain with standard criticality (e.g., OC and SOC) while it does not diverge in the quasi-critical region of a brain with non-standard criticality (e.g., qC and SOqC). In \textbf{Fig. 1F}, we summarize the commonness and difference between these four types of brain criticality discussed in our review. From a neuroscience perspective, a brain with critical neural dynamics is expected to be near the critical point and prepared for tremendous changes in cortical states during a short duration. This intriguing property coincides with the experimentally observed efficient transitions between cortical states (e.g., \cite{jercog2017up,holcman2006emergence,reimer2014pupil,lee2020state,cardin2019functional}) and, therefore, interests researchers for the potential existence of brain criticality. The importance of identifying brain criticality in neural dynamics is beyond brain criticality itself because it implies an opportunity to explain and predict brain function characteristics by various statistical physics theories built on non-equilibrium criticality. 

\subsection{Neural avalanches and their phases}
To identify potential non-equilibrium criticality in the brain, researchers actually characterize neural dynamics as a physical system with absorbing states and avalanche behaviors \cite{hinrichsen2000non,lubeck2004universal,larremore2012statistical}. In general, one need to consider the propagation of neural dynamics where neurons are either activated (``on" state) or silent (''off" state) \cite{dalla2019modeling}. An silent neuron may be activated with a probability defined by the number of activated pre-synaptic neurons and the coupling strength $\theta$ among neurons (e.g., neural correlation \cite{franke2016structures}). An activated neuron spontaneously becomes silent at a constant rate (e.g., after the refractory period \cite{squire2012fundamental,kinouchi2006optimal}). These definitions naturally support to distinguish between different phases of neural dynamics. Here we review two kinds of phase partition that are active in neuroscience.

\subsubsection{Absorbing versus active} The first group of phases are absorbing and active phases \cite{larremore2012statistical}. The absorbing phase refers to cases where couplings between neurons are weak and all neurons eventually become silent (neural dynamics vanishes). Once a neural dynamics process vanishes, it can not reappear by itself. The brain requires new drives (e.g., neurons activated spontaneously or by stimuli) to trigger new neural dynamics. The active phase, on the other hand, correspond to cases where ``on" state propagates among neurons with strong couplings, leading to stable self-sustained neural dynamics (e.g., non-zero time- and ensemble-averaged density of active neurons in the brain). In \textbf{Fig. 2A}, we show conceptual instances of neural avalanches, self-sustained neural dynamics, and vanished neural dynamics. Denoting $\rho\left(t\right)$ as the density of active neurons at moment $t$, we can simply represent the absorbing (Eq. (\ref{EQ1})) and active (Eq. (\ref{EQ2})) phases of a neural dynamics process triggered by an active neuron at moment $0$ as
\begin{align}
    \rho\left(t\right)=0,\;\exists t>0,\label{EQ1}\\
    \rho\left(t\right)>0,\;\forall t>0.\label{EQ2}
\end{align}

\subsubsection{Synchronous versus asynchronous} The second group of phases are synchronous and asynchronous phases \cite{di2018landau,fontenele2019criticality,girardi2021unified}. As their names suggest, these two phases correspond to the situations where synchronization emerges or disappears in neural activities, respectively. Synchronization refers to the cases where ``on" states appear in an oscillatory, although not strictly periodic, manner. To quantify its potential existence, we can measure the variability of neural dynamics using the coefficient of variation (CV) \cite{di2018landau,fontenele2019criticality,girardi2021unified} or the Kuramoto order parameter \cite{arenas2008synchronization,acebron2005Kuramoto}. CV can be defined from diverse perspectives, yet the most common definition is the ratio between the standard deviation and the mean of the inter-spike interval length \cite{di2018landau,fontenele2019criticality,girardi2021unified}. A higher value of CV implies the reduction of synchronization. For most neural dynamics data, an empirical choice of the CV threshold that separates between synchronous and asynchronous phases may be $\simeq 1$ \cite{fontenele2019criticality} or $\simeq\frac{3}{2}$ \cite{fontenele2019criticality}. The Kuramoto order parameter $\omega\in\left[0,1\right]$ measures the coherent degree of neural dynamics based on the Kuramoto model of oscillators (see \cite{arenas2008synchronization,acebron2005Kuramoto} for detailed definitions). Perfect synchronization emerges when $\omega=1$ and vanishes when $\omega=0$ \cite{arenas2008synchronization,acebron2005Kuramoto}.

\subsubsection{Critical point or quasi-critical region} The boundary between these two phases is the critical point, at which the brain is on the edge of exhibiting self-sustained (for absorbing and active phases) or synchronous (for synchronous and asynchronous phases) neural dynamics. Perturbations (e.g., the propagation of ``on" state among neurons) to the absorbing or asynchronous phase do not have characteristic lifetime and size. These perturbations, referred to as neural avalanches, are expected to exhibit power-law properties in their lifetime (time difference between the first and last activation of neurons in between complete quiescent epochs) and size (number of active neurons along with the excursion) distributions \cite{hinrichsen2000non,lubeck2004universal,larremore2012statistical,hesse2014self}. In general, the emergence of neural avalanches implies the slowing down of neural dynamics, i.e., the brain state recovery process towards the baseline state after fluctuations changes from fast (exponential) to slow (power-law) \cite{hesse2014self,cocchi2017criticality}. The dynamic stability of neural dynamics is limited by the slow recovery and, therefore, can not robustly counteract perturbations. Consequently, small perturbations initiated on the microscopic scale may still make the brain change sharply on the macroscopic scale \cite{hesse2014self,cocchi2017criticality}. In \textbf{Fig. 2B}, we conceptually illustrate how the recovery process slows down when the brain is close to the critical point or the quasi-critical region.

\subsection{General relations between neural avalanches and brain criticality}
The relation between neural avalanches and brain criticality is frequently neglected or misunderstood. Neural avalanche data alone is not sufficient to determine the concrete type of brain criticality (i.e., OC, qC, SOC, and SOqC) unless additional information about the mechanisms underlying neural avalanche emergence is provided (e.g., if neural dynamics is conserved or self-organizing). To explore a concrete type of brain criticality, researchers need to explicitly present its definition depending on different control parameters (e.g., the balance between excitatory and inhibitory neurons in CROS models \cite{poil2012critical,hardstone2014neuronal}) and order parameters (e.g., active neuron density and synchronous degree \cite{dalla2019modeling}). A brain criticality hypothesis without strict definitions of control and order parameters is not informative \cite{cocchi2017criticality,girardi2021brain}. To present conceptual instances, we illustrate four possible critical phenomena in \textbf{Fig. 2}, each of which corresponds to a concrete brain criticality type. 

\subsubsection{Instance of ordinary criticality} To produce ordinary criticality (OC), we can control neural dynamics and manipulate $\langle\theta\rangle$, the expectation of coupling strength $\theta$ among all neurons (e.g., averaged neural correlation), by some top-down and global biological effects. These effects, for instance, may be anesthesia effects (e.g., by ketamine-xylazine \cite{ribeiro2010spike} and isoflurane \cite{hahn2017spontaneous}) or sleep restoration effects \cite{meisel2013fading}. We use the Kuramoto order parameter $\omega$ \cite{arenas2008synchronization,acebron2005Kuramoto} as the order parameter to define synchronous and asynchronous phases \cite{di2018landau,fontenele2019criticality}. As $\langle\theta\rangle$ increases, we may see transitions from asynchronous to synchronous phase in some situations (see a similar instance in \cite{villegas2014frustrated}). One can see \textbf{Fig. 2C} for conceptual illustrations.

\begin{figure}
\begin{adjustwidth}{-0cm}{}
\centering
\includegraphics[width=1\hsize]{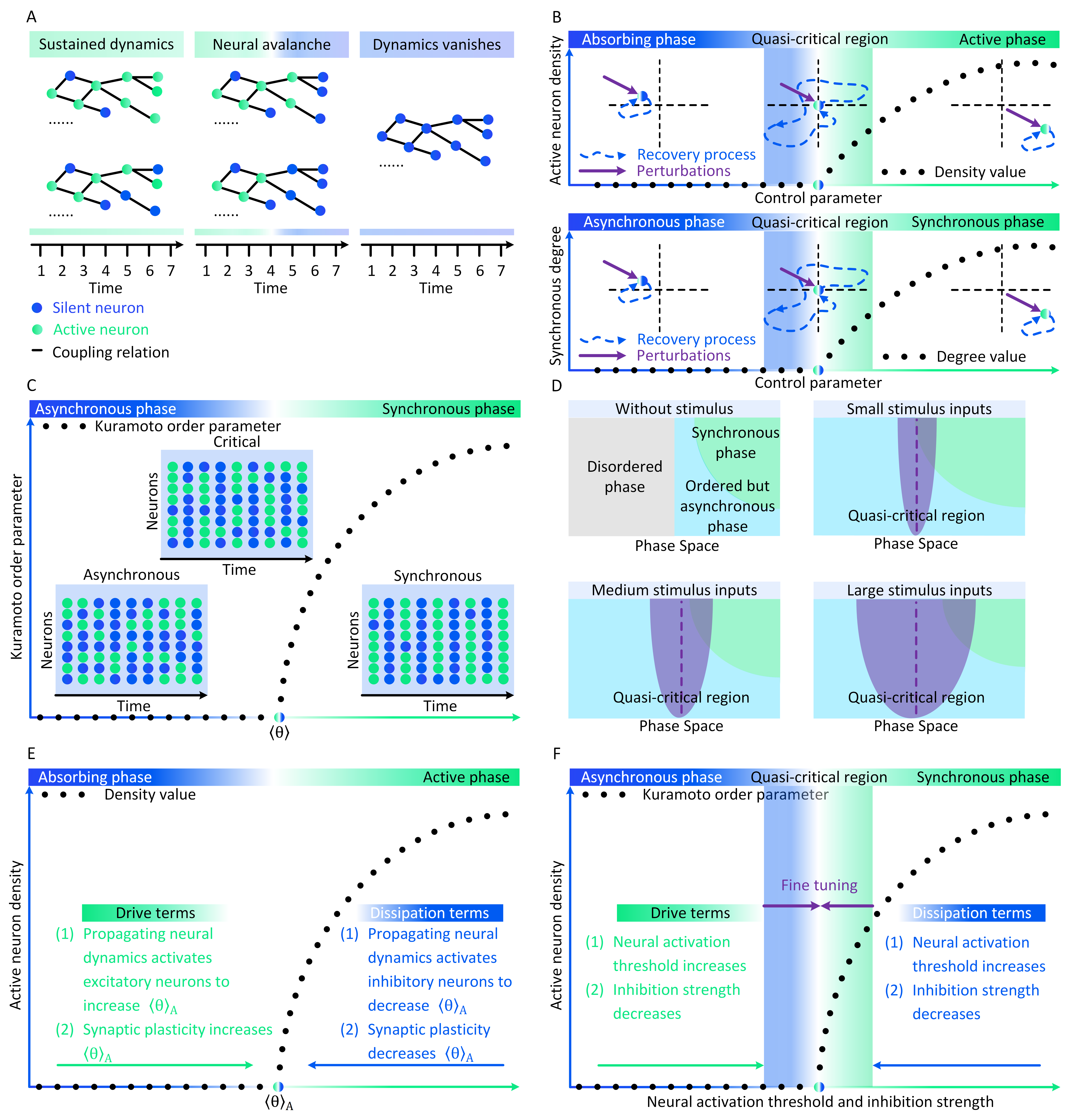}
\caption{Please see the caption on the next page.}
\end{adjustwidth}
\end{figure}
\newpage
\addtocounter{figure}{-1}
\begin{figure}[t!]
\begin{adjustwidth}{-0cm}{}
    \caption[]{Conceptual illustrations of the relations between neural avalanches and brain criticality. \textbf{A}, Instances of neural avalanche, self-sustained neural dynamics, and vanished neural dynamics. \textbf{B}, The recovery processes of brain states after the same perturbation in the space of absorbing and active phases (upper parallel) and the space of synchronous and asynchronous phases (bottom parallel). The recovery processes after perturbations are relatively fast when the brain is far from the critical point or the quasi-critical region, These recovery processes slow down when the brain is close to the critical point or the quasi-critical region. \textbf{C} The conceptual illustrations of neural dynamics when the brain state is asynchronous, synchronous, or at ordinary criticality. \textbf{D} Without stimuli, there initially exist disordered (gray), ordered but asynchronous (light blue), synchronous (green) phases in the phase space of the brain. Stimulus inputs imply quasi-criticality in the brain. An increasing stimulus intensity enlarges the quasi-critical region (purple) around the Widom line (purple dashed line). \textbf{E} The conceptual illustrations of how endogenous mechanisms in conserved neural dynamics can function as drive or dissipation terms to create self-organized criticality between absorbing and active phases in the brain. \textbf{F} In the self-organized quasi-critical brain, endogenous mechanisms in non-conserved neural dynamics only support the self-organization towards a quasi-critical region between asynchronous and synchronous phases. Extra exogenous mechanisms are required to fine tune the brain towards the critical point.}
    \end{adjustwidth}
\end{figure}

\subsubsection{Instance of quasi-criticality} To produce quasi-criticality (qC), we can manipulate refractory period length $\tau$, branching ratio $\kappa$, and stimulus intensity $\upsilon$ as control parameters (e.g., control $\tau$ and $\kappa$ by pharmacological perfusion or ionic concentration adjustment \cite{shew2011information,chiappalone2003networks}). There exist a disordered phase (sub-critical), an ordered but asynchronous phase (super-critical), and a synchronous (quasi-periodic) phase in the space of $\left(\upsilon,\tau,\kappa\right)$ \cite{williams2014quasicritical,fosque2021evidence}. These phases can be characterized by specific order parameters related to synchronization. As $\upsilon$ increases, a qC phenomenon emerges in the space, where the quasi-critical region is defined by all combinations of $\left(\upsilon,\tau,\kappa\right)$ whose susceptibility values are at least half-maximum. Cross-over behaviours (i.e., a generalization of phase transition with finite susceptibility) emerge when the quasi-critical region has overlaps with at least two phases \cite{williams2014quasicritical,fosque2021evidence}. In \textbf{Fig. 2D}, we show this qC phenomenon in details.

\subsubsection{Instance of self-organized criticality} To study self-organized criticality (SOC), we consider the conserved neural dynamics generated by integrate-and-fire neurons \cite{levina2007dynamical}. The order parameter is active neuron density $\rho$, whose dynamics is controlled by parameter $\langle\theta\rangle_{A}$, the averaged coupling strength $\theta$ between activated neurons and their post-synaptic neurons (here $A$ denotes the set of activated neurons). In specific cases, the considered neural dynamics may self-organize to the critical point under the joint effects of excitatory and inhibitory neurons, neural spiking processes (activation and silence), as well as neural plasticity. In \textbf{Fig. 2E}, we conceptually illustrate a case where these endogenous mechanisms enable the brain to self-organize to the criticality between absorbing and active phases.

\subsubsection{Instance of self-organized quasi-criticality} To analyze self-organized quasi-criticality (SOqC), we consider the non-conserved neural dynamics affected by two homeostatic adaptation processes, i.e., the short-term depression of inhibition and the spike-dependent threshold increase. These processes are controlled by $\widehat{y}$, the maximum inhibitory coupling strength, as well as $\tau_{x}$ and $\tau_{y}$, the decay time scales of neural activation threshold increase and synaptic depression. These control parameters affect neural activation threshold $x$ and inhibition strength $y$ to shape neural dynamics states (e.g., the active neuron density $\rho$). With appropriate $x$, $y$, and $\rho$, neural avalanches with power-law behaviours will occur to indicate the criticality between an asynchronous phase (stochastic oscillations) and a synchronous phase (periodic oscillations). According to \cite{girardi2021unified}, $x$ and $\rho$ self-organize to their appropriate values through quasi-critical fluctuations under biologically reasonable conditions (i.e., $\tau_{x}\gg 1$) while $y$ hovers around the expected value. Additional fine tuning of $y$ based on exogenous mechanisms are necessary to place neural dynamics at the perfect criticality. Meanwhile, synaptic homeostasis is discovered as constantly canceled by the variation of the activation threshold, impeding neural dynamics from self-organizing to perfect criticality. In \textbf{Fig. 2F}, we conceptually illustrate the defined SOqC phenomenon in a similar manner of \textbf{Fig. 2D} and \textbf{Fig. 2E}. As for the precise description of quasi-critical fluctuations, one can see \cite{girardi2021unified} for details.

To this point, we have conceptually introduced the phenomenological properties of brain criticality. To verify the hypothetical brain criticality, one need to learn about analytic brain criticality theories and the properties of neural avalanche predicted by them. Below, we present accessible expositions of these theoretical foundations.

\section{Brain criticality: physical theories}
\subsection{Mean field and stochastic field theories of brain criticality}
One of the main challenges faced by neuroscientists in studying ordinary criticality (OC), quasi-criticality (qC), self-organized criticality (SOC), and self-organized quasi-criticality (SOqC) is how to understand their theoretical relations \cite{girardi2021brain}. Overcoming this challenge is crucial for understanding why we can verify the existence of different types of brain criticality with certain theoretical tools. To present a concise and thorough review, we first focus on brain criticality between absorbing and active phases, where we generalize the idea in \cite{bonachela2009self,buendia2020feedback} to present a possible framework for unification. 

\subsubsection{Langevin formulation of ordinary criticality} In general, brain criticality in the space of absorbing and active phases are related to directed percolation \cite{dalla2019modeling}, a universality class of continuous phase transitions into absorbing states \cite{hinrichsen2000non,lubeck2004universal}. Here a universality class can be understood as the set of all systems with the same scaling properties \cite{hinrichsen2000non,lubeck2004universal,sethna2001crackling}. Directed percolation theory initially covers OC phenomena \cite{hinrichsen2000non,lubeck2004universal}. Let us begin with a variant of the classic Reggeon field theory, the simplest description of absorbing phase transitions \cite{henkel2008non}. The Langevin equation of the activity neuron field $\rho\left(\vec{x},t\right)$ is defined as
\begin{align}
\frac{\partial}{\partial t} \rho\left(\vec{x},t\right)&=\left(a+b\nu\left(\vec{x},t\right)\right)\rho\left(\vec{x},t\right)-c\rho^2\left(\vec{x},t\right)+d\nabla^2\rho\left(\vec{x},t\right)+e\sqrt{\rho\left(\vec{x},t\right)}\sigma\left(\vec{x},t\right),\label{EQ3}\\
\frac{\partial}{\partial t} \nu\left(\vec{x},t\right)&=\nabla^2\nu\left(\vec{x},t\right)+f\left(\vec{x},t\right)-g\left(\vec{x},t\right)\rho\left(\vec{x},t\right),\label{EQ4}
\end{align}
where $\vec{x}$ represents spatial coordinates, $a\in\mathbb{R}$, $b\in\left(0,\infty\right)$, $c\in\left(0,\infty\right)$, $d\in\mathbb{R}$ is the diffusion factor, and $e\in\mathbb{R}$ is the noise factor. Function $\sigma\left(\cdot,\cdot\right)$ defines a zero-mean Gaussian noise with a spatio-temporal correlation $\langle \rho\left(\vec{x},t\right)\rho\left(\vec{x}^{\prime},t^{\prime}\right)\rangle=\delta\left(\vec{x}-\vec{x}^{\prime}\right)\delta\left(t-t^{\prime}\right)$, where $\delta\left(\cdot\right)$ is the delta function. In general, $\sigma\left(\cdot,\cdot\right)$ reflects the collective fluctuations in neural activities that vanish in the absorbing phase $\rho\left(\vec{x},t\right)=0$ under the effects of factor $\sqrt{\rho\left(\vec{x},t\right)}$. The term $\nabla^2\rho\left(\vec{x},t\right)$ reflects the propagation of neural dynamics. The function $\nu\left(\vec{x},t\right)$ defines the energy (i.e., membrane potential) that propagates according to $\nabla^2\nu\left(\vec{x},t\right)$, increases with external drives $f\left(\vec{x},t\right)$, and decreases with bulk dissipation $g\left(\vec{x},t\right)$. Please note that $\rho\left(\vec{x},t\right)\geq 0$ and $\nu\left(\vec{x},t\right)\geq 0$ always hold. The initial active neuron density and energy are assumed as non-zero. It is clear that $a+b\nu\left(\vec{x},t\right)<0$ makes the neural dynamics eventually vanish (i.e., absorbing phase) while $a+b\nu\left(\vec{x},t\right)>0$ does not  (i.e., active phase). Therefore, we can fine tune the control parameter $\nu\left(\vec{x},t\right)$ to make the brain exhibit OC dynamics at $a+b\nu_{c}\left(\vec{x},t\right)=0$, a critical point defined by $\nu_{c}$. The fine tuning relies on manipulating $f\left(\vec{x},t\right)$ and $g\left(\vec{x},t\right)$ by exogenous mechanisms.

\subsubsection{Langevin formulation of quasi-criticality} Then we turn to analyzing qC, whose mean field approximation is initially derived based on the cortical branching model \cite{williams2014quasicritical,fosque2021evidence}. A cortical branching model with no stimulus input belongs to directed percolation universality class according to the Janssen-Grassberger conjecture \cite{williams2014quasicritical}. Non-zero stimulus inputs make the cortical branching model depart from directed percolation universality class to create qC \cite{williams2014quasicritical}. Nevertheless, the above mean field theory is defined in the space of synchronous and asynchronous phases. To derive a qC phenomenon between absorbing and active phases, we can provisionally analyze a mean field approximation of Eqs. (\ref{EQ3}-\ref{EQ4})
\begin{align}
\frac{\partial}{\partial t} \rho\left(\vec{x},t\right)&=\left(a+b\nu\left(\vec{x},t\right)\right)\rho\left(\vec{x},t\right)-c\rho^2\left(\vec{x},t\right),\label{EQ5}\\
\frac{\partial}{\partial t} \nu\left(\vec{x},t\right)&=f\left(\vec{x},t\right)-g\left(\vec{x},t\right)\rho\left(\vec{x},t\right),\label{EQ6}
\end{align}
where $\nabla^2\rho\left(\vec{x},t\right)$, $\nabla^2\nu\left(\vec{x},t\right)$, and $\sigma\left(\vec{x},t\right)$ in Eqs. (\ref{EQ3}-\ref{EQ4}) are neglected under the mean field assumption. We consider the cases where stimulus inputs vanish, i.e., $f\left(\vec{x},t\right)\equiv 0$. The critical point between active and absorbing phase becomes $\nu_{c}=-\frac{a}{b}$. The steady state solutions of Eqs. (\ref{EQ5}-\ref{EQ6}) are
\begin{align}
\rho\left(\vec{x},t\right)&=0,\label{EQ7}\\
\nu\left(\vec{x},t\right)&=r\in\left(0,\infty\right),\label{EQ8}
\end{align}
respectively. Therefore, OC is one of the steady states of neural dynamics when there is no stimulus. In the cases where stimulus inputs become increasingly strong, there exists no steady state solution of Eqs. (\ref{EQ5}-\ref{EQ6}) unless $\frac{f\left(\vec{x},t\right)}{g\left(\vec{x},t\right)}\rightarrow r\in\left(0,\infty\right)$. If $\frac{f\left(\vec{x},t\right)}{g\left(\vec{x},t\right)}\rightarrow r\in\left(0,\infty\right)$ holds, we can derive
\begin{align}
\rho\left(\vec{x},t\right)&=\frac{f\left(\vec{x},t\right)}{g\left(\vec{x},t\right)}\rightarrow r,\label{EQ9}\\
\nu\left(\vec{x},t\right)&=\frac{1}{b}\left(c\frac{f\left(\vec{x},t\right)}{g\left(\vec{x},t\right)}-a\right)\rightarrow\frac{1}{b}\left(cr-a\right).\label{EQ10}
\end{align}
Because the critical point $\nu_{c}=-\frac{a}{b}$ is not necessarily a steady state, it can be disturbed by diverse factors (e.g., by stimuli). Unless there exist certain ideal exogenous mechanisms that persistently enlarge $g\left(\vec{x},t\right)$ whenever $f\left(\vec{x},t\right)$ increases, the fine tuning of neural dynamics can not cancel the effects of $f\left(\vec{x},t\right)$. Consequently, the fine tuning process may only enable the brain to reach a quasi-critical region where the susceptibility of neural dynamics is relatively large. The initial OC vanishes and is replaced by qC.

\subsubsection{Langevin formulation of self-organized criticality} Although SOC is treated as a rather isolated concept after its first discovery in statistical physics \cite{bak1987self}, subsequent analyses demonstrate SOC as relevant with ordinary continuous phase transitions into infinitely many absorbing states \cite{narayan1994avalanches,sornette1995mapping,dickman1998self,dickman2000paths}. Specifically, SOC models can be subdivided into two families, which we refer to as external dynamics family (e.g., Bak-Sneppen model \cite{bak1993punctuated}) and conserved field family (e.g., sandpile models such as Manna model \cite{manna1991two} and Bak-Tang-
Wiesenfeld model \cite{bak1987self}). The second family, being the main theoretical source of studying SOC in neural dynamics, corresponds to absorbing-state transitions since it can represent any system with conserved local dynamics and continuous transitions to absorbing states \cite{dickman2000paths,lubeck2004universal}. Although the universality class of the second family should be precisely referred to as conserved directed percolation, the explicit behaviours (e.g., avalanche exponents and scaling relations) of conserved directed percolation are similar to those of directed percolation in high-dimensional systems (e.g., neural dynamics) \cite{buendia2020feedback,bonachela2008confirming,bonachela2009self}. Therefore, SOC and OC share some identification criteria in practice. To understand the connections between SOC and OC more precisely, we can consider the cases where $f\left(\vec{x},t\right)\rightarrow 0$ such that $\frac{f\left(\vec{x},t\right)}{g\left(\vec{x},t\right)}\rightarrow 0$ (i.e., infinite separation of timescales). The steady state solutions of Eqs. (\ref{EQ5}-\ref{EQ6}) become
\begin{align}
\rho\left(\vec{x},t\right)&=\frac{f\left(\vec{x},t\right)}{g\left(\vec{x},t\right)}\rightarrow 0,\label{EQ11}\\
\nu\left(\vec{x},t\right)&=\frac{1}{b}\left(c\frac{f\left(\vec{x},t\right)}{g\left(\vec{x},t\right)}-a\right)\rightarrow \nu_{c},\label{EQ12}
\end{align}
respectively. Self-organization properties are reflected by the following processes: if the brain is in the absorbing phase because neural dynamics vanishes, i.e, $\rho\left(\vec{x},t\right)\rightarrow 0$, Eq. (\ref{EQ6}) becomes $\frac{\partial}{\partial t} \nu\left(\vec{x},t\right)=f\left(\vec{x},t\right)$ to shift the brain towards the active phase; if the brain is in the active phase, Eq. (\ref{EQ6}) becomes $\frac{\partial}{\partial t} \nu\left(\vec{x},t\right)\simeq -g\left(\vec{x},t\right)\rho\left(\vec{x},t\right)$ to reduce neural dynamics since $f\left(\vec{x},t\right)\ll g\left(\vec{x},t\right)$. These feedback-control loops drive the brain to the critical point. One may be curious about why energy conservation, i.e., $g\left(\vec{x},t\right)\rightarrow 0$, is necessary for SOC since the above derivations seem to be independent of $g\left(\vec{x},t\right)\rightarrow 0$. Later we show that the absence of $g\left(\vec{x},t\right)\rightarrow 0$ in Eq. (\ref{EQ14}) makes the active phase no longer exist. In other words, the non-conserved energy implies a kind of continuous phase transition that does not belong to conserved directed percolation or directed percolation when the infinite separation of timescales is satisfied. Therefore, energy conservation is necessary for SOC.

\subsubsection{Langevin formulation of self-organized quasi-criticality} As for SOqC, non-zero bulk dissipation breaks the conservation law to generate non-Markovian components in neural dynamics \cite{bonachela2009self,buendia2020feedback}. In the ideal cases where the drive terms (e.g., stimulus inputs) of a sufficiently large neural dynamics system occur at an arbitrarily slow timescale (i.e., only occur in the interval between neural avalanches), the brain exhibits pure dynamical percolation behaviours \cite{buendia2020feedback}. To understand this property, let us consider a variant of Eqs. (\ref{EQ5}-\ref{EQ6}) where the dissipation term $g\left(\vec{x},t\right)$ is non-negligible
\begin{align}
\frac{\partial}{\partial t} \rho\left(\vec{x},t\right)&=\left(a+b\nu\left(\vec{x},t\right)\right)\rho\left(\vec{x},t\right)-c\rho^2\left(\vec{x},t\right),\label{EQ13}\\
\frac{\partial}{\partial t} \nu\left(\vec{x},t\right)&=-g\left(\vec{x},t\right)\rho\left(\vec{x},t\right).\label{EQ14}
\end{align}
By integrating Eq. (\ref{EQ14}) and plugging the integral into Eq. (\ref{EQ13}), we can derive
\begin{align}
\frac{\partial}{\partial t} \rho\left(\vec{x},t\right)&=\left(a+b\nu\left(\vec{x},0\right)\right)\rho\left(\vec{x},t\right)-c\rho^2\left(\vec{x},t\right)-b\rho\left(\vec{x},t\right)\int_{0}^{t}g\left(\vec{x},t\right)\rho\left(\vec{x},\tau\right)\mathsf{d}\tau,\label{EQ15}
\end{align}
The non-Markovian term $-b\rho\left(\vec{x},t\right)\int_{0}^{t}g\left(\vec{x},t\right)\rho\left(\vec{x},\tau\right)\mathsf{d}\tau$ in Eq. (\ref{EQ15}) makes the regions already visited by neural dynamics become more unlikely to be activated \cite{bonachela2009self,buendia2020feedback}. Therefore, the pure self-sustained active phase vanishes and is replaced by a spreading phase, where local perturbations can transiently propagate across the whole system without reaching a self-sustained state, and a non-spreading phase, where local perturbations can never span the entire system \cite{bonachela2009self,buendia2020feedback}. The phase transition and corresponding critical point $\nu_{d}>\nu_{c}$ between spreading and non-spreading phases belong to the universality class of dynamical percolation rather than conserved directed percolation \cite{bonachela2009self,buendia2020feedback}. The initial neural dynamics can be created by random shifts at moment $0$ \cite{bonachela2009self,buendia2020feedback}
\begin{align}
 \rho\left(\vec{x}^{*},0\right)&\rightarrow \epsilon,\label{EQ16}\\
 \nu\left(\vec{x}^{*},0\right)&\rightarrow \nu\left(\vec{x}^{*},0\right)+h\left(\vec{x}^{*},0\right),\label{EQ17}
\end{align}
where $\vec{x}^{*}$ is a randomly selected coordinate, and function $h\left(\cdot,\cdot\right)$ is a driving function of energy at moment $0$. Every time a neural avalanche occurs after random shifts, the strong dissipation term $g\left(\vec{x},t\right)$ pushes the brain towards the sub-critical phase. Consequently, the brain can not exactly self-organize to the perfect criticality. Instead, the brain just hovers around the critical point $\nu_{d}$ to form a quasi-critical region, exhibiting finite fluctuations to the both sides of $\nu_{d}$. In the more realistic cases where the drive terms do not necessarily occur at an arbitrarily slow timescale (i.e., can occur at an arbitrary moment), however, neural dynamics may be phenomenology controlled by conserved directed percolation transitions and hover around the critical point. Let us add a drive term in Eq. (\ref{EQ14}) 
\begin{align}
\frac{\partial}{\partial t} \nu\left(\vec{x},t\right)&=f\left(\vec{x},t\right)-g\left(\vec{x},t\right)\rho\left(\vec{x},t\right).\label{EQ18}
\end{align}
Then Eq. (\ref{EQ15}) becomes 
\begin{align}
\frac{\partial}{\partial t} \rho\left(\vec{x},t\right)&=\left(a+b\nu\left(\vec{x},0\right)\right)\rho\left(\vec{x},t\right)-c\rho^2\left(\vec{x},t\right)-b\rho\left(\vec{x},t\right)\int_{0}^{t}\left(f\left(\vec{x},\tau\right)-g\left(\vec{x},\tau\right)\rho\left(\vec{x},\tau\right)\right)\mathsf{d}\tau,\label{EQ19}
\end{align}
If we can ideally fine tune the drive term $f\left(\vec{x},t\right)$ to ensure that $\frac{f\left(\vec{x},t\right)}{g\left(\vec{x},t\right)}\rightarrow r\in\left(0,\infty\right)$, the steady state solutions of Eqs. (\ref{EQ18}-\ref{EQ19}) are
\begin{align}
\rho\left(\vec{x},t\right)&=\frac{f\left(\vec{x},t\right)}{g\left(\vec{x},t\right)}\rightarrow r,\label{EQ20}\\
\nu\left(\vec{x},0\right)&=\frac{1}{b}\left(c\frac{f\left(\vec{x},t\right)}{g\left(\vec{x},t\right)}-a\right)\rightarrow \frac{1}{b}\left(cr-a\right),\label{EQ21}\\
\nu\left(\vec{x},t\right)&=\nu\left(\vec{x},0\right)+\int_{0}^{t}\left(f\left(\vec{x},\tau\right)-g\left(\vec{x},\tau\right)\rho\left(\vec{x},\tau\right)\right)\mathsf{d}\tau\rightarrow \frac{1}{b}\left(cr-a\right)\label{EQ22}.
\end{align}
Eqs. (\ref{EQ20}-\ref{EQ22}) correspond to a steady state of the brain with $\rho\left(\vec{x},t\right)\rightarrow r$ and conserved energy, which is similar to SOC. Therefore, the brain may self-organize to a quasi-critical region around $\nu_{c}$, the critical point of SOC. Reaching the critical point requires ideal fine tuning. These emerged conserved-directed-percolation behaviours enable scientists to recognize SOqC in a similar manner of SOC in practice (i.e., when stimulus inputs can occur at any moment) \cite{bonachela2009self,buendia2020feedback}. 

\subsubsection{Summary of theoretical relations} Taken together, neuroscientists can approximately verify the existence of brain criticality in the space of absorbing and active phases with specific tools coming from directed percolation theory. This is because OC, qC, SOC, and SOqC exhibit or approximately exhibit directed percolation behaviours under certain conditions. The verification may be inaccurate since the approximation holds conditionally. As for the brain criticality between asynchronous and synchronous phases, however, the universality class properties become rather elusive because an analytic and complete theory of synchronous phase transitions in the brain remains absent yet (see \cite{di2018landau,buendia2021hybrid} for early attempts). Although some behaviours of absorbing phase transitions can be observed in synchronous phase transitions (e.g., see \cite{di2018landau,girardi2021unified,buendia2021hybrid,fontenele2019criticality}), there also exist numerous differences between them (e.g., see \cite{fontenele2019criticality,buendia2021hybrid,girardi2021unified}). As suggested by \cite{dalla2019modeling}, it remains elusive if directed percolation properties are applicable, at least conditionally applicable, to analyzing synchronous phase transitions. More explorations are necessary in the future. 

There are numerous properties of brain criticality predicted by directed percolation theory, among which, neural avalanche exponents (the power-law exponents of lifetime and size distributions), scaling relation, universal collapse shape, and slow decay of auto-correlation are applicable in both analytic derivations and statistical estimations from empirical data. These properties are our main focuses. For convenience, we summarize important glossaries and symbol conventions before we discuss theoretical details (\textbf{Table 2}).

\begin{table}[!t]
\begin{adjustwidth}{-0cm}{}
\caption{Glossaries and symbol conventions. Please note that \textbf{Table 2} mainly contains important glossaries with fixed symbol definitions. There are many symbols uncovered by \textbf{Table 2} since they are only used for mathematical derivations.}
\label{T2}
\centering
\begin{tabular}{cc}
\hline
  Variable & Meaning \\
\hline
$T$  &  The lifetime of the neural avalanche \\
$S$  &  The size of the neural avalanche \\
$A$  &  The area of the neural avalanche \\
$\langle S\left(T\right)\rangle$ & The averaged size of neural avalanches with lifetime $T$\\
$\langle S\left(t\mid T\right)\rangle$  &  The averaged time-dependent avalanche size at moment $t$ during neural avalanches with the lifetime $T$ \\
 $\mathcal{P}_{T}\left(t\right)$  &  The probability distribution of neural avalanche lifetime \\
   $\mathcal{P}_{S}\left(s\right)$  &  The probability distribution of neural avalanche size \\
      $\alpha$  &  Power-law exponent of the neural avalanche lifetime distribution $\mathcal{P}_{T}\left(t\right)\propto t^{-\alpha}$ \\
      $\beta$  &  Power-law exponent of the neural avalanche size distribution $\mathcal{P}_{S}\left(s\right)\propto s^{-\beta}$ \\
      $\gamma$  &  Power exponent of the neural avalanche area $A\propto T^{\gamma}$ \\
      $\mathcal{H}\left(\cdot\right)$  &   Universal scaling function \\
      $\operatorname{Cov}\left(\cdot,\cdot\right)$  &   Auto-correlation function \\
       $\chi$  &  Power-law decay rate of auto-correlation \\
      $\xi$  &  Exponential decay rate of auto-correlation \\
\hline
\end{tabular}
\end{adjustwidth}
\end{table}

\subsection{Neural avalanche exponents}
As we have mentioned above, neural avalanches are expected to exhibit power-law properties in their lifetime and size distributions when the brain is at the critical point \cite{hinrichsen2000non,lubeck2004universal,larremore2012statistical}. Therefore, it is pivotal to confirm the detailed values of neural avalanche exponents. To analytically derive these exponents, one can consider critical branching process \cite{garcia1993branching,harris1963theory,otter1949multiplicative,di2017simple,gros2010complex}, neural field theory \cite{robinson2021neural}, mean-field Abelian sandpile models \cite{janowsky1993exact,lee2004branching}, and avalanches in networks \cite{larremore2012statistical}. The key idea to derive neural avalanche exponents shared by these existing theories is to confirm the explicit forms of $\mathcal{P}_{T}\left(t\right)$ and $\mathcal{P}_{S}\left(s\right)$, the probability distributions of the lifetime and size of neural avalanches, under ideal conditions (e.g., when the maximum lifetime and size are unlimited and can be infinitely large). In real cases where lifetime and size are restricted because the brain is a finite system, slight deviations from idea values may be observed but theoretical derivations of neural avalanche exponents principally hold. 

To present accessible expositions, we consider a critical branching process in Eqs. (\ref{EQ23}-\ref{EQ34}) to describe related backgrounds. More importantly, we present a novel and simple idea to calculate target exponents in the context of neuroscience in \textbf{Box 1}. Abstractly, one can define $\mathcal{P}\left(n,t\right)$ as the probability for an active neuron at moment $t$ to activate $n$ post-synaptic neurons subsequently and define $\mathcal{Z}\left(n,t\right)$ as the probability of finding $n$ active neurons at moment $t$. Meanwhile, one denote 
\begin{align}
    \mathcal{F}\left(x,t\right)=\sum_{n=0}^{\infty}\mathcal{P}\left(n,t\right)x^{n},\label{EQ23}\\
    \mathcal{G}\left(x,t\right)=\sum_{n=0}^{\infty}\mathcal{Z}\left(n,t\right)x^{n}\label{EQ24}
\end{align}
as the corresponding generating functions \cite{fristedt2013modern,rao2006probability}. Then, one can readily see the recursion relation
\begin{align}
    \mathcal{G}\left(x,t\right)&=\sum_{n=0}^{\infty}\mathcal{Z}\left(n,t-\delta t\right)\mathcal{F}\left(x,t-\delta t\right)^{n},\label{EQ25}\\
    &=\mathcal{G}\left(\mathcal{F}\left(x,t-\delta t\right),t-\delta t\right),\label{EQ26}
\end{align}
where $\delta t$ denotes the minimum time step. Eq. (\ref{EQ26}) implies that branching processes are Markovian. Similarly, one can measure the expectations
\begin{align}
    \mu\left(t\right)&=\frac{\partial}{\partial x}\mathcal{F}\left(x,t\right)\Big\vert_{x=1}, \label{EQ27}\\
    \phi\left(t\right)&=\frac{\partial}{\partial x}\mathcal{G}\left(x,t\right)\Big\vert_{x=1} \label{EQ28}
\end{align}
to derive another recursion relation
\begin{align}
    \phi\left(t\right)&=\frac{\partial}{\partial x}\mathcal{F}\left(x,t-1\right)\Big\vert_{x=1}\frac{\partial}{\partial x}\mathcal{G}\left(x,t-1\right)\Big\vert_{x=1}, \label{EQ29}\\
    &=\phi\left(t-\delta\right)\mu\left(t-\delta\right), \label{EQ30}\\
    &=\prod_{\tau=0}^{t-\delta}\mu\left(\tau\right).\label{EQ31}
\end{align}
Note that Eq. (\ref{EQ31}) is derived from the fact that $\phi\left(0\right)=1$ (one neuron is activated at moment $0$ to trigger neural avalanches). Please see \cite{markovic2014power} for more explanations of Eqs. (\ref{EQ25}-\ref{EQ31}). Assuming that $\phi\left(t\right)$ scales as $\exp\left(\lambda t\right)$ for large $t$, we know that $\phi\left(t\right)$ converges to $0$ given a negative Lyapunov exponent $\lambda$ (the branching process is sub-critical \cite{garcia1993branching,harris1963theory,otter1949multiplicative,di2017simple,gros2010complex}) and diverges with a positive Lyapunov exponent $\lambda$ (the branching process is super-critical \cite{garcia1993branching,harris1963theory,otter1949multiplicative,di2017simple,gros2010complex}). Here $\lambda$ can be defined according to Eq. (\ref{EQ32})
\begin{align}
    \lambda&=\lim_{t\rightarrow \infty}\ln \left(\frac{1}{t}\phi\left(t\right)\right)=\lim_{t\rightarrow \infty} \frac{1}{t}\sum_{\tau=0}^{t-\delta}\ln\left(\mu\left(\tau\right)\right)\label{EQ32}.
\end{align}
If the branching process is homogeneous, namely $\mathcal{P}\left(n,t\right)=\mathcal{P}\left(n\right)$, $\mathcal{Z}\left(n,t\right)=\mathcal{Z}\left(n\right)$, $\mu\left(\tau\right)=\mu$, and $\phi\left(\tau\right)=\phi$ for every moment $\tau$, then $\mu=1$ is the condition for the branching process to be critical. To relate these results with neural avalanches, one only need to consider the avalanche size $S=\sum_{t}z\left(t\right)$, where $z\left(t\right)\sim \mathcal{Z}$ denotes the number of active neurons at moment $t$, and the avalanche life time $T=\min\{t\mid z\left(t\right)>0\;\text{and}\;z\left(t+\delta t\right)=0\}$. It has been analytically proved that in terms of fixed environments and a Poisson generating function $\mathcal{F}$ one can derive \cite{otter1949multiplicative}
\begin{align}
    \mathcal{P}_{S}\left(s\right)&\sim s^{-3/2}\mu^{s-1}\exp\left[s\left(1-\mu\right)\right]\label{EQ33},\\
    \mathcal{P}_{T}\left(t\right)&\sim t^{-2}\mu^{t-1}\exp\left[t\left(1-\mu\right)\right]\label{EQ34}.
\end{align}
In the case with $\mu=1$, one can obtain $\mathcal{P}_{S}\left(s\right)\sim s^{-3/2}$ and $\mathcal{P}_{T}\left(t\right)\sim t^{-2}$, the power-law distributions of neural avalanche size and neural avalanche life time \cite{garcia1993branching,harris1963theory,otter1949multiplicative,di2017simple,gros2010complex,robinson2021neural,janowsky1993exact,lee2004branching,larremore2012statistical,lombardi2017balance,jung2020avalanche}, from Eqs. (\ref{EQ33}-\ref{EQ34}). 

The derivations of avalanche exponents $\alpha=2$ and $\beta=\frac{3}{2}$ are non-trivial. However, few neuroscience studies elaborate on these details, impeding researchers from understanding the theoretical foundations of brain criticality in the brain. The importance of these derivations is beyond the detailed values of avalanche exponents since they reveal the fundamental properties of neural dynamics  \cite{di2017simple,cocchi2017criticality,girardi2021brain}. In \textbf{Box 1}, we sketch an original idea to derive these avalanche exponents in the terminology of neuroscience. In \textbf{Fig. 3A}, we present graphical illustrations of our idea in \textbf{Box 1}.

\begin{boxedtext}{Derivations of neural avalanche exponents}
Consider a time-continuous neural dynamics process, where an active neuron implies three possibilities: becoming absorbed with probability $\varsigma$, activating another neuron with probability $\eta$, or remaining effect-free with probability $1-\left(\varsigma+\eta\right)$. In critical states, we have $\varsigma=\eta$ \cite{garcia1993branching}. We define $\mathcal{A}_{n}\left(t\right)$ as the probability for $n$ active neurons to exist at $t^{*}+t$ given that $1$ active neuron exists at $t^{*}$. Assuming the independence of neuron activation, we have 
\begin{align}
    \mathcal{A}_{n}\left(t\right)=\sum_{\substack{n_{1}+\ldots+n_{k}=n}}\mathcal{A}_{n_{1}}\left(t\right)\ldots\mathcal{A}_{n_{k}}\left(t\right).\label{EQ35}
\end{align}
If $\mathcal{A}_n(t),n\in \mathbb{N}^{+}$ admits a Maclaurin expansion $\mathcal{A}_{n}\left(t\right)=a_{n}t+o\left(t^2\right)$ (when $n\neq 1$) or $\mathcal{A}_{n}\left(t\right)=a_{n}t+1+o\left(t^2\right)$ (when $n=1$) where $a_{n}=\mathrm{d}\mathcal{A}_{n}\left(0\right)/\mathrm{d}t$, we can readily derive $a_{0}=a_{2}=\varsigma$ and $a_{1}=-2\varsigma$ \cite{garcia1993branching}.  Meanwhile, we can know
\begin{align}
\mathcal{A}_{n}\left(t+\mathrm{d} t\right)-\mathcal{A}_{n}\left(t\right)=\sum_{k=0}^{\infty}a_{k}\mathcal{A}_{n-k}\left(t\right)\mathrm{d}t.\label{EQ36}
\end{align}
Eqs. (\ref{EQ15}-\ref{EQ16}) readily lead to
\begin{align}
\frac{\partial}{\partial t}\mathcal{W}\left(x,t\right)=\sum_{k=0}^{\infty}a_{k}\sum_{n=0}^{\infty}\left(\sum_{\substack{n_{1}+\ldots+n_{k}=n-k}}\prod_{i=1}^{k}\mathcal{A}_{n_{i}}\left(t\right)\right)x^{n}
=\sum_{k=0}^{\infty}a_{k}\mathcal{W}\left(x,t\right)^{k},\label{EQ37}
\end{align}
where $\mathcal{W}\left(x,t\right)=\sum_{n=0}^{\infty}\mathcal{A}_{n}\left(t\right)x^{n},\;x\in\left[0,1\right]$ denotes the generating function. Applying a trick introduced in \cite{garcia1993branching}, we define $\mathcal{H}\left(x\right)=\frac{\partial}{\partial t}\mathcal{W}\left(x,0\right)$, which naturally leads to $\frac{\partial}{\partial t}\mathcal{W}\left(x,t\right)=\mathcal{H}\left(\mathcal{W}\left(x,t\right)\right)$. Meanwhile, $\mathcal{H}\left(x\right)=\varsigma\left(1-x\right)^{2}$ can be derived based on $a_{0}$, $a_{1}$, and $a_{2}$ \cite{garcia1993branching}. Taken together, we have
\begin{align}
    \frac{\partial}{\partial t}\mathcal{W}\left(x,t\right)=\varsigma\left(1-\mathcal{W}\left(x,t\right)\right)^{2}.\label{EQ38}
\end{align}
Note that the initial condition is $\mathcal{W}\left(x,0\right)=x$ since one neuron is activated at $t^{*}$. Solving Eq. (\ref{EQ38}), we derive that
\begin{align}
    \mathcal{W}\left(x,t\right)=\frac{\varsigma\left(1-x\right)t}{\varsigma\left(1-x\right)t+1}.\label{EQ39}
\end{align}
Therefore, we have $\mathcal{A}_{0}\left(t\right)=\mathcal{W}\left(0,t\right)=\frac{\varsigma t}{\varsigma t+1}$, supporting a calculation of lifetime distribution $\mathcal{P}_T\left(t\right)$
\begin{align}
    \lim_{t\rightarrow\infty}\mathcal{P}_T\left(t\right)= \lim_{t\rightarrow\infty}\frac{\mathrm{d}}{\mathrm{d}t}\mathcal{W}\left(0,t\right)\sim t^{-2}.\label{EQ40}
\end{align}
Following \cite{garcia1993branching,harris1963theory,otter1949multiplicative}, one can similarily calculate 
\begin{align}
    \lim_{s\rightarrow\infty}\mathcal{P}_S\left(s\right)\sim s^{-\frac{3}{2}}.\label{EQ41}
\end{align}
\end{boxedtext}

There are three important things to remind. First, the lifetime exponent $\alpha=2$ and size exponent $\beta=\frac{3}{2}$ can only be treated as ideal exponents under mean field assumptions of directed percolation. There are numerous factors, such as granularity, network topology, and neural dynamics variability, can be considered in derivations to affect the detailed values of avalanche exponents \cite{girardi2021brain,bonachela2009self}. In \textbf{Table 3}, we summarize the possible intervals of $\alpha$ and $\beta$ in empirical neural data. Second, $\alpha$ and $\beta$ alone are not sufficient to verify the existence of brain criticality. Even when the actual values of $\alpha$ and $\beta$ in empirical data are exactly equal to theoretical predictions, they may still not satisfy the scaling relation and universal collapse. Meanwhile, as we shall discuss later, estimating $\alpha$ and $\beta$ in practice is statistically error-pone. Third, one can not confirm or disprove a detail type of brain criticality based on $\alpha$ and $\beta$ unless additional information is provided. Although four types of brain criticality exhibit (e.g., OC) or approximatively exhibit (e.g., qC, SOC, and SOqC) directed percolation behaviours under certain conditions, these preconditions are difficult to verify in practice. 

\subsection{Scaling relation}
In the previous section, we discuss how the neural avalanche lifetime and size distributions exhibit power-law properties when the brain is at the critical point \cite{hinrichsen2000non,lubeck2004universal,larremore2012statistical}. Apart from lifetime $T$ and size $S$, there are several other quantities that characterize neural avalanches, such as area $A$ (number of distinct active neurons, measured as $A\simeq \langle S\left(T\right)\rangle$ where the expectation $\langle \cdot\rangle$ is averaged across all neural avalanches with the same lifetime $T$) and radius exponent $R$ (radius of gyration) \cite{lubeck2003universal,lubeck2004universal}. In general, the corresponding probability distributions of these four quantities decay algebraically 
\begin{align}
    \mathcal{P}_X\left(x\right)\propto x^{-\lambda_{X}},\label{EQ42}
\end{align}
where random variable $X\in\{S,T,A,R\}$ can be an arbitrary quantity to characterize neural avalanches. The avalanche exponent $\lambda_{X}$ is defined according to the selected meaning of $X$ (e.g., $\lambda_{T}=2$ and $\lambda_{T}=\frac{3}{2}$ under mean filed assumptions). Assuming that variables $\{S,T,A,R\}$ scale as a power of each other
\begin{align}
  X^{\prime}\propto X^{\psi_{X^{\prime}X}},\; \forall X,\;X^{\prime}\in\{S,T,A,R\},\label{EQ43}
\end{align}
we can derive the scaling relation from Eqs. (\ref{EQ42}-\ref{EQ43})
\begin{align}
  \psi_{X^{\prime}X}=\frac{\lambda_{X}-1}{\lambda_{X^{\prime}}-1}.\label{EQ44}
\end{align}
If we let $X^{\prime}=A$ and $X=T$, we can specify Eq. (\ref{EQ44}) as 
\begin{align}
  \gamma&=\frac{\alpha-1}{\beta-1},\label{EQ45}
\end{align}
where $\mathcal{P}_T\left(t\right)\propto t^{-\alpha}$, $ \mathcal{P}_S\left(s\right)\propto s^{-\beta}$, and $A\propto T^{\gamma}$. Eq. (\ref{EQ45}) leads to $\gamma=2$ in the mean field theory of directed percolation. In \textbf{Table 3}, one can see the possible interval of $\gamma$ in empirical neural data. Eq. (\ref{EQ45}) is widely used as a criterion to verify if the brain is at the critical point in neuroscience studies (e.g., \cite{dalla2019modeling,friedman2012universal,fontenele2019criticality,ponce2018whole}). Once the scaling relation is confirmed among observed neural avalanche exponents, it indicates key features of the universality class (please note that $\alpha$, $\beta$, and $\gamma$ should be derived independently). For neuroscientists, the importance of Eq. (\ref{EQ45}) lies in that it provides extra verification of the validity of estimated neural avalanche exponents. This verification is necessary given that neural avalanche exponent estimation is frequently inaccurate \cite{fontenele2019criticality}. In \textbf{Fig. 3B}, we illustrate the scaling relation in Eq. (\ref{EQ45}) under mean field assumptions.

In \cite{lubeck2004universal}, one can further learn about how brain criticality is mapped to an directed percolation transition characterized by ordinary critical exponents. Meanwhile, one can see how to connect these neural avalanche exponents with second order phase transition exponents \cite{lubeck2003universal}.

\begin{figure}[!t]
\begin{adjustwidth}{-0cm}{}
\includegraphics[width=1\hsize]{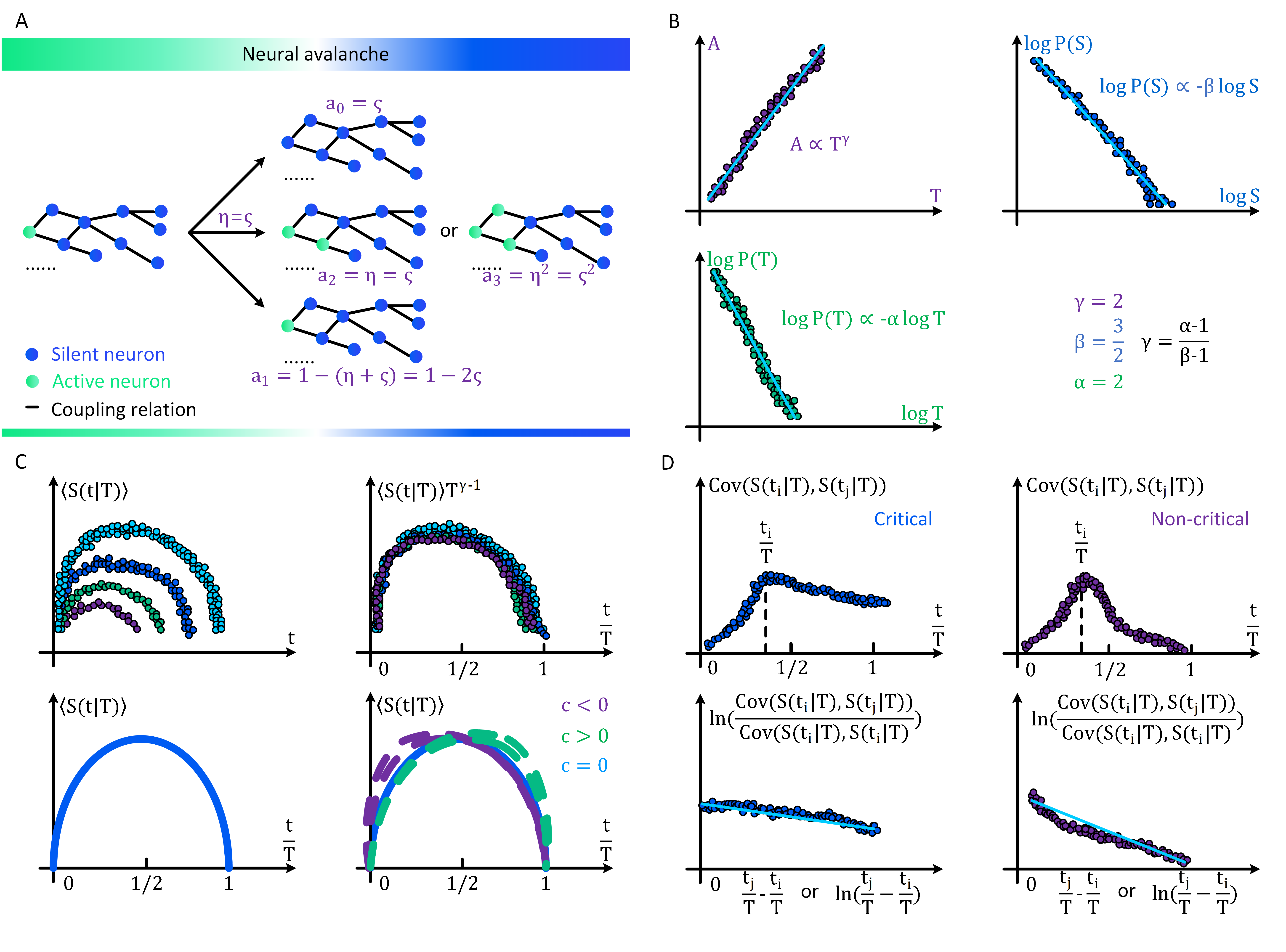}
\caption{Conceptual illustrations of the neural avalanche properties predicted by analytic theories of brain criticality. \textbf{A}, Illustrations of the framework to derive neural avalanche exponents in \textbf{Box 3}. \textbf{B}, Illustrations of the scaling relation satisfied by neural avalanches under mean field assumptions. \textbf{C}, Illustrations of the universal collapse shape of neural avalanches. The un-scaled plot of $t$ vs. $\langle S\left(t\mid T\right)\rangle$ (upper left) and the scaled plot $\langle S\left(t\mid T\right)\rangle T^{1-\gamma}$ vs. $\frac{t}{T}$ (upper right) are shown for comparison. Here terms $\frac{1}{T}$ and $T^{1-\gamma}$ respectively serve as scale factors on x-axis and y-axis to create a universal collapse shape. Meanwhile, the symmetric collapse shape in Eq. (\ref{EQ51}) (bottom left) and the asymmetric collapse shape controlled by skewness $c$ in Eq. (\ref{EQ52}) (bottom right) are also presented. \textbf{D}, Auto-correlations and their decays in critical (left) and non-critical (right) cases are shown. Auto-correlations are calculated after a $t_{i}\in\left[0,T\right)$ is randomly selected as a reference (upper left and right). Meanwhile, the auto-correlation decays measured on $t_{j}\in\left[t_{i},T\right]$ is fitted to derive the coefficients $\chi$ (x-axis corresponds to $\ln\left(\frac{t_{j}-t_{i}}{T}\right)$) and $\xi$ (x-axis corresponds to $\frac{t_{j}-t_{i}}{T}$) in Eq. (\ref{EQ53}) and Eq. (\ref{EQ55}) (bottom left and right). One can see that auto-correlations in the critical case have slower decays (smaller $\chi$ and $\xi$) than those in the non-critical case.}
\end{adjustwidth}
\end{figure}

\begin{table}[!b]
 \begin{adjustwidth}{-0cm}{}
\caption{Neural avalanche exponents with scaling relation in empirical data. The data is acquired from \cite{girardi2021brain}, where 45 experimental observations of neuronal avalanches reported by 30 studies are summarized. These observations can be classified according to the recording techniques of neural avalanches. Detailed data classification criteria (e.g., details of spike sorting and thresholding) can be seen in \cite{girardi2021brain}. We only include the data where $\alpha$, $\beta$, and $\gamma$ are all recorded and satisfy the scaling relation in Eq. (\ref{EQ45}). For LFP recordings filtered by spike sorting, included observations are reported by \cite{carvalho2021subsampled,fontenele2019criticality,senzai2019layer,fosque2021evidence,ma2019cortical,mariani2021beyond}. For LFP recordings with thresholding, included observations come from \cite{mariani2021beyond,shew2015adaptation}. For Ca and voltage imaging, observations are provided by \cite{ponce2018whole,yaghoubi2018neuronal}. Although numerous studies report neural avalanches in whole-brain imaging (e.g., MEG, M/EEG, and invasive ECoG), these studies either do not report three exponents together \cite{shriki2013neuronal,zhigalov2015relationship,palva2013neuronal} or have not observed the scaling relation in Eq. (\ref{EQ45}) \cite{varley2020differential}. One can see \cite{girardi2021brain} for a summary of these results.}
\label{T3}
\begin{tabular}{llll}
\hline
Data type     & Observed interval of $\alpha$  &  Observed interval of $\beta$  &  Observed interval of $\gamma$                                 \\\hline  
LFP recordings filtered by spike sorting  & $\alpha\in\left[1.35,2.67\right]$  &  $\beta\in\left[1.3,2.5\right]$  &  $\gamma\in\left[1.16,1.48\right]$  \\
LFP recordings with thresholding   & $\alpha\in\left[1.82,2.84\right]$  &  $\beta\in\left[1.57,2.59\right]$  &  $\gamma\in\left[1.12,1.39\right]$ \\
Ca and voltage imaging  & $\alpha\in\left[2.15,3.5\right]$  &  $\beta\in\left[1.5,2.3\right]$  &  $\gamma\in\left[1.75,2.5\right]$ \\
\hline
\end{tabular}
 \end{adjustwidth}
\end{table}

\subsection{Universal collapse shape}
\subsubsection{Universal collapse with an implicit scaling function} Apart from the scaling relation discussed above, the average temporal shape of bursts, a fundamental signature of avalanches \cite{baldassarri2003average,laurson2013evolution,papanikolaou2011universality}, can also be used to verify the existence of brain criticality in a more precise manner. This approach has been previously applied on diverse physical systems, such as plastically deforming crystals \cite{laurson20061} and Barkhausen noise \cite{mehta2002universal,papanikolaou2011universality}, and is recently introduced into neuroscience \cite{ponce2018whole,fontenele2019criticality,pausch2020time,dalla2019modeling,friedman2012universal}. To understand this approach, let us step back to the power relation in Eq. (\ref{EQ43}) and specify that $X^{\prime}=A$ and $X=T$. These settings naturally lead to 
\begin{align}
  A\simeq \langle S\left(T\right)\rangle\equiv\int_{0}^{T}\langle S\left(t\mid T\right)\rangle dt\propto T^{\gamma},\label{EQ46}
\end{align}
where $\langle S\left(t\mid T\right)\rangle$ measures the averaged time-dependent avalanche size during an avalanche and the expectation $\langle \cdot\rangle$ is averaged across all neural avalanches with the same lifetime $T$. Eq. (\ref{EQ46}) can be readily reformulated as 
\begin{align}
  \langle S\left(t\mid T\right)\rangle\propto T^{\gamma-1}.\label{EQ47}
\end{align}
The general form of Eq. (\ref{EQ47}) is usually given by \cite{baldassarri2003average,laurson2013evolution,papanikolaou2011universality}
\begin{align}
  \langle S\left(t\mid T\right)\rangle= T^{\gamma-1}\mathcal{H}\left(\frac{t}{T}\right),\label{EQ48}
\end{align}
where $\mathcal{H}\left(\cdot\right)$ denotes a universal scaling function. When the brain is at the critical point, all data of $\langle S\left(t\mid T\right)\rangle T^{1-\gamma}$ is expected to collapse onto $\mathcal{H}\left(\cdot\right)$ with reasonable errors \cite{baldassarri2003average,laurson2013evolution,papanikolaou2011universality}. Here the terminology ``collapse onto" means that all data generally exhibits a similar pattern in a plot of $\langle S\left(t\mid T\right)\rangle T^{1-\gamma}$ vs. $\frac{t}{T}$ (e.g., all data follows function $\mathcal{H}\left(\cdot\right)$). Meanwhile, scaling function $\mathcal{H}\left(\cdot\right)$ is expected to be a parabolic function \cite{baldassarri2003average,laurson2013evolution,papanikolaou2011universality}. By testing these properties, neuroscientists can verify whether the brain is at criticality (e.g., \cite{dalla2019modeling,friedman2012universal,fontenele2019criticality,ponce2018whole}). In \textbf{Fig. 3C}, we graphically illustrate these properties.

\subsubsection{Universal collapse with an explicit scaling function} Under specific conditions, researchers can further consider an explicit form of scaling function $\mathcal{H}\left(\cdot\right)$ \cite{laurson2013evolution}. Assuming that the early-time growth of neural avalanches averagely follows a power-law of time, one can derive that $\langle S\left(t\mid T\right)\rangle\propto t^{\kappa}$ for certain $\frac{t}{T}\leq \varepsilon\ll1$. Meanwhile, one knows that $\langle S\left(\varepsilon T\mid T\right)\rangle\propto T^{\gamma-1}$ should hold according to Eq. (\ref{EQ28}). To ensure these two properties, one needs to have $\langle S\left(\varepsilon T\mid T\right)\rangle\propto \left(\varepsilon T\right)^{\kappa}\propto T^{\gamma-1}$, which readily leads to $\kappa=\gamma-1$. Based on these derivations, one can know
\begin{align}
  \langle S\left(t\mid T\right)\rangle\propto t^{\gamma-1},\;t\ll T.\label{EQ49}
\end{align}
To find an explicit form of $\mathcal{H}\left(\cdot\right)$ that satisfies Eqs. (\ref{EQ48}-\ref{EQ49}), one can consider a possible answer \cite{laurson2013evolution}
\begin{align}
  \mathcal{H}\left(\frac{t}{T}\right)=\left[\frac{t}{T}\left(1-\frac{t}{T}\right)\right]^{\gamma-1},\label{EQ50}
\end{align}
which can be analytically derived by multiplying Eq. (\ref{EQ49}) by $\left(1-\frac{t}{T}\right)^{\gamma-1}$. Here $\left(1-\frac{t}{T}\right)^{\gamma-1}$ is a term to characterize the deceleration at the ends of neural avalanches \cite{laurson2013evolution}. Because $\gamma=2$ is expected for critical neural avalanches under mean field assumptions, Eq. (\ref{EQ48}) and Eq. (\ref{EQ50}) imply that 
\begin{align}
  \langle S\left(t\mid T\right)\rangle\propto t\left(1-\frac{t}{T}\right).\label{EQ51}
\end{align}
This result is consistent with the prediction by the ABBM model in the limit of vanishing drive rate and demagnetizing factor \cite{friedman2012universal,laurson2013evolution}.

A potential limitation of Eq. (\ref{EQ51}) in applications lies in its internal symmetry property \cite{laurson2013evolution}. Although avalanches under mean-field frameworks have a symmetric average shape \cite{friedman2012universal}, it does not mean that symmetry generally holds in real complex systems \cite{laurson2013evolution}. Applying Eq. (\ref{EQ48}) on neural data, researchers may observe a nonstandard parabolic function $\mathcal{H}\left(\cdot\right)$ with specific skewness. This does not necessarily mean that neural dynamics is not at criticality. When neural avalanches are time-irreversible (this is generally true in the brain since the detailed balance of neural dynamics is frequently broken \cite{lynn2021broken}), one can consider small temporal asymmetry in the collapse shape \cite{laurson2013evolution}. To characterize potential asymmetry, one can add a correction term controlled by skewness degree $c$ into Eq. (\ref{EQ51})
\begin{align}
  \langle S\left(t\mid T\right)\rangle\propto t\left(1-\frac{t}{T}\right)\left[1-c\left(\frac{t}{T}-\frac{1}{2}\right)\right].\label{EQ52}
\end{align}
If $c=0$, then Eq. (\ref{EQ52}) reduces to Eq. (\ref{EQ51}). Otherwise, neural avalanches can have a temporally asymmetric collapse shape with a positive ($c>0$) or negative ($c<0$) skewness \cite{laurson2013evolution}. We suggest that Eq. (\ref{EQ52}) may be more applicable to real data of neural dynamics. In \textbf{Fig. 3C}, we show examples of Eqs. (\ref{EQ51}-\ref{EQ52}).

\subsection{Slow decay of auto-correlation}
In applications, researchers can also consider a more practical verification of the potential brain criticality. When the brain is at the critical point, a slow decay of auto-correlation is expected to occur in neural avalanches, corresponding to long-range correlations \cite{schaworonkow2015power,smit2011scale,dalla2019modeling,erdos2018power}. This slow decay property is initially derived from the power-law decay of auto-correlation, which can be analytically derived as a part of the scaling relation if ordinary critical exponents of directed percolation transition are considered (see \cite{girardi2021brain} for details). The power-law decay is expressed as
\begin{align}
    \ln\left[\frac{\operatorname{Cov}\left(S\left(t_{i}\mid T\right),S\left(t_{j}\mid T\right)\right)}{\operatorname{Cov}\left(S\left(t_{i}\mid T\right),S\left(t_{i}\mid T\right)\right)}\right]=-\chi\ln\left(\frac{t_{j}-t_{i}}{T}\right)+r,\label{EQ53}
\end{align}
where $t_{i}\in\left[0,T\right)$ is used as a reference and $t_{j}\in\left[t_{i},T\right]$ traverses the entire interval \cite{schaworonkow2015power,smit2011scale}. According to the Wiener–Khinchin theorem, coefficient $\chi$ is related to $\mathcal{S}\left(f\right)$, the power spectrum of neural avalanches (notion $f$ denotes frequency) \cite{bak1987self,linkenkaer2001long,girardi2021brain}. One may expect $\mathcal{S}\left(f\right)\sim f^{-\upsilon}$ at the critical point, where $\chi=1-\upsilon$ \cite{bak1987self,linkenkaer2001long,girardi2021brain}. The power-law decay of auto-correlation in Eq. (\ref{EQ53}) breaks down when $\upsilon=1$, leading to infinitely long temporal correlations. Therefore, $\chi\in\left[0,\infty\right)$ in Eq. (\ref{EQ53}) is expected to be sufficiently small. Certainly, the actual value of $\chi$ may not be perfectly zero in empirical data. For instance, $\chi\in\left[0.58\pm 0.23,0.73\pm 0.31\right]$ and $\chi\in\left[0.52\pm 0.35,0.81\pm 0.32\right]$ are observed in spontaneous alpha oscillations in MEG and EEG data, respectively \cite{linkenkaer2001long}.

Apart from verifying power-law decay directly, one can also consider the exponential decay, which is active in neuroscience as well \cite{pausch2020time,wilting2019between,miller2006power}. The exponential decay can described by 
\begin{align}
\frac{\partial}{\partial t}\operatorname{Cov}\left(S\left(t_{i}\mid T\right),S\left(t_{j}\mid T\right)\right)=-\xi\operatorname{Cov}\left(S\left(t_{i}\mid T\right),S\left(t_{j}\mid T\right)\right).\label{EQ54}
\end{align}
Eq. (\ref{EQ54}) directly leads to
\begin{align}
    \ln\left[\frac{\operatorname{Cov}\left(S\left(t_{i}\mid T\right),S\left(t_{j}\mid T\right)\right)}{\operatorname{Cov}\left(S\left(t_{i}\mid T\right),S\left(t_{i}\mid T\right)\right)}\right]=-\xi\left(\frac{t_{j}-t_{i}}{T}\right)+r.\label{EQ55}
\end{align}
The exponential decay can be seen in the dynamics with short-term correlations (i.e., correlations have a characteristic time scale). Mathematically, the exponential decay can be related to power-law decay in a form of $x^{-y}=\Gamma\left(y\right)\int_{0}^{\infty}z^{y-1}\exp\left(-xz\right)\mathsf{d}z$, where $\Gamma\left(\cdot\right)$ denotes the Gamma function. When $\xi\in\left[0,\infty\right)$ is sufficiently small, Eq. (\ref{EQ55}) can be treated as a looser criterion that approximately verifies the slow decay of auto-correlation and may be more applicable to non-standard brain criticality (e.g., qC and SOqC) \cite{wilting2019between}. Despite of its practicality, this looser criterion should be used with cautions since it is not analytically derived from criticality theories.

In \textbf{Fig. 3D}, we illustrate examples of auto-correlation slow decay in critical cases and compare them with non-critical cases. Compared with other properties previously discussed, a slow auto-correlation decay can be readily verified by conventional data fitting. However, we need to note that one should not confirm or reject the possibility of brain criticality only based on the decay characteristic of auto-correlation in Eqs. (\ref{EQ53}-\ref{EQ55}).
This is because Eqs. (\ref{EQ53}-\ref{EQ55}) only serve as the approximate descriptions of long-range correlations at criticality. The strict criterion $\chi,\xi\rightarrow 0$ is rarely seen in empirical data while the determination of whether $\chi$ and $\xi$ are sufficiently small in the looser criterion is relatively subjective.

In summary, we have reviewed the physical foundations of identifying and characterizing criticality in the brain. Based on these analytic derivations, we attempt to present systematic explanations of what is brain criticality and how to identify potential criticality in neural dynamics. Nevertheless, physical theories alone are not sufficient to support neuroscience studies because the implementation of these theories on empirical data is even more challenging than the theories themselves. To overcome these challenges, one needs to learn about statistic techniques to computationally estimate brain criticality from empirical data.

\section{Brain criticality: statistic techniques}
 While most properties of neural avalanches analytically predicted by the physical theories of brain criticality can be estimated by conventional statistic techniques, there exist several properties that frequently imply serious validity issues and deserve special attention. Below, we discuss them in detail.

\subsection{Estimating neural avalanche exponents}
Perhaps the estimation of neural avalanche exponents from empirical data is the most error-prone step in brain criticality analysis. The least-square approach is abused in fitting power-law data and frequently derives highly inaccurate results \cite{clauset2009power,virkar2014power}. To derive neural avalanche exponents $\alpha$ and $\beta$ in Eq. (\ref{EQ25}) with reasonable errors, one need to consider the maximum likelihood estimation (MLE) approach and corresponding statistic tests (see MLE on un-binned data \cite{clauset2009power} and binned data \cite{virkar2014power}). Taking the avalanche size distribution as an instance, the estimator $\widehat{\beta}$ of distribution exponent $\beta$ is expected to maximize the log-likelihood function
\begin{align}
    \mathcal{L}\left(\beta\right)&=-n\ln\left[\zeta\left(\beta,s^{\prime}\right)\right]-\beta\sum_{i=1}^{n}\ln\left(s_{i}\right),\label{EQ56}\\
    \mathcal{L}\left(\beta\right)&=n\left(\beta-1\right)\ln b^{\prime}+\sum_{i=1}^{k}h_{i}\ln\left(b_{i}^{1-\beta}-b_{i+1}^{1-\beta}\right).\label{EQ57}
\end{align}
Here Eq. (\ref{EQ56}) and Eq. (\ref{EQ57}) denote the log-likelihood functions on un-binned and binned data, respectively. Function $\zeta\left(\cdot,\cdot\right)$ is the generalized zeta function \cite{clauset2009power,bauke2007parameter}. Notion $s$ denotes avalanche size samples in Eq. (\ref{EQ33}) and Eq. (\ref{EQ41}) \cite{clauset2009power}. Notion $b$ denotes bin boundaries defined on these samples and $h$ counts the number of samples within each bin \cite{virkar2014power}. Notions $s^{\prime}$ and $b^{\prime}$ are the lower cutoffs of un-binned and binned power-law distributions \cite{clauset2009power,virkar2014power}. They are necessary because few empirical data exhibits power-law properties on the entire distribution \cite{clauset2009power}. Notions $n$ and $k$ measure the numbers of samples and bins above cutoffs, respectively \cite{clauset2009power,virkar2014power}. To estimate $\widehat{\beta}$ precisely, researchers are suggested to follow several indispensable steps \cite{clauset2009power,virkar2014power}: (1) for each potential choice of $s^{\prime}$ or $b^{\prime}$, estimate the power-law model on the distribution tail above the cutoff. Compute the Kolmogorov–Smirnov (KS) goodness-of-fit statistic between the cumulative probability distributions of power-law model and empirical data. Find the ideal choice of $s^{\prime}$ or $b^{\prime}$ that minimizes KS statistic; (2) derive the corresponding estimator $\widehat{\beta}$ and KS statistic based on the chosen cutoff; (3) use the semi-parametric bootstrap to generate numerous synthetic data distributions that follow the estimated power-law model above the cutoff but follow the empirical data distribution below the cutoff. Estimate new power-law models on these synthetic data distributions and measure the goodness-of-fit by KS statistic. Define a $p$-value, the fraction of these KS statistics in step (3) that are no less than the KS statistic in step (2). Rule out the estimated power-law model in steps (1-2) if $p<0.1$ (conservative criterion). Apart from these necessary steps, one can further consider Vuong's likelihood ratio test for alternative distribution checking \cite{vuong1989likelihood,clauset2009power,virkar2014power} and information loss measurement of binning approach \cite{virkar2014power}. During the above process, we measure the goodness-of-fit by KS statistic instead of the well-known $\chi^{2}$ statistic because the latter has less statistic power \cite{bauke2007parameter,clauset2009power,virkar2014power}. Meanwhile, KS statistic is measured on cumulative probability distributions rather than probability distributions to control the effects of extreme values in empirical data \cite{clauset2009power,virkar2014power}. Except for the above approach \cite{clauset2009power,virkar2014power}, one can also consider the BIC method (for un-binned data) \cite{schwarz1978estimating} and the RT method (for binned data) \cite{reiss2007statistical} for comparisons. In practice, the approaches proposed by Clauset \emph{et al.} are more robust \cite{clauset2009power,virkar2014power} and have attracted numerous follow-up studies for improvements (e.g., \cite{yu2014scale,marshall2016analysis,deluca2013fitting}).

\subsection{Estimating universal collapse shape}
Another error-prone step is the calculation and evaluation of the universal collapse shape, which is closely related to scaling relation analysis. Deriving the collapse shape from empirical data may be problematic because the goodness evaluation of collapse shape is rather subjective (e.g., depends on personal opinions about whether all data follows function $\mathcal{H}\left(\cdot\right)$ in Eq. (\ref{EQ48})) in most cases \cite{marshall2016analysis}. Although important efforts have been devoted to quantify if a given data set exhibits shape collapse \cite{shaukat2016statistical,bhattacharjee2001measure}, common approaches in practice still depend on specific shape collapse algorithms that search potential scaling parameters (e.g., $\gamma$ in Eq. (\ref{EQ48})) in a data-driven manner \cite{marshall2016analysis}. In these algorithms, thresholding on neural avalanches before analyzing the shape collapse is a standard pre-processing scheme to control noises (e.g., set an avalanche size threshold and remove all data below the threshold) \cite{marshall2016analysis,papanikolaou2011universality}. While experimental noises are partly limited, unexpected excursions of scaling parameters away from theoretical predictions may occur after thresholding as well \cite{laurson2009effect}. To our best knowledge, the effects of thresholding on brain criticality analysis are non-negligible. Although being highly practical, thresholding may lead to significant transient effects to cloud the true scaling property \cite{villegas2019time}. Therefore, any qualitative evaluation of collapse shape after thresholding is questionable regardless of its practicability. Although an ideal approach requires further explorations, we suggest researchers to consider following methods: (1) estimate $\gamma$ by area fitting (e.g., follow Eq. (\ref{EQ47}) in scaling relation analysis) and collapse shape fitting (e.g., follow Eq. (\ref{EQ48}) in collapse shape analysis), respectively; (2) compare between $\gamma$ derived by these two kinds of fitting and measure the difference. Search for a threshold that minimizes the difference (e.g., makes variation amplitude $<1\%$) and maintains a reasonable sample size (e.g., maintains $>80\%$ samples); (3) Given the chosen threshold and corresponding $\gamma$, measure the difference (e.g., the dynamic time warping \cite{keogh2001derivative}) between $\langle S\left(t\mid T\right)\rangle T^{1-\gamma}$ derived on neural avalanches with different lifetime $T$ in the plot of $\langle S\left(t\mid T\right)\rangle T^{1-\gamma}$ vs. $\frac{t}{T}$. Denote the shape collapse error as the averaged difference. Combining these three steps, researchers may partly avoid the errors implied by subjective judgment. Similar ideas can be seen in \cite{marshall2016analysis}.

\subsection{Estimating the slow decay of auto-correlation}
Finally, the analysis of slow decay of auto-correlation is also error-prone in practice. Although this approach is practical and has been extensively applied (e.g., \cite{pausch2020time,wilting2019between}), the criterion to determine if the decay is truly slow (e.g., $\chi>0$ in Eq. (\ref{EQ54}) and $\xi>0$ in Eq. (\ref{EQ54}) are sufficiently small) remains ambiguous. A fixed criterion (e.g., $\chi,\xi<0.5$) may serve as an explicit condition of a slow decay. However, this presupposed criterion may deviate from real situations. For instance, the baseline of decay rate in a non-critical brain may be essentially high (e.g., $\chi,\xi>10$). Even though the decay rate drops significantly when the brain becomes critical (e.g., $\chi,\xi\simeq 1$), the presupposed criterion is still unsatisfied and leads to unnecessary controversies on criticality hypothesis. Given that $\xi$ is principally independent from spatial sub-sampling on neurons or brain regions at criticality \cite{pausch2020time}, we suggest researchers to consider the following approaches: (1) do spatial sub-sampling in both critical and non-critical brains to derive two groups of $\chi$ or $\xi$ (one group for criticality and another group for non-criticality); (2) use appropriate statistic tests (e.g., choose $t$-test \cite{kanji2006100}, Kolmogorov-Smirnov test \cite{berger2014kolmogorov}, or Wilcoxon-Mann-Whitney test \cite{fay2010wilcoxon} according to sample distribution properties) to verify if two groups of $\chi$ or $\xi$ belong to different distributions. Test if the expectation and variance of $\chi$ or $\xi$ drops significantly from the non-critical group to the critical group according to certain effect sizes. 

In summary, statistic techniques bridge between brain criticality theory and empirical data. However, misconception and misuse of statistic analyses of neural avalanche properties still occasionally appear in practice. Although existing techniques remain imperfect in brain criticality analysis, we wish that our discussion may inspire future studies.

\section{Brain criticality and other neuroscience theories}
Ever since brain criticality is introduced into neuroscience, it is frequently speculated as contradictory with other traditional neuroscience hypotheses, such as the conjectured hierarchical processing characteristic of neural information \cite{felleman1991distributed} and the asynchronous-irregular characteristic of neural dynamics (e.g., neurons spike independently in Poisson manners \cite{burns1976spontaneous,softky1993highly,stein2005neuronal}). Meanwhile, the differences between brain criticality and scale-free neural dynamics \cite{chialvo2010emergent,martinello2017neutral,he2014scale} are frequently neglected. Before we put an end to our review, we discuss the relations between brain criticality and these neuroscience theories.  

\subsection{Brain criticality and hierarchical processing}
The hierarchical processing of neural information \cite{felleman1991distributed} is initially speculated to contradict critical neural dynamics since hierarchical topology has not been used as a explicit condition to analytically derive criticality (e.g., see derivations in \cite{garcia1993branching,harris1963theory,otter1949multiplicative,di2017simple,gros2010complex,robinson2021neural,janowsky1993exact,lee2004branching,larremore2012statistical}). On the contrary, random graphs without strict hierarchical structures seem to be more widespread in criticality derivations. Recently, this speculation has been challenged by numerous discoveries of the facilitation effects of hierarchical modular structures on critical dynamics \cite{friedman2013hierarchical,kaiser2010optimal,wang2012hierarchical,rubinov2011neurobiologically}. Meanwhile, computational analysis suggests that information transmission in standard feed-forward networks is maximized by critical neural dynamics \cite{beggs2003neuronal}. Parallel to neuroscience, a recent machine learning study empirically observes and analytically demonstrates that artificial neural networks, a kind of hierarchical structure, self-organize to criticality during learning \cite{katsnelson2021self}. Therefore, brain criticality is not necessarily contradictory with hierarchical information processing, yet more analyses are required to understand how brain criticality affects hierarchical processing schemes.

\subsection{Brain criticality and asynchronous-irregular characteristic}
Brain criticality and the asynchronous-irregular (AI) characteristic may correspond to distinct encoding schemes in the brain \cite{wilting2019between,girardi2021unified}. While AI characteristic can minimize redundancy \cite{atick1992could,bell1997independent,van1998independent,barlow1961possible} to improve neural encoding \cite{van1996chaos}, brain criticality may optimize encoding performance utilizing a series of reverberations of neural activities \cite{bertschinger2004real,haldeman2005critical,kinouchi2006optimal,wang2011fisher,boedecker2012information,shew2013functional,del2017criticality}. The coexistence of empirical evidence of AI and brain criticality characteristics initially confuses researchers since these characteristics are hypothesized as contradictory with each other \cite{wilting2019between,girardi2021unified}. In experiments, AI characteristic is supported by small correlations between the spike rates of different neurons in cortical microcircuits \cite{ecker2010decorrelated,cohen2011measuring} and exponential distributions of inter-spike intervals \cite{kara2000low,carandini2004amplification} while brain criticality characteristic is observed in neural dynamics recorded from multiple species (e.g., awake monkeys \cite{petermann2009spontaneous}, anesthetized rats \cite{gireesh2008neuronal}, slices of rat cortices \cite{shew2009neuronal,beggs2003neuronal}, and humans \cite{poil2008avalanche}). A recent study demonstrates that cortical spikes may propagate at somewhere between perfect criticality (e.g., OC or SOC depending on whether underlying mechanisms are exogenous or endogenous) and full irregularity \cite{wilting2019between}, similar to the cases of qC and SOqC. Meanwhile, it is known that stimulus drives suppress irregularity in neural activities \cite{molgedey1992suppressing}. These results imply that brain criticality may not necessarily contradict AI characteristic. On the contrary, they may coexist when stimulus drives are too weak to disturb brain criticality (e.g., OC or SOC) and suppress AI characteristic. In our previous discussions, we have analytically proven that neural avalanche exponents, the fundamental properties of brain criticality, can still be derived under the condition of independent neuron activation, a key feature of AI characteristic \cite{wilting2019between}. This result suggests that brain criticality and AI characteristic do not contradict each other. As for the case where stimulus drives are non-negligible, a recent study presents an elegant theory to prove that two homeostatic adaptation mechanisms (i.e., the short-term depression of inhibition and the spike-dependent threshold increase) enable synaptic excitation/inhibition balance, AI characteristic, and SOqC to appear simultaneously in the same neural dynamics \cite{girardi2021unified}. Similarly, it is suggested that neural dynamics with criticality or with AI characteristic can be generated by the same neural populations if the synaptic excitation/inhibition balance is fine tuned appropriately \cite{li2020tuning}.

\subsection{Brain criticality and power-law behaviours in neural dynamics}
Neural dynamics with power-law behaviours is a necessary but insufficient condition of brain criticality. This property is frequently neglected in practice. Power-law behaviours are widespread in the nature because it can be generated by diverse mechanisms, such as exponential curve summation and preferential attachment \cite{reed2002gene,mitzenmacher2004brief}. It has been reported that the aggregate behaviours of non-critical stochastic systems may also create scale-free dynamics within a limited range \cite{touboul2010can,touboul2017power}. In the brain, the generic scale-free properties can be implied by neutral dynamics, a kind of dynamics where the population size of neutral individuals (or dynamically homogeneous individuals) does not tend to increase or decrease after adding a new individual that is neutral to existing ones (see neutral theories for further explanations \cite{blythe2007stochastic,liggett2006interacting}). This generic property can generate power-law neural avalanches without criticality \cite{martinello2017neutral}. Meanwhile, bistability phenomena, a kind of fine tuned or self-organized discontinuous phase transitions with limit cycles rather than critical points, can also create neural dynamics with power-law properties \cite{buendia2020feedback,di2016self,cocchi2017criticality}. Consequently, we emphasize that neural avalanches exponents alone are insufficient to prove or disprove any brain criticality hypothesis. These power-law exponents are meaningless for brain criticality hypothesis unless they satisfy the scaling relation.

\section{Brain criticality: conclusions on current progresses and limitations}
Given what have been reviewed above, we arrive at a point to conclude on the current progresses and limitations in establishing theoretical foundations of different types of brain criticality, i.e., ordinary criticality (OC), quasi-criticality (qC), self-organized criticality (SOC), and self-organized quasi-criticality (SOqC). As we have suggested, an inescapable cause of various controversies is the non-triviality of physical theories that analytically derive brain criticality and statistic techniques that estimate brain criticality from empirical data. Immoderate omitting of these theoretical foundations, especially their imperfection, in practice may lead to confusions on the precise meaning, identification criteria, and biological corollaries of brain criticality. To address these problems, we have introduced mainstream theoretical foundations of brain criticality, reformulated them in the terminology of neuroscience, and discussed their mistakable details. 

Thanks to the increasing efforts devoted to improving theoretical frameworks of criticality in the brain, researchers have seen substantial progresses in explaining various important neuroscience problems, including but not limited to efficient cortical state transitions \cite{fontenele2019criticality}, dynamic range maximization in neural responses \cite{kinouchi2006optimal,shew2009neuronal}, and optimization of information transmission and representation \cite{shew2011information}. These advances have been comprehensively reviewed by existing works \cite{shew2013functional,chialvo2010emergent,beggs2007build,hesse2014self,cocchi2017criticality,munoz2018colloquium} and are no longer discussed in details in our review. The benefits of studying brain criticality, as we have suggested, lay in the possibility to analyze brain function characteristics with numerous statistical physics theories relevant to brain criticality, such as directed percolation \cite{hinrichsen2000non,lubeck2004universal}, conserved directed percolation \cite{bonachela2008confirming,bonachela2010self}, and dynamical percolation theories \cite{steif2009survey,bonachela2010self}. These theories characterize the brain as a physical system with avalanche behaviors, enabling researchers to analyze various propagation, synchronization, and correlation properties of neural dynamics (e.g., continuous phase transitions). These properties intrinsically shape neural information processing (e.g., encoding \cite{bertschinger2004real,haldeman2005critical,kinouchi2006optimal,wang2011fisher,boedecker2012information,shew2013functional,del2017criticality}, transmission \cite{shew2011information}, and memory \cite{krotov2020large,haldeman2005critical}) and can be readily recorded in neuroscience experiments. Therefore, the non-equilibrium dynamic processes and potential criticality defined by statistical physics theories are highly applicable to characterizing brain functions. As we have discussed in \textbf{Fig. 2}, researchers can consider diverse brain criticality phenomena in neural dynamics by defining different control (e.g., the balance between excitatory and inhibitory neurons \cite{poil2012critical,hardstone2014neuronal}) and order (e.g., active neuron density \cite{dalla2019modeling}) parameters, corresponding to multifarious biological mechanisms underlying neural dynamics (e.g., synaptic depression \cite{levina2007dynamical}). Meanwhile, the definition of neural avalanches can flexibly change from neural spikes, local field potentials, to global cortical oscillations. The flexibility of brain criticality and neural avalanche definitions enables researchers to analyze different functional properties on distinct organizational levels in the brain.

The limited theoretical foundations of brain criticality in the brain, however, have become irreconcilable with their increasingly widespread applications. Although the analytic theories of brain criticality have solid physics backgrounds, they needlessly become black boxes for neuroscientists in practice. On the one hand, the details of brain criticality theory frequently experience immoderate neglecting in neuroscience studies. On the other hand, to our best knowledge, there is no accessible and systematic introduction of the statistical physics foundations of brain criticality in the terminology of neuroscience yet. These obstacles severely impede neuroscientists from comprehensively understanding brain criticality, eventually motivating us to present this review. When we turn to bridging between brain criticality theories and experiments, one can find non-negligible gaps separating between theories and experiments. Although numerous biological factors (e.g., neural plasticity \cite{levina2007dynamical,levina2009phase,de2006self}, membrane potential leakage \cite{millman2010self,levina2007dynamical,rubinov2011neurobiologically,stepp2015synaptic}, retro-synaptic signals \cite{hernandez2017self}, spatial heterogeneity \cite{moretti2013griffiths,girardi2016griffiths}, and refractory period \cite{williams2014quasicritical,fosque2021evidence}) have been considered in brain criticality characterization, existing theories more or less suffer from deviations from actual neural system properties. For instance, the requirements of conserved neural dynamics and an infinite time scale separation between the dissipation and drive processes required by SOC may not be biologically realistic \cite{munoz2018colloquium}. The implicit requirement of a sufficiently large system size by the mean field theories of brain criticality may not always be satisfied during neural avalanche recording, implying non-negligible finite size effects \cite{girardi2021brain}. Meanwhile, precisely verify the existence of a detailed type of brain criticality (e.g., confirm the actual universality class) in empirical neural data is principally infeasible. As we have explained, the common criteria used for brain criticality hypothesis verification, such as neural avalanche exponents \cite{clauset2009power,bauke2007parameter,yu2014scale,marshall2016analysis,deluca2013fitting}, scaling relation \cite{lubeck2003universal,lubeck2004universal}, universal collapse shape \cite{marshall2016analysis,papanikolaou2011universality,laurson2009effect,bhattacharjee2001measure}, and slow decay of auto-correlation \cite{pausch2020time,wilting2019between}, are derived according to directed percolation theory under mean field assumptions. Among four types of brain criticality in absorbing phase transitions, only OC originally belongs to directed percolation universality class while qC, SOC, and SOqC conditionally exhibit directed percolation behaviours. In most cases, one can only verify if the brain is plausibly at criticality (e.g., whether neural avalanches obey universal collapse and have the power-law exponents that satisfy the scaling relation). When observed neural avalanche exponents depart from their mean field approximation results but still satisfy the scaling relation, there may exist an OC phenomenon affected by non-mean-field factors (e.g., network topology \cite{girardi2021brain}) or exist a certain qC, SOC, or SOqC phenomenon caused by diverse mechanisms. Additional information of neural dynamics properties is inevitably required to determine the category belonging of the hypothesized brain criticality, which poses daunting challenges to neuroscience experiment designs. Moreover, the potential validity issues of applying the theoretical tools derived from directed percolation theory to verify brain criticality in synchronous phase transitions deserve special attention (see \cite{dalla2019modeling} for similar opinions). It remains controversial if absorbing and synchronous phase transitions robustly share specific features (see reported similarities \cite{di2018landau,girardi2021unified,buendia2021hybrid,fontenele2019criticality} and differences \cite{fontenele2019criticality,buendia2021hybrid,girardi2021unified}). Any speculated relations between these two kinds of critical phenomena should be tested with cautions. Furthermore, statistic techniques to estimate and verify brain criticality from empirical data are yet imperfect. The estimation of some properties of neural avalanches is error-prone in practice and may lead to serious validity issues. Although we suggest compromised solutions to these issues, more optimal approaches are required in future studies.

\section{Brain criticality: suggestions of future direction}
We pursue that this review not only summarizes latest developments in the field of studying criticality in the brain, but also serves as a blueprint for further explorations. Below, we offer concrete recommendations of future directions.

First, we suggest researchers to carefully rethink the theoretical foundations of criticality in the brain. Immoderately omitting these foundations in neuroscience needlessly muddies an already complex scientific field and leads to potential validity issues. While we have presented a self-contained framework of brain criticality to characterize neural dynamics as a physical system with avalanches, plentiful details are uncovered in this article (e.g., the Landau–Ginzburg theory \cite{di2018landau}) because the statistical physics theories of brain criticality are essentially grand. We recommend researchers to further improve our work and explore a more accessible and systematic reformulation of related physics theories, such as directed percolation, conserved directed percolation, dynamic percolation, non-equilibrium dynamics, in the context of neuroscience. Moreover, we note that these theories are not initially proposed for brain analysis. It is normal to see gaps between these theories and real situations of the brain. We urge researchers to develop new variants of criticality formalism that is more applicable to the brain or even explore new universality classes of continuous phase transitions. 

Second, neuroscience is in urgent need of new physical theories and statistic techniques to bridge between brain criticality hypotheses and experiments. Although existing theories and techniques have become increasingly booming and covered most of the pivotal details of brain criticality, there remain various limitations as we have suggested. Specifically, we suggest five potential directions to resolve these problems: (1) combine brain criticality theories with large-scale neural dynamics recording or computation to include more realistic biological details into brain criticality theories and establish a closer connection with experimental observations; (2) try to summarize, standardize, and subdivide these theories according to the concrete biological meanings of brain criticality phenomena, prerequisites of model definitions, and scopes of application. Try to avoid abusing or misusing of different brain criticality theories; (3) develop open-source toolboxes of theoretical models and statistic techniques to routinize brain criticality analysis in neuroscience studies (one can see existing efforts to achieve this objective \cite{marshall2016analysis}); (4) establish open-source, multi-species, and large-scale data sets of neural dynamics recorded from both critical and non-critical brains. Validate different statistic techniques of brain criticality estimation and testing on these data sets and, more importantly, confirm appropriate baselines to define the criteria of brain criticality identification (see notable contributions in \cite{girardi2021brain}); (5) explore new non-equilibrium statistical physics theories for synchronous phase transitions or analytically verify the theoretical validity of directed percolation formulation of synchronous phase transitions. 

Third, parallel to neuroscience, the discoveries of critical phenomena in other learning and computation systems also merit attention. Learning or computing at the edge of chaos has been proven as a mechanism to optimize the performance of learners (e.g., recurrent neural networks \cite{bertschinger2004real}). The well-known residual connections can control the performance degradation of artificial neural networks because they enable networks to self-organize to criticality between stability and chaos to preserve gradient information flows \cite{yang2017mean}. It is recently demonstrated that any artificial neural network generally self-organizes to criticality during the learning process \cite{katsnelson2021self}. In the future, it would be interesting to explore whether information processing processes in brains and artificial neural networks can be universally characterized by a unified criticality theory.

Overall, we anticipate the potential of well-validated studies of criticality in the brain to greatly deepen our understanding of neural dynamics characteristics and their roles in neural information processing. Laying solid theoretical foundations of studies is the most effective and indispensable path to contributing to this booming research area.  

\acknowledgments
Correspondence of this paper should be addressed to P.S. Author Y.T. conceptualizes the idea, develops theoretical frameworks, and writes the manuscript. Authors Z.R.T. and H.D.H. contribute equally to developing mathematics of neural avalanche exponents. Author G.Q.L. contributes to reviewing and revising the manuscript. Authors A.H.C., Y.K.Q., and K.Y.W. contribute equally to the proof reading of Langevin formulation. Author C.C. contributes to literature collection and summarizing. Author P.S. contributes to idea conceptualization, manuscript writing, and project supervision. This project is supported by the Artificial and General Intelligence Research Program of Guo Qiang Research Institute at Tsinghua University (2020GQG1017) as well as the Tsinghua University Initiative Scientific Research Program. Authors are grateful for discussions and assistance of Drs. Yaoyuan Wang and Ziyang Zhang from the Laboratory of Advanced Computing and Storage, Central Research Institute, 2012 Laboratories, Huawei Technologies Co. Ltd., Beijing, 100084, China. 


\bibliography{NETNbibsamp}

\begin{thebibliography}{}

\bibitem [\protect \citeauthoryear {%
Abbott%
}{%
Abbott%
}{%
{\protect \APACyear {2008}}%
}]{%
abbott2008theoretical}
\APACinsertmetastar {%
abbott2008theoretical}%
\begin{APACrefauthors}%
Abbott, L\BPBI F.%
\end{APACrefauthors}%
\unskip\
\newblock
\APACrefYearMonthDay{2008}{}{}.
\newblock
{\BBOQ}\APACrefatitle {Theoretical neuroscience rising} {Theoretical
  neuroscience rising}.{\BBCQ}
\newblock
\APACjournalVolNumPages{Neuron}{60}{3}{489--495}.
\PrintBackRefs{\CurrentBib}

\bibitem [\protect \citeauthoryear {%
Acebr\'on%
, Bonilla%
, P\'erez~Vicente%
, Ritort%
\BCBL {}\ \BBA {} Spigler%
}{%
Acebr\'on%
\ \protect \BOthers {.}}{%
{\protect \APACyear {2005}}%
}]{%
acebron2005Kuramoto}
\APACinsertmetastar {%
acebron2005Kuramoto}%
\begin{APACrefauthors}%
Acebr\'on, J\BPBI A.%
, Bonilla, L\BPBI L.%
, P\'erez~Vicente, C\BPBI J.%
, Ritort, F.%
\BCBL {}\ \BBA {} Spigler, R.%
\end{APACrefauthors}%
\unskip\
\newblock
\APACrefYearMonthDay{2005}{Apr}{}.
\newblock
{\BBOQ}\APACrefatitle {The Kuramoto model: A simple paradigm for
  synchronization phenomena} {The kuramoto model: A simple paradigm for
  synchronization phenomena}.{\BBCQ}
\newblock
\APACjournalVolNumPages{Rev. Mod. Phys.}{77}{}{137--185}.
\newblock
\begin{APACrefURL} \url{https://link.aps.org/doi/10.1103/RevModPhys.77.137}
  \end{APACrefURL}
\newblock
\begin{APACrefDOI} \doi{10.1103/RevModPhys.77.137} \end{APACrefDOI}
\PrintBackRefs{\CurrentBib}

\bibitem [\protect \citeauthoryear {%
Antonopoulos%
}{%
Antonopoulos%
}{%
{\protect \APACyear {2016}}%
}]{%
antonopoulos2016dynamic}
\APACinsertmetastar {%
antonopoulos2016dynamic}%
\begin{APACrefauthors}%
Antonopoulos, C\BPBI G.%
\end{APACrefauthors}%
\unskip\
\newblock
\APACrefYearMonthDay{2016}{}{}.
\newblock
{\BBOQ}\APACrefatitle {Dynamic range in the C. elegans brain network} {Dynamic
  range in the c. elegans brain network}.{\BBCQ}
\newblock
\APACjournalVolNumPages{Chaos: An Interdisciplinary Journal of Nonlinear
  Science}{26}{1}{013102}.
\PrintBackRefs{\CurrentBib}

\bibitem [\protect \citeauthoryear {%
Arenas%
, D{\'\i}az-Guilera%
, Kurths%
, Moreno%
\BCBL {}\ \BBA {} Zhou%
}{%
Arenas%
\ \protect \BOthers {.}}{%
{\protect \APACyear {2008}}%
}]{%
arenas2008synchronization}
\APACinsertmetastar {%
arenas2008synchronization}%
\begin{APACrefauthors}%
Arenas, A.%
, D{\'\i}az-Guilera, A.%
, Kurths, J.%
, Moreno, Y.%
\BCBL {}\ \BBA {} Zhou, C.%
\end{APACrefauthors}%
\unskip\
\newblock
\APACrefYearMonthDay{2008}{}{}.
\newblock
{\BBOQ}\APACrefatitle {Synchronization in complex networks} {Synchronization in
  complex networks}.{\BBCQ}
\newblock
\APACjournalVolNumPages{Physics reports}{469}{3}{93--153}.
\PrintBackRefs{\CurrentBib}

\bibitem [\protect \citeauthoryear {%
Atick%
}{%
Atick%
}{%
{\protect \APACyear {1992}}%
}]{%
atick1992could}
\APACinsertmetastar {%
atick1992could}%
\begin{APACrefauthors}%
Atick, J\BPBI J.%
\end{APACrefauthors}%
\unskip\
\newblock
\APACrefYearMonthDay{1992}{}{}.
\newblock
{\BBOQ}\APACrefatitle {Could information theory provide an ecological theory of
  sensory processing?} {Could information theory provide an ecological theory
  of sensory processing?}{\BBCQ}
\newblock
\APACjournalVolNumPages{Network: Computation in neural
  systems}{3}{2}{213--251}.
\PrintBackRefs{\CurrentBib}

\bibitem [\protect \citeauthoryear {%
Bak%
}{%
Bak%
}{%
{\protect \APACyear {2013}}%
}]{%
bak2013nature}
\APACinsertmetastar {%
bak2013nature}%
\begin{APACrefauthors}%
Bak, P.%
\end{APACrefauthors}%
\unskip\
\newblock
\APACrefYear{2013}.
\newblock
\APACrefbtitle {How nature works: the science of self-organized criticality}
  {How nature works: the science of self-organized criticality}.
\newblock
\APACaddressPublisher{}{Springer Science \& Business Media}.
\PrintBackRefs{\CurrentBib}

\bibitem [\protect \citeauthoryear {%
Bak%
\ \BBA {} Sneppen%
}{%
Bak%
\ \BBA {} Sneppen%
}{%
{\protect \APACyear {1993}}%
}]{%
bak1993punctuated}
\APACinsertmetastar {%
bak1993punctuated}%
\begin{APACrefauthors}%
Bak, P.%
\BCBT {}\ \BBA {} Sneppen, K.%
\end{APACrefauthors}%
\unskip\
\newblock
\APACrefYearMonthDay{1993}{}{}.
\newblock
{\BBOQ}\APACrefatitle {Punctuated equilibrium and criticality in a simple model
  of evolution} {Punctuated equilibrium and criticality in a simple model of
  evolution}.{\BBCQ}
\newblock
\APACjournalVolNumPages{Physical review letters}{71}{24}{4083}.
\PrintBackRefs{\CurrentBib}

\bibitem [\protect \citeauthoryear {%
Bak%
, Tang%
\BCBL {}\ \BBA {} Wiesenfeld%
}{%
Bak%
\ \protect \BOthers {.}}{%
{\protect \APACyear {1987}}%
}]{%
bak1987self}
\APACinsertmetastar {%
bak1987self}%
\begin{APACrefauthors}%
Bak, P.%
, Tang, C.%
\BCBL {}\ \BBA {} Wiesenfeld, K.%
\end{APACrefauthors}%
\unskip\
\newblock
\APACrefYearMonthDay{1987}{}{}.
\newblock
{\BBOQ}\APACrefatitle {Self-organized criticality: An explanation of the 1/f
  noise} {Self-organized criticality: An explanation of the 1/f noise}.{\BBCQ}
\newblock
\APACjournalVolNumPages{Physical review letters}{59}{4}{381}.
\PrintBackRefs{\CurrentBib}

\bibitem [\protect \citeauthoryear {%
Baldassarri%
, Colaiori%
\BCBL {}\ \BBA {} Castellano%
}{%
Baldassarri%
\ \protect \BOthers {.}}{%
{\protect \APACyear {2003}}%
}]{%
baldassarri2003average}
\APACinsertmetastar {%
baldassarri2003average}%
\begin{APACrefauthors}%
Baldassarri, A.%
, Colaiori, F.%
\BCBL {}\ \BBA {} Castellano, C.%
\end{APACrefauthors}%
\unskip\
\newblock
\APACrefYearMonthDay{2003}{}{}.
\newblock
{\BBOQ}\APACrefatitle {Average shape of a fluctuation: Universality in
  excursions of stochastic processes} {Average shape of a fluctuation:
  Universality in excursions of stochastic processes}.{\BBCQ}
\newblock
\APACjournalVolNumPages{Physical review letters}{90}{6}{060601}.
\PrintBackRefs{\CurrentBib}

\bibitem [\protect \citeauthoryear {%
Barlow%
\ \protect \BOthers {.}}{%
Barlow%
\ \protect \BOthers {.}}{%
{\protect \APACyear {1961}}%
}]{%
barlow1961possible}
\APACinsertmetastar {%
barlow1961possible}%
\begin{APACrefauthors}%
Barlow, H\BPBI B.%
\BCBT {}\ \BOthersPeriod {.}
\end{APACrefauthors}%
\unskip\
\newblock
\APACrefYearMonthDay{1961}{}{}.
\newblock
{\BBOQ}\APACrefatitle {Possible principles underlying the transformation of
  sensory messages} {Possible principles underlying the transformation of
  sensory messages}.{\BBCQ}
\newblock
\APACjournalVolNumPages{Sensory communication}{1}{01}{}.
\PrintBackRefs{\CurrentBib}

\bibitem [\protect \citeauthoryear {%
Bassett%
\ \protect \BOthers {.}}{%
Bassett%
\ \protect \BOthers {.}}{%
{\protect \APACyear {2010}}%
}]{%
bassett2010efficient}
\APACinsertmetastar {%
bassett2010efficient}%
\begin{APACrefauthors}%
Bassett, D\BPBI S.%
, Greenfield, D\BPBI L.%
, Meyer-Lindenberg, A.%
, Weinberger, D\BPBI R.%
, Moore, S\BPBI W.%
\BCBL {}\ \BBA {} Bullmore, E\BPBI T.%
\end{APACrefauthors}%
\unskip\
\newblock
\APACrefYearMonthDay{2010}{}{}.
\newblock
{\BBOQ}\APACrefatitle {Efficient physical embedding of topologically complex
  information processing networks in brains and computer circuits} {Efficient
  physical embedding of topologically complex information processing networks
  in brains and computer circuits}.{\BBCQ}
\newblock
\APACjournalVolNumPages{PLoS computational biology}{6}{4}{e1000748}.
\PrintBackRefs{\CurrentBib}

\bibitem [\protect \citeauthoryear {%
Bauke%
}{%
Bauke%
}{%
{\protect \APACyear {2007}}%
}]{%
bauke2007parameter}
\APACinsertmetastar {%
bauke2007parameter}%
\begin{APACrefauthors}%
Bauke, H.%
\end{APACrefauthors}%
\unskip\
\newblock
\APACrefYearMonthDay{2007}{}{}.
\newblock
{\BBOQ}\APACrefatitle {Parameter estimation for power-law distributions by
  maximum likelihood methods} {Parameter estimation for power-law distributions
  by maximum likelihood methods}.{\BBCQ}
\newblock
\APACjournalVolNumPages{The European Physical Journal B}{58}{2}{167--173}.
\PrintBackRefs{\CurrentBib}

\bibitem [\protect \citeauthoryear {%
Beggs%
}{%
Beggs%
}{%
{\protect \APACyear {2007}}%
}]{%
beggs2007build}
\APACinsertmetastar {%
beggs2007build}%
\begin{APACrefauthors}%
Beggs, J\BPBI M.%
\end{APACrefauthors}%
\unskip\
\newblock
\APACrefYearMonthDay{2007}{}{}.
\newblock
{\BBOQ}\APACrefatitle {How to build a critical mind} {How to build a critical
  mind}.{\BBCQ}
\newblock
\APACjournalVolNumPages{Nature Physics}{3}{12}{835--835}.
\PrintBackRefs{\CurrentBib}

\bibitem [\protect \citeauthoryear {%
Beggs%
\ \BBA {} Plenz%
}{%
Beggs%
\ \BBA {} Plenz%
}{%
{\protect \APACyear {2003}}%
}]{%
beggs2003neuronal}
\APACinsertmetastar {%
beggs2003neuronal}%
\begin{APACrefauthors}%
Beggs, J\BPBI M.%
\BCBT {}\ \BBA {} Plenz, D.%
\end{APACrefauthors}%
\unskip\
\newblock
\APACrefYearMonthDay{2003}{}{}.
\newblock
{\BBOQ}\APACrefatitle {Neuronal avalanches in neocortical circuits} {Neuronal
  avalanches in neocortical circuits}.{\BBCQ}
\newblock
\APACjournalVolNumPages{Journal of neuroscience}{23}{35}{11167--11177}.
\PrintBackRefs{\CurrentBib}

\bibitem [\protect \citeauthoryear {%
Beggs%
\ \BBA {} Timme%
}{%
Beggs%
\ \BBA {} Timme%
}{%
{\protect \APACyear {2012}}%
}]{%
beggs2012being}
\APACinsertmetastar {%
beggs2012being}%
\begin{APACrefauthors}%
Beggs, J\BPBI M.%
\BCBT {}\ \BBA {} Timme, N.%
\end{APACrefauthors}%
\unskip\
\newblock
\APACrefYearMonthDay{2012}{}{}.
\newblock
{\BBOQ}\APACrefatitle {Being critical of criticality in the brain} {Being
  critical of criticality in the brain}.{\BBCQ}
\newblock
\APACjournalVolNumPages{Frontiers in physiology}{3}{}{163}.
\PrintBackRefs{\CurrentBib}

\bibitem [\protect \citeauthoryear {%
Bell%
\ \BBA {} Sejnowski%
}{%
Bell%
\ \BBA {} Sejnowski%
}{%
{\protect \APACyear {1997}}%
}]{%
bell1997independent}
\APACinsertmetastar {%
bell1997independent}%
\begin{APACrefauthors}%
Bell, A\BPBI J.%
\BCBT {}\ \BBA {} Sejnowski, T\BPBI J.%
\end{APACrefauthors}%
\unskip\
\newblock
\APACrefYearMonthDay{1997}{}{}.
\newblock
{\BBOQ}\APACrefatitle {The “independent components” of natural scenes are
  edge filters} {The “independent components” of natural scenes are edge
  filters}.{\BBCQ}
\newblock
\APACjournalVolNumPages{Vision research}{37}{23}{3327--3338}.
\PrintBackRefs{\CurrentBib}

\bibitem [\protect \citeauthoryear {%
Berger%
\ \BBA {} Zhou%
}{%
Berger%
\ \BBA {} Zhou%
}{%
{\protect \APACyear {2014}}%
}]{%
berger2014kolmogorov}
\APACinsertmetastar {%
berger2014kolmogorov}%
\begin{APACrefauthors}%
Berger, V\BPBI W.%
\BCBT {}\ \BBA {} Zhou, Y.%
\end{APACrefauthors}%
\unskip\
\newblock
\APACrefYearMonthDay{2014}{}{}.
\newblock
{\BBOQ}\APACrefatitle {Kolmogorov--smirnov test: Overview} {Kolmogorov--smirnov
  test: Overview}.{\BBCQ}
\newblock
\APACjournalVolNumPages{Wiley statsref: Statistics reference online}{}{}{}.
\PrintBackRefs{\CurrentBib}

\bibitem [\protect \citeauthoryear {%
Bertschinger%
\ \BBA {} Natschl{\"a}ger%
}{%
Bertschinger%
\ \BBA {} Natschl{\"a}ger%
}{%
{\protect \APACyear {2004}}%
}]{%
bertschinger2004real}
\APACinsertmetastar {%
bertschinger2004real}%
\begin{APACrefauthors}%
Bertschinger, N.%
\BCBT {}\ \BBA {} Natschl{\"a}ger, T.%
\end{APACrefauthors}%
\unskip\
\newblock
\APACrefYearMonthDay{2004}{}{}.
\newblock
{\BBOQ}\APACrefatitle {Real-time computation at the edge of chaos in recurrent
  neural networks} {Real-time computation at the edge of chaos in recurrent
  neural networks}.{\BBCQ}
\newblock
\APACjournalVolNumPages{Neural computation}{16}{7}{1413--1436}.
\PrintBackRefs{\CurrentBib}

\bibitem [\protect \citeauthoryear {%
Betzel%
\ \protect \BOthers {.}}{%
Betzel%
\ \protect \BOthers {.}}{%
{\protect \APACyear {2016}}%
}]{%
betzel2016generative}
\APACinsertmetastar {%
betzel2016generative}%
\begin{APACrefauthors}%
Betzel, R\BPBI F.%
, Avena-Koenigsberger, A.%
, Go{\~n}i, J.%
, He, Y.%
, De~Reus, M\BPBI A.%
, Griffa, A.%
\BDBL {}others%
\end{APACrefauthors}%
\unskip\
\newblock
\APACrefYearMonthDay{2016}{}{}.
\newblock
{\BBOQ}\APACrefatitle {Generative models of the human connectome} {Generative
  models of the human connectome}.{\BBCQ}
\newblock
\APACjournalVolNumPages{Neuroimage}{124}{}{1054--1064}.
\PrintBackRefs{\CurrentBib}

\bibitem [\protect \citeauthoryear {%
Betzel%
\ \BBA {} Bassett%
}{%
Betzel%
\ \BBA {} Bassett%
}{%
{\protect \APACyear {2017}}%
{\protect \APACexlab {{\protect \BCnt {1}}}}}]{%
betzel2017generative}
\APACinsertmetastar {%
betzel2017generative}%
\begin{APACrefauthors}%
Betzel, R\BPBI F.%
\BCBT {}\ \BBA {} Bassett, D\BPBI S.%
\end{APACrefauthors}%
\unskip\
\newblock
\APACrefYearMonthDay{2017{\protect \BCnt {1}}}{}{}.
\newblock
{\BBOQ}\APACrefatitle {Generative models for network neuroscience: prospects
  and promise} {Generative models for network neuroscience: prospects and
  promise}.{\BBCQ}
\newblock
\APACjournalVolNumPages{Journal of The Royal Society
  Interface}{14}{136}{20170623}.
\PrintBackRefs{\CurrentBib}

\bibitem [\protect \citeauthoryear {%
Betzel%
\ \BBA {} Bassett%
}{%
Betzel%
\ \BBA {} Bassett%
}{%
{\protect \APACyear {2017}}%
{\protect \APACexlab {{\protect \BCnt {2}}}}}]{%
betzel2017multi}
\APACinsertmetastar {%
betzel2017multi}%
\begin{APACrefauthors}%
Betzel, R\BPBI F.%
\BCBT {}\ \BBA {} Bassett, D\BPBI S.%
\end{APACrefauthors}%
\unskip\
\newblock
\APACrefYearMonthDay{2017{\protect \BCnt {2}}}{}{}.
\newblock
{\BBOQ}\APACrefatitle {Multi-scale brain networks} {Multi-scale brain
  networks}.{\BBCQ}
\newblock
\APACjournalVolNumPages{Neuroimage}{160}{}{73--83}.
\PrintBackRefs{\CurrentBib}

\bibitem [\protect \citeauthoryear {%
Betzel%
, Medaglia%
\BCBL {}\ \BBA {} Bassett%
}{%
Betzel%
\ \protect \BOthers {.}}{%
{\protect \APACyear {2018}}%
}]{%
betzel2018diversity}
\APACinsertmetastar {%
betzel2018diversity}%
\begin{APACrefauthors}%
Betzel, R\BPBI F.%
, Medaglia, J\BPBI D.%
\BCBL {}\ \BBA {} Bassett, D\BPBI S.%
\end{APACrefauthors}%
\unskip\
\newblock
\APACrefYearMonthDay{2018}{}{}.
\newblock
{\BBOQ}\APACrefatitle {Diversity of meso-scale architecture in human and
  non-human connectomes} {Diversity of meso-scale architecture in human and
  non-human connectomes}.{\BBCQ}
\newblock
\APACjournalVolNumPages{Nature communications}{9}{1}{1--14}.
\PrintBackRefs{\CurrentBib}

\bibitem [\protect \citeauthoryear {%
Bhattacharjee%
\ \BBA {} Seno%
}{%
Bhattacharjee%
\ \BBA {} Seno%
}{%
{\protect \APACyear {2001}}%
}]{%
bhattacharjee2001measure}
\APACinsertmetastar {%
bhattacharjee2001measure}%
\begin{APACrefauthors}%
Bhattacharjee, S\BPBI M.%
\BCBT {}\ \BBA {} Seno, F.%
\end{APACrefauthors}%
\unskip\
\newblock
\APACrefYearMonthDay{2001}{}{}.
\newblock
{\BBOQ}\APACrefatitle {A measure of data collapse for scaling} {A measure of
  data collapse for scaling}.{\BBCQ}
\newblock
\APACjournalVolNumPages{Journal of Physics A: Mathematical and
  General}{34}{33}{6375}.
\PrintBackRefs{\CurrentBib}

\bibitem [\protect \citeauthoryear {%
Blythe%
\ \BBA {} McKane%
}{%
Blythe%
\ \BBA {} McKane%
}{%
{\protect \APACyear {2007}}%
}]{%
blythe2007stochastic}
\APACinsertmetastar {%
blythe2007stochastic}%
\begin{APACrefauthors}%
Blythe, R\BPBI A.%
\BCBT {}\ \BBA {} McKane, A\BPBI J.%
\end{APACrefauthors}%
\unskip\
\newblock
\APACrefYearMonthDay{2007}{}{}.
\newblock
{\BBOQ}\APACrefatitle {Stochastic models of evolution in genetics, ecology and
  linguistics} {Stochastic models of evolution in genetics, ecology and
  linguistics}.{\BBCQ}
\newblock
\APACjournalVolNumPages{Journal of Statistical Mechanics: Theory and
  Experiment}{2007}{07}{P07018}.
\PrintBackRefs{\CurrentBib}

\bibitem [\protect \citeauthoryear {%
Boedecker%
, Obst%
, Lizier%
, Mayer%
\BCBL {}\ \BBA {} Asada%
}{%
Boedecker%
\ \protect \BOthers {.}}{%
{\protect \APACyear {2012}}%
}]{%
boedecker2012information}
\APACinsertmetastar {%
boedecker2012information}%
\begin{APACrefauthors}%
Boedecker, J.%
, Obst, O.%
, Lizier, J\BPBI T.%
, Mayer, N\BPBI M.%
\BCBL {}\ \BBA {} Asada, M.%
\end{APACrefauthors}%
\unskip\
\newblock
\APACrefYearMonthDay{2012}{}{}.
\newblock
{\BBOQ}\APACrefatitle {Information processing in echo state networks at the
  edge of chaos} {Information processing in echo state networks at the edge of
  chaos}.{\BBCQ}
\newblock
\APACjournalVolNumPages{Theory in Biosciences}{131}{3}{205--213}.
\PrintBackRefs{\CurrentBib}

\bibitem [\protect \citeauthoryear {%
Bonachela%
, De~Franciscis%
, Torres%
\BCBL {}\ \BBA {} Munoz%
}{%
Bonachela%
\ \protect \BOthers {.}}{%
{\protect \APACyear {2010}}%
}]{%
bonachela2010self}
\APACinsertmetastar {%
bonachela2010self}%
\begin{APACrefauthors}%
Bonachela, J\BPBI A.%
, De~Franciscis, S.%
, Torres, J\BPBI J.%
\BCBL {}\ \BBA {} Munoz, M\BPBI A.%
\end{APACrefauthors}%
\unskip\
\newblock
\APACrefYearMonthDay{2010}{}{}.
\newblock
{\BBOQ}\APACrefatitle {Self-organization without conservation: are neuronal
  avalanches generically critical?} {Self-organization without conservation:
  are neuronal avalanches generically critical?}{\BBCQ}
\newblock
\APACjournalVolNumPages{Journal of Statistical Mechanics: Theory and
  Experiment}{2010}{02}{P02015}.
\PrintBackRefs{\CurrentBib}

\bibitem [\protect \citeauthoryear {%
Bonachela%
\ \BBA {} Mu{\~n}oz%
}{%
Bonachela%
\ \BBA {} Mu{\~n}oz%
}{%
{\protect \APACyear {2008}}%
}]{%
bonachela2008confirming}
\APACinsertmetastar {%
bonachela2008confirming}%
\begin{APACrefauthors}%
Bonachela, J\BPBI A.%
\BCBT {}\ \BBA {} Mu{\~n}oz, M\BPBI A.%
\end{APACrefauthors}%
\unskip\
\newblock
\APACrefYearMonthDay{2008}{}{}.
\newblock
{\BBOQ}\APACrefatitle {Confirming and extending the hypothesis of universality
  in sandpiles} {Confirming and extending the hypothesis of universality in
  sandpiles}.{\BBCQ}
\newblock
\APACjournalVolNumPages{Physical Review E}{78}{4}{041102}.
\PrintBackRefs{\CurrentBib}

\bibitem [\protect \citeauthoryear {%
Bonachela%
\ \BBA {} Munoz%
}{%
Bonachela%
\ \BBA {} Munoz%
}{%
{\protect \APACyear {2009}}%
}]{%
bonachela2009self}
\APACinsertmetastar {%
bonachela2009self}%
\begin{APACrefauthors}%
Bonachela, J\BPBI A.%
\BCBT {}\ \BBA {} Munoz, M\BPBI A.%
\end{APACrefauthors}%
\unskip\
\newblock
\APACrefYearMonthDay{2009}{}{}.
\newblock
{\BBOQ}\APACrefatitle {Self-organization without conservation: true or just
  apparent scale-invariance?} {Self-organization without conservation: true or
  just apparent scale-invariance?}{\BBCQ}
\newblock
\APACjournalVolNumPages{Journal of Statistical Mechanics: Theory and
  Experiment}{2009}{09}{P09009}.
\PrintBackRefs{\CurrentBib}

\bibitem [\protect \citeauthoryear {%
Breskin%
, Soriano%
, Moses%
\BCBL {}\ \BBA {} Tlusty%
}{%
Breskin%
\ \protect \BOthers {.}}{%
{\protect \APACyear {2006}}%
}]{%
breskin2006percolation}
\APACinsertmetastar {%
breskin2006percolation}%
\begin{APACrefauthors}%
Breskin, I.%
, Soriano, J.%
, Moses, E.%
\BCBL {}\ \BBA {} Tlusty, T.%
\end{APACrefauthors}%
\unskip\
\newblock
\APACrefYearMonthDay{2006}{}{}.
\newblock
{\BBOQ}\APACrefatitle {Percolation in living neural networks} {Percolation in
  living neural networks}.{\BBCQ}
\newblock
\APACjournalVolNumPages{Physical review letters}{97}{18}{188102}.
\PrintBackRefs{\CurrentBib}

\bibitem [\protect \citeauthoryear {%
Buend{\'\i}a%
, Di~Santo%
, Bonachela%
\BCBL {}\ \BBA {} Mu{\~n}oz%
}{%
Buend{\'\i}a%
, Di~Santo%
\BCBL {}\ \protect \BOthers {.}}{%
{\protect \APACyear {2020}}%
}]{%
buendia2020feedback}
\APACinsertmetastar {%
buendia2020feedback}%
\begin{APACrefauthors}%
Buend{\'\i}a, V.%
, Di~Santo, S.%
, Bonachela, J\BPBI A.%
\BCBL {}\ \BBA {} Mu{\~n}oz, M\BPBI A.%
\end{APACrefauthors}%
\unskip\
\newblock
\APACrefYearMonthDay{2020}{}{}.
\newblock
{\BBOQ}\APACrefatitle {Feedback mechanisms for self-organization to the edge of
  a phase transition} {Feedback mechanisms for self-organization to the edge of
  a phase transition}.{\BBCQ}
\newblock
\APACjournalVolNumPages{Frontiers in physics}{8}{}{333}.
\PrintBackRefs{\CurrentBib}

\bibitem [\protect \citeauthoryear {%
Buend{\'\i}a%
, di Santo%
, Villegas%
, Burioni%
\BCBL {}\ \BBA {} Mu{\~n}oz%
}{%
Buend{\'\i}a%
, di Santo%
\BCBL {}\ \protect \BOthers {.}}{%
{\protect \APACyear {2020}}%
}]{%
buendia2020self}
\APACinsertmetastar {%
buendia2020self}%
\begin{APACrefauthors}%
Buend{\'\i}a, V.%
, di Santo, S.%
, Villegas, P.%
, Burioni, R.%
\BCBL {}\ \BBA {} Mu{\~n}oz, M\BPBI A.%
\end{APACrefauthors}%
\unskip\
\newblock
\APACrefYearMonthDay{2020}{}{}.
\newblock
{\BBOQ}\APACrefatitle {Self-organized bistability and its possible relevance
  for brain dynamics} {Self-organized bistability and its possible relevance
  for brain dynamics}.{\BBCQ}
\newblock
\APACjournalVolNumPages{Physical Review Research}{2}{1}{013318}.
\PrintBackRefs{\CurrentBib}

\bibitem [\protect \citeauthoryear {%
Buend{\'\i}a%
, Villegas%
, Burioni%
\BCBL {}\ \BBA {} Munoz%
}{%
Buend{\'\i}a%
\ \protect \BOthers {.}}{%
{\protect \APACyear {2021}}%
}]{%
buendia2021hybrid}
\APACinsertmetastar {%
buendia2021hybrid}%
\begin{APACrefauthors}%
Buend{\'\i}a, V.%
, Villegas, P.%
, Burioni, R.%
\BCBL {}\ \BBA {} Munoz, M\BPBI A.%
\end{APACrefauthors}%
\unskip\
\newblock
\APACrefYearMonthDay{2021}{}{}.
\newblock
{\BBOQ}\APACrefatitle {Hybrid-type synchronization transitions: Where incipient
  oscillations, scale-free avalanches, and bistability live together}
  {Hybrid-type synchronization transitions: Where incipient oscillations,
  scale-free avalanches, and bistability live together}.{\BBCQ}
\newblock
\APACjournalVolNumPages{Physical Review Research}{3}{2}{023224}.
\PrintBackRefs{\CurrentBib}

\bibitem [\protect \citeauthoryear {%
Bullmore%
\ \BBA {} Sporns%
}{%
Bullmore%
\ \BBA {} Sporns%
}{%
{\protect \APACyear {2012}}%
}]{%
bullmore2012economy}
\APACinsertmetastar {%
bullmore2012economy}%
\begin{APACrefauthors}%
Bullmore, E.%
\BCBT {}\ \BBA {} Sporns, O.%
\end{APACrefauthors}%
\unskip\
\newblock
\APACrefYearMonthDay{2012}{}{}.
\newblock
{\BBOQ}\APACrefatitle {The economy of brain network organization} {The economy
  of brain network organization}.{\BBCQ}
\newblock
\APACjournalVolNumPages{Nature reviews neuroscience}{13}{5}{336--349}.
\PrintBackRefs{\CurrentBib}

\bibitem [\protect \citeauthoryear {%
Burns%
\ \BBA {} Webb%
}{%
Burns%
\ \BBA {} Webb%
}{%
{\protect \APACyear {1976}}%
}]{%
burns1976spontaneous}
\APACinsertmetastar {%
burns1976spontaneous}%
\begin{APACrefauthors}%
Burns, B\BPBI D.%
\BCBT {}\ \BBA {} Webb, A.%
\end{APACrefauthors}%
\unskip\
\newblock
\APACrefYearMonthDay{1976}{}{}.
\newblock
{\BBOQ}\APACrefatitle {The spontaneous activity of neurones in the cat’s
  cerebral cortex} {The spontaneous activity of neurones in the cat’s
  cerebral cortex}.{\BBCQ}
\newblock
\APACjournalVolNumPages{Proceedings of the Royal Society of London. Series B.
  Biological Sciences}{194}{1115}{211--223}.
\PrintBackRefs{\CurrentBib}

\bibitem [\protect \citeauthoryear {%
Capolupo%
, Freeman%
\BCBL {}\ \BBA {} Vitiello%
}{%
Capolupo%
\ \protect \BOthers {.}}{%
{\protect \APACyear {2013}}%
}]{%
capolupo2013dissipation}
\APACinsertmetastar {%
capolupo2013dissipation}%
\begin{APACrefauthors}%
Capolupo, A.%
, Freeman, W\BPBI J.%
\BCBL {}\ \BBA {} Vitiello, G.%
\end{APACrefauthors}%
\unskip\
\newblock
\APACrefYearMonthDay{2013}{}{}.
\newblock
{\BBOQ}\APACrefatitle {Dissipation of ‘dark energy’by cortex in knowledge
  retrieval} {Dissipation of ‘dark energy’by cortex in knowledge
  retrieval}.{\BBCQ}
\newblock
\APACjournalVolNumPages{Physics of life reviews}{10}{1}{85--94}.
\PrintBackRefs{\CurrentBib}

\bibitem [\protect \citeauthoryear {%
Carandini%
\ \BBA {} Stevens%
}{%
Carandini%
\ \BBA {} Stevens%
}{%
{\protect \APACyear {2004}}%
}]{%
carandini2004amplification}
\APACinsertmetastar {%
carandini2004amplification}%
\begin{APACrefauthors}%
Carandini, M.%
\BCBT {}\ \BBA {} Stevens, C.%
\end{APACrefauthors}%
\unskip\
\newblock
\APACrefYearMonthDay{2004}{}{}.
\newblock
{\BBOQ}\APACrefatitle {Amplification of trial-to-trial response variability by
  neurons in visual cortex} {Amplification of trial-to-trial response
  variability by neurons in visual cortex}.{\BBCQ}
\newblock
\APACjournalVolNumPages{PLoS biology}{2}{9}{e264}.
\PrintBackRefs{\CurrentBib}

\bibitem [\protect \citeauthoryear {%
Cardin%
}{%
Cardin%
}{%
{\protect \APACyear {2019}}%
}]{%
cardin2019functional}
\APACinsertmetastar {%
cardin2019functional}%
\begin{APACrefauthors}%
Cardin, J\BPBI A.%
\end{APACrefauthors}%
\unskip\
\newblock
\APACrefYearMonthDay{2019}{}{}.
\newblock
{\BBOQ}\APACrefatitle {Functional flexibility in cortical circuits} {Functional
  flexibility in cortical circuits}.{\BBCQ}
\newblock
\APACjournalVolNumPages{Current opinion in neurobiology}{58}{}{175--180}.
\PrintBackRefs{\CurrentBib}

\bibitem [\protect \citeauthoryear {%
Carvalho%
\ \protect \BOthers {.}}{%
Carvalho%
\ \protect \BOthers {.}}{%
{\protect \APACyear {2021}}%
}]{%
carvalho2021subsampled}
\APACinsertmetastar {%
carvalho2021subsampled}%
\begin{APACrefauthors}%
Carvalho, T\BPBI T.%
, Fontenele, A\BPBI J.%
, Girardi-Schappo, M.%
, Feliciano, T.%
, Aguiar, L\BPBI A.%
, Silva, T\BPBI P.%
\BDBL {}Copelli, M.%
\end{APACrefauthors}%
\unskip\
\newblock
\APACrefYearMonthDay{2021}{}{}.
\newblock
{\BBOQ}\APACrefatitle {Subsampled directed-percolation models explain scaling
  relations experimentally observed in the brain} {Subsampled
  directed-percolation models explain scaling relations experimentally observed
  in the brain}.{\BBCQ}
\newblock
\APACjournalVolNumPages{Frontiers in neural circuits}{}{}{83}.
\PrintBackRefs{\CurrentBib}

\bibitem [\protect \citeauthoryear {%
Chialvo%
}{%
Chialvo%
}{%
{\protect \APACyear {2010}}%
}]{%
chialvo2010emergent}
\APACinsertmetastar {%
chialvo2010emergent}%
\begin{APACrefauthors}%
Chialvo, D\BPBI R.%
\end{APACrefauthors}%
\unskip\
\newblock
\APACrefYearMonthDay{2010}{}{}.
\newblock
{\BBOQ}\APACrefatitle {Emergent complex neural dynamics} {Emergent complex
  neural dynamics}.{\BBCQ}
\newblock
\APACjournalVolNumPages{Nature physics}{6}{10}{744--750}.
\PrintBackRefs{\CurrentBib}

\bibitem [\protect \citeauthoryear {%
Chiappalone%
\ \protect \BOthers {.}}{%
Chiappalone%
\ \protect \BOthers {.}}{%
{\protect \APACyear {2003}}%
}]{%
chiappalone2003networks}
\APACinsertmetastar {%
chiappalone2003networks}%
\begin{APACrefauthors}%
Chiappalone, M.%
, Vato, A.%
, Marcoli, M.%
, Davide, F.%
, Martinoia, S.%
\BCBL {}\ \BOthersPeriod {.}\end{APACrefauthors}%
\unskip\
\newblock
\APACrefYearMonthDay{2003}{}{}.
\newblock
{\BBOQ}\APACrefatitle {Networks of neurons coupled to microelectrode arrays: a
  neuronal sensory system for pharmacological applications} {Networks of
  neurons coupled to microelectrode arrays: a neuronal sensory system for
  pharmacological applications}.{\BBCQ}
\newblock
\APACjournalVolNumPages{Biosensors and Bioelectronics}{18}{5-6}{627--634}.
\PrintBackRefs{\CurrentBib}

\bibitem [\protect \citeauthoryear {%
Clauset%
, Shalizi%
\BCBL {}\ \BBA {} Newman%
}{%
Clauset%
\ \protect \BOthers {.}}{%
{\protect \APACyear {2009}}%
}]{%
clauset2009power}
\APACinsertmetastar {%
clauset2009power}%
\begin{APACrefauthors}%
Clauset, A.%
, Shalizi, C\BPBI R.%
\BCBL {}\ \BBA {} Newman, M\BPBI E.%
\end{APACrefauthors}%
\unskip\
\newblock
\APACrefYearMonthDay{2009}{}{}.
\newblock
{\BBOQ}\APACrefatitle {Power-law distributions in empirical data} {Power-law
  distributions in empirical data}.{\BBCQ}
\newblock
\APACjournalVolNumPages{SIAM review}{51}{4}{661--703}.
\PrintBackRefs{\CurrentBib}

\bibitem [\protect \citeauthoryear {%
Cocchi%
, Gollo%
, Zalesky%
\BCBL {}\ \BBA {} Breakspear%
}{%
Cocchi%
\ \protect \BOthers {.}}{%
{\protect \APACyear {2017}}%
}]{%
cocchi2017criticality}
\APACinsertmetastar {%
cocchi2017criticality}%
\begin{APACrefauthors}%
Cocchi, L.%
, Gollo, L\BPBI L.%
, Zalesky, A.%
\BCBL {}\ \BBA {} Breakspear, M.%
\end{APACrefauthors}%
\unskip\
\newblock
\APACrefYearMonthDay{2017}{}{}.
\newblock
{\BBOQ}\APACrefatitle {Criticality in the brain: A synthesis of neurobiology,
  models and cognition} {Criticality in the brain: A synthesis of neurobiology,
  models and cognition}.{\BBCQ}
\newblock
\APACjournalVolNumPages{Progress in neurobiology}{158}{}{132--152}.
\PrintBackRefs{\CurrentBib}

\bibitem [\protect \citeauthoryear {%
Cohen%
\ \BBA {} Kohn%
}{%
Cohen%
\ \BBA {} Kohn%
}{%
{\protect \APACyear {2011}}%
}]{%
cohen2011measuring}
\APACinsertmetastar {%
cohen2011measuring}%
\begin{APACrefauthors}%
Cohen, M\BPBI R.%
\BCBT {}\ \BBA {} Kohn, A.%
\end{APACrefauthors}%
\unskip\
\newblock
\APACrefYearMonthDay{2011}{}{}.
\newblock
{\BBOQ}\APACrefatitle {Measuring and interpreting neuronal correlations}
  {Measuring and interpreting neuronal correlations}.{\BBCQ}
\newblock
\APACjournalVolNumPages{Nature neuroscience}{14}{7}{811--819}.
\PrintBackRefs{\CurrentBib}

\bibitem [\protect \citeauthoryear {%
Collell%
\ \BBA {} Fauquet%
}{%
Collell%
\ \BBA {} Fauquet%
}{%
{\protect \APACyear {2015}}%
}]{%
collell2015brain}
\APACinsertmetastar {%
collell2015brain}%
\begin{APACrefauthors}%
Collell, G.%
\BCBT {}\ \BBA {} Fauquet, J.%
\end{APACrefauthors}%
\unskip\
\newblock
\APACrefYearMonthDay{2015}{}{}.
\newblock
{\BBOQ}\APACrefatitle {Brain activity and cognition: a connection from
  thermodynamics and information theory} {Brain activity and cognition: a
  connection from thermodynamics and information theory}.{\BBCQ}
\newblock
\APACjournalVolNumPages{Frontiers in psychology}{6}{}{818}.
\PrintBackRefs{\CurrentBib}

\bibitem [\protect \citeauthoryear {%
Dalla~Porta%
\ \BBA {} Copelli%
}{%
Dalla~Porta%
\ \BBA {} Copelli%
}{%
{\protect \APACyear {2019}}%
}]{%
dalla2019modeling}
\APACinsertmetastar {%
dalla2019modeling}%
\begin{APACrefauthors}%
Dalla~Porta, L.%
\BCBT {}\ \BBA {} Copelli, M.%
\end{APACrefauthors}%
\unskip\
\newblock
\APACrefYearMonthDay{2019}{}{}.
\newblock
{\BBOQ}\APACrefatitle {Modeling neuronal avalanches and long-range temporal
  correlations at the emergence of collective oscillations: Continuously
  varying exponents mimic M/EEG results} {Modeling neuronal avalanches and
  long-range temporal correlations at the emergence of collective oscillations:
  Continuously varying exponents mimic m/eeg results}.{\BBCQ}
\newblock
\APACjournalVolNumPages{PLoS computational biology}{15}{4}{e1006924}.
\PrintBackRefs{\CurrentBib}

\bibitem [\protect \citeauthoryear {%
David%
\ \BBA {} Friston%
}{%
David%
\ \BBA {} Friston%
}{%
{\protect \APACyear {2003}}%
}]{%
david2003neural}
\APACinsertmetastar {%
david2003neural}%
\begin{APACrefauthors}%
David, O.%
\BCBT {}\ \BBA {} Friston, K\BPBI J.%
\end{APACrefauthors}%
\unskip\
\newblock
\APACrefYearMonthDay{2003}{}{}.
\newblock
{\BBOQ}\APACrefatitle {A neural mass model for MEG/EEG:: coupling and neuronal
  dynamics} {A neural mass model for meg/eeg:: coupling and neuronal
  dynamics}.{\BBCQ}
\newblock
\APACjournalVolNumPages{NeuroImage}{20}{3}{1743--1755}.
\PrintBackRefs{\CurrentBib}

\bibitem [\protect \citeauthoryear {%
de Andrade~Costa%
, Copelli%
\BCBL {}\ \BBA {} Kinouchi%
}{%
de Andrade~Costa%
\ \protect \BOthers {.}}{%
{\protect \APACyear {2015}}%
}]{%
de2015can}
\APACinsertmetastar {%
de2015can}%
\begin{APACrefauthors}%
de Andrade~Costa, A.%
, Copelli, M.%
\BCBL {}\ \BBA {} Kinouchi, O.%
\end{APACrefauthors}%
\unskip\
\newblock
\APACrefYearMonthDay{2015}{}{}.
\newblock
{\BBOQ}\APACrefatitle {Can dynamical synapses produce true self-organized
  criticality?} {Can dynamical synapses produce true self-organized
  criticality?}{\BBCQ}
\newblock
\APACjournalVolNumPages{Journal of Statistical Mechanics: Theory and
  Experiment}{2015}{6}{P06004}.
\PrintBackRefs{\CurrentBib}

\bibitem [\protect \citeauthoryear {%
de Arcangelis%
\ \BBA {} Herrmann%
}{%
de Arcangelis%
\ \BBA {} Herrmann%
}{%
{\protect \APACyear {2010}}%
}]{%
de2010learning}
\APACinsertmetastar {%
de2010learning}%
\begin{APACrefauthors}%
de Arcangelis, L.%
\BCBT {}\ \BBA {} Herrmann, H\BPBI J.%
\end{APACrefauthors}%
\unskip\
\newblock
\APACrefYearMonthDay{2010}{}{}.
\newblock
{\BBOQ}\APACrefatitle {Learning as a phenomenon occurring in a critical state}
  {Learning as a phenomenon occurring in a critical state}.{\BBCQ}
\newblock
\APACjournalVolNumPages{Proceedings of the National Academy of
  Sciences}{107}{9}{3977--3981}.
\PrintBackRefs{\CurrentBib}

\bibitem [\protect \citeauthoryear {%
De~Arcangelis%
, Perrone-Capano%
\BCBL {}\ \BBA {} Herrmann%
}{%
De~Arcangelis%
\ \protect \BOthers {.}}{%
{\protect \APACyear {2006}}%
}]{%
de2006self}
\APACinsertmetastar {%
de2006self}%
\begin{APACrefauthors}%
De~Arcangelis, L.%
, Perrone-Capano, C.%
\BCBL {}\ \BBA {} Herrmann, H\BPBI J.%
\end{APACrefauthors}%
\unskip\
\newblock
\APACrefYearMonthDay{2006}{}{}.
\newblock
{\BBOQ}\APACrefatitle {Self-organized criticality model for brain plasticity}
  {Self-organized criticality model for brain plasticity}.{\BBCQ}
\newblock
\APACjournalVolNumPages{Physical review letters}{96}{2}{028107}.
\PrintBackRefs{\CurrentBib}

\bibitem [\protect \citeauthoryear {%
Deco%
, Tononi%
, Boly%
\BCBL {}\ \BBA {} Kringelbach%
}{%
Deco%
\ \protect \BOthers {.}}{%
{\protect \APACyear {2015}}%
}]{%
deco2015rethinking}
\APACinsertmetastar {%
deco2015rethinking}%
\begin{APACrefauthors}%
Deco, G.%
, Tononi, G.%
, Boly, M.%
\BCBL {}\ \BBA {} Kringelbach, M\BPBI L.%
\end{APACrefauthors}%
\unskip\
\newblock
\APACrefYearMonthDay{2015}{}{}.
\newblock
{\BBOQ}\APACrefatitle {Rethinking segregation and integration: contributions of
  whole-brain modelling} {Rethinking segregation and integration: contributions
  of whole-brain modelling}.{\BBCQ}
\newblock
\APACjournalVolNumPages{Nature Reviews Neuroscience}{16}{7}{430--439}.
\PrintBackRefs{\CurrentBib}

\bibitem [\protect \citeauthoryear {%
Del~Papa%
, Priesemann%
\BCBL {}\ \BBA {} Triesch%
}{%
Del~Papa%
\ \protect \BOthers {.}}{%
{\protect \APACyear {2017}}%
}]{%
del2017criticality}
\APACinsertmetastar {%
del2017criticality}%
\begin{APACrefauthors}%
Del~Papa, B.%
, Priesemann, V.%
\BCBL {}\ \BBA {} Triesch, J.%
\end{APACrefauthors}%
\unskip\
\newblock
\APACrefYearMonthDay{2017}{}{}.
\newblock
{\BBOQ}\APACrefatitle {Criticality meets learning: Criticality signatures in a
  self-organizing recurrent neural network} {Criticality meets learning:
  Criticality signatures in a self-organizing recurrent neural network}.{\BBCQ}
\newblock
\APACjournalVolNumPages{PloS one}{12}{5}{e0178683}.
\PrintBackRefs{\CurrentBib}

\bibitem [\protect \citeauthoryear {%
Del~Pozo%
\ \protect \BOthers {.}}{%
Del~Pozo%
\ \protect \BOthers {.}}{%
{\protect \APACyear {2021}}%
}]{%
del2021unconsciousness}
\APACinsertmetastar {%
del2021unconsciousness}%
\begin{APACrefauthors}%
Del~Pozo, S\BPBI M.%
, Laufs, H.%
, Bonhomme, V.%
, Laureys, S.%
, Balenzuela, P.%
\BCBL {}\ \BBA {} Tagliazucchi, E.%
\end{APACrefauthors}%
\unskip\
\newblock
\APACrefYearMonthDay{2021}{}{}.
\newblock
{\BBOQ}\APACrefatitle {Unconsciousness reconfigures modular brain network
  dynamics} {Unconsciousness reconfigures modular brain network
  dynamics}.{\BBCQ}
\newblock
\APACjournalVolNumPages{Chaos: An Interdisciplinary Journal of Nonlinear
  Science}{31}{9}{093117}.
\PrintBackRefs{\CurrentBib}

\bibitem [\protect \citeauthoryear {%
Deluca%
\ \BBA {} Corral%
}{%
Deluca%
\ \BBA {} Corral%
}{%
{\protect \APACyear {2013}}%
}]{%
deluca2013fitting}
\APACinsertmetastar {%
deluca2013fitting}%
\begin{APACrefauthors}%
Deluca, A.%
\BCBT {}\ \BBA {} Corral, {\'A}.%
\end{APACrefauthors}%
\unskip\
\newblock
\APACrefYearMonthDay{2013}{}{}.
\newblock
{\BBOQ}\APACrefatitle {Fitting and goodness-of-fit test of non-truncated and
  truncated power-law distributions} {Fitting and goodness-of-fit test of
  non-truncated and truncated power-law distributions}.{\BBCQ}
\newblock
\APACjournalVolNumPages{Acta Geophysica}{61}{6}{1351--1394}.
\PrintBackRefs{\CurrentBib}

\bibitem [\protect \citeauthoryear {%
Dickman%
, Mu{\~n}oz%
, Vespignani%
\BCBL {}\ \BBA {} Zapperi%
}{%
Dickman%
\ \protect \BOthers {.}}{%
{\protect \APACyear {2000}}%
}]{%
dickman2000paths}
\APACinsertmetastar {%
dickman2000paths}%
\begin{APACrefauthors}%
Dickman, R.%
, Mu{\~n}oz, M\BPBI A.%
, Vespignani, A.%
\BCBL {}\ \BBA {} Zapperi, S.%
\end{APACrefauthors}%
\unskip\
\newblock
\APACrefYearMonthDay{2000}{}{}.
\newblock
{\BBOQ}\APACrefatitle {Paths to self-organized criticality} {Paths to
  self-organized criticality}.{\BBCQ}
\newblock
\APACjournalVolNumPages{Brazilian Journal of Physics}{30}{1}{27--41}.
\PrintBackRefs{\CurrentBib}

\bibitem [\protect \citeauthoryear {%
Dickman%
, Vespignani%
\BCBL {}\ \BBA {} Zapperi%
}{%
Dickman%
\ \protect \BOthers {.}}{%
{\protect \APACyear {1998}}%
}]{%
dickman1998self}
\APACinsertmetastar {%
dickman1998self}%
\begin{APACrefauthors}%
Dickman, R.%
, Vespignani, A.%
\BCBL {}\ \BBA {} Zapperi, S.%
\end{APACrefauthors}%
\unskip\
\newblock
\APACrefYearMonthDay{1998}{}{}.
\newblock
{\BBOQ}\APACrefatitle {Self-organized criticality as an absorbing-state phase
  transition} {Self-organized criticality as an absorbing-state phase
  transition}.{\BBCQ}
\newblock
\APACjournalVolNumPages{Physical Review E}{57}{5}{5095}.
\PrintBackRefs{\CurrentBib}

\bibitem [\protect \citeauthoryear {%
di Santo%
, Burioni%
, Vezzani%
\BCBL {}\ \BBA {} Munoz%
}{%
di Santo%
\ \protect \BOthers {.}}{%
{\protect \APACyear {2016}}%
}]{%
di2016self}
\APACinsertmetastar {%
di2016self}%
\begin{APACrefauthors}%
di Santo, S.%
, Burioni, R.%
, Vezzani, A.%
\BCBL {}\ \BBA {} Munoz, M\BPBI A.%
\end{APACrefauthors}%
\unskip\
\newblock
\APACrefYearMonthDay{2016}{}{}.
\newblock
{\BBOQ}\APACrefatitle {Self-organized bistability associated with first-order
  phase transitions} {Self-organized bistability associated with first-order
  phase transitions}.{\BBCQ}
\newblock
\APACjournalVolNumPages{Physical review letters}{116}{24}{240601}.
\PrintBackRefs{\CurrentBib}

\bibitem [\protect \citeauthoryear {%
di Santo%
, Villegas%
, Burioni%
\BCBL {}\ \BBA {} Mu{\~n}oz%
}{%
di Santo%
\ \protect \BOthers {.}}{%
{\protect \APACyear {2017}}%
}]{%
di2017simple}
\APACinsertmetastar {%
di2017simple}%
\begin{APACrefauthors}%
di Santo, S.%
, Villegas, P.%
, Burioni, R.%
\BCBL {}\ \BBA {} Mu{\~n}oz, M\BPBI A.%
\end{APACrefauthors}%
\unskip\
\newblock
\APACrefYearMonthDay{2017}{}{}.
\newblock
{\BBOQ}\APACrefatitle {Simple unified view of branching process statistics:
  Random walks in balanced logarithmic potentials} {Simple unified view of
  branching process statistics: Random walks in balanced logarithmic
  potentials}.{\BBCQ}
\newblock
\APACjournalVolNumPages{Physical Review E}{95}{3}{032115}.
\PrintBackRefs{\CurrentBib}

\bibitem [\protect \citeauthoryear {%
Di~Santo%
, Villegas%
, Burioni%
\BCBL {}\ \BBA {} Mu{\~n}oz%
}{%
Di~Santo%
\ \protect \BOthers {.}}{%
{\protect \APACyear {2018}}%
}]{%
di2018landau}
\APACinsertmetastar {%
di2018landau}%
\begin{APACrefauthors}%
Di~Santo, S.%
, Villegas, P.%
, Burioni, R.%
\BCBL {}\ \BBA {} Mu{\~n}oz, M\BPBI A.%
\end{APACrefauthors}%
\unskip\
\newblock
\APACrefYearMonthDay{2018}{}{}.
\newblock
{\BBOQ}\APACrefatitle {Landau--Ginzburg theory of cortex dynamics: Scale-free
  avalanches emerge at the edge of synchronization} {Landau--ginzburg theory of
  cortex dynamics: Scale-free avalanches emerge at the edge of
  synchronization}.{\BBCQ}
\newblock
\APACjournalVolNumPages{Proceedings of the National Academy of
  Sciences}{115}{7}{E1356--E1365}.
\PrintBackRefs{\CurrentBib}

\bibitem [\protect \citeauthoryear {%
Ecker%
\ \protect \BOthers {.}}{%
Ecker%
\ \protect \BOthers {.}}{%
{\protect \APACyear {2010}}%
}]{%
ecker2010decorrelated}
\APACinsertmetastar {%
ecker2010decorrelated}%
\begin{APACrefauthors}%
Ecker, A\BPBI S.%
, Berens, P.%
, Keliris, G\BPBI A.%
, Bethge, M.%
, Logothetis, N\BPBI K.%
\BCBL {}\ \BBA {} Tolias, A\BPBI S.%
\end{APACrefauthors}%
\unskip\
\newblock
\APACrefYearMonthDay{2010}{}{}.
\newblock
{\BBOQ}\APACrefatitle {Decorrelated neuronal firing in cortical microcircuits}
  {Decorrelated neuronal firing in cortical microcircuits}.{\BBCQ}
\newblock
\APACjournalVolNumPages{science}{327}{5965}{584--587}.
\PrintBackRefs{\CurrentBib}

\bibitem [\protect \citeauthoryear {%
Effenberger%
, Jost%
\BCBL {}\ \BBA {} Levina%
}{%
Effenberger%
\ \protect \BOthers {.}}{%
{\protect \APACyear {2015}}%
}]{%
effenberger2015self}
\APACinsertmetastar {%
effenberger2015self}%
\begin{APACrefauthors}%
Effenberger, F.%
, Jost, J.%
\BCBL {}\ \BBA {} Levina, A.%
\end{APACrefauthors}%
\unskip\
\newblock
\APACrefYearMonthDay{2015}{}{}.
\newblock
{\BBOQ}\APACrefatitle {Self-organization in balanced state networks by STDP and
  homeostatic plasticity} {Self-organization in balanced state networks by stdp
  and homeostatic plasticity}.{\BBCQ}
\newblock
\APACjournalVolNumPages{PLoS computational biology}{11}{9}{e1004420}.
\PrintBackRefs{\CurrentBib}

\bibitem [\protect \citeauthoryear {%
Erdos%
, Kruger%
\BCBL {}\ \BBA {} Renfrew%
}{%
Erdos%
\ \protect \BOthers {.}}{%
{\protect \APACyear {2018}}%
}]{%
erdos2018power}
\APACinsertmetastar {%
erdos2018power}%
\begin{APACrefauthors}%
Erdos, L.%
, Kruger, T.%
\BCBL {}\ \BBA {} Renfrew, D.%
\end{APACrefauthors}%
\unskip\
\newblock
\APACrefYearMonthDay{2018}{}{}.
\newblock
{\BBOQ}\APACrefatitle {Power law decay for systems of randomly coupled
  differential equations} {Power law decay for systems of randomly coupled
  differential equations}.{\BBCQ}
\newblock
\APACjournalVolNumPages{SIAM Journal on Mathematical
  Analysis}{50}{3}{3271--3290}.
\PrintBackRefs{\CurrentBib}

\bibitem [\protect \citeauthoryear {%
Fay%
\ \BBA {} Proschan%
}{%
Fay%
\ \BBA {} Proschan%
}{%
{\protect \APACyear {2010}}%
}]{%
fay2010wilcoxon}
\APACinsertmetastar {%
fay2010wilcoxon}%
\begin{APACrefauthors}%
Fay, M\BPBI P.%
\BCBT {}\ \BBA {} Proschan, M\BPBI A.%
\end{APACrefauthors}%
\unskip\
\newblock
\APACrefYearMonthDay{2010}{}{}.
\newblock
{\BBOQ}\APACrefatitle {Wilcoxon-Mann-Whitney or t-test? On assumptions for
  hypothesis tests and multiple interpretations of decision rules}
  {Wilcoxon-mann-whitney or t-test? on assumptions for hypothesis tests and
  multiple interpretations of decision rules}.{\BBCQ}
\newblock
\APACjournalVolNumPages{Statistics surveys}{4}{}{1}.
\PrintBackRefs{\CurrentBib}

\bibitem [\protect \citeauthoryear {%
Felleman%
\ \BBA {} Van~Essen%
}{%
Felleman%
\ \BBA {} Van~Essen%
}{%
{\protect \APACyear {1991}}%
}]{%
felleman1991distributed}
\APACinsertmetastar {%
felleman1991distributed}%
\begin{APACrefauthors}%
Felleman, D\BPBI J.%
\BCBT {}\ \BBA {} Van~Essen, D\BPBI C.%
\end{APACrefauthors}%
\unskip\
\newblock
\APACrefYearMonthDay{1991}{}{}.
\newblock
{\BBOQ}\APACrefatitle {Distributed hierarchical processing in the primate
  cerebral cortex.} {Distributed hierarchical processing in the primate
  cerebral cortex.}{\BBCQ}
\newblock
\APACjournalVolNumPages{Cerebral cortex (New York, NY: 1991)}{1}{1}{1--47}.
\PrintBackRefs{\CurrentBib}

\bibitem [\protect \citeauthoryear {%
Fontenele%
\ \protect \BOthers {.}}{%
Fontenele%
\ \protect \BOthers {.}}{%
{\protect \APACyear {2019}}%
}]{%
fontenele2019criticality}
\APACinsertmetastar {%
fontenele2019criticality}%
\begin{APACrefauthors}%
Fontenele, A\BPBI J.%
, de Vasconcelos, N\BPBI A.%
, Feliciano, T.%
, Aguiar, L\BPBI A.%
, Soares-Cunha, C.%
, Coimbra, B.%
\BDBL {}others%
\end{APACrefauthors}%
\unskip\
\newblock
\APACrefYearMonthDay{2019}{}{}.
\newblock
{\BBOQ}\APACrefatitle {Criticality between cortical states} {Criticality
  between cortical states}.{\BBCQ}
\newblock
\APACjournalVolNumPages{Physical review letters}{122}{20}{208101}.
\PrintBackRefs{\CurrentBib}

\bibitem [\protect \citeauthoryear {%
Fosque%
, Williams-Garc{\'\i}a%
, Beggs%
\BCBL {}\ \BBA {} Ortiz%
}{%
Fosque%
\ \protect \BOthers {.}}{%
{\protect \APACyear {2021}}%
}]{%
fosque2021evidence}
\APACinsertmetastar {%
fosque2021evidence}%
\begin{APACrefauthors}%
Fosque, L\BPBI J.%
, Williams-Garc{\'\i}a, R\BPBI V.%
, Beggs, J\BPBI M.%
\BCBL {}\ \BBA {} Ortiz, G.%
\end{APACrefauthors}%
\unskip\
\newblock
\APACrefYearMonthDay{2021}{}{}.
\newblock
{\BBOQ}\APACrefatitle {Evidence for quasicritical brain dynamics} {Evidence for
  quasicritical brain dynamics}.{\BBCQ}
\newblock
\APACjournalVolNumPages{Physical Review Letters}{126}{9}{098101}.
\PrintBackRefs{\CurrentBib}

\bibitem [\protect \citeauthoryear {%
Franke%
\ \protect \BOthers {.}}{%
Franke%
\ \protect \BOthers {.}}{%
{\protect \APACyear {2016}}%
}]{%
franke2016structures}
\APACinsertmetastar {%
franke2016structures}%
\begin{APACrefauthors}%
Franke, F.%
, Fiscella, M.%
, Sevelev, M.%
, Roska, B.%
, Hierlemann, A.%
\BCBL {}\ \BBA {} da Silveira, R\BPBI A.%
\end{APACrefauthors}%
\unskip\
\newblock
\APACrefYearMonthDay{2016}{}{}.
\newblock
{\BBOQ}\APACrefatitle {Structures of neural correlation and how they favor
  coding} {Structures of neural correlation and how they favor coding}.{\BBCQ}
\newblock
\APACjournalVolNumPages{Neuron}{89}{2}{409--422}.
\PrintBackRefs{\CurrentBib}

\bibitem [\protect \citeauthoryear {%
E\BPBI J.~Friedman%
\ \BBA {} Landsberg%
}{%
E\BPBI J.~Friedman%
\ \BBA {} Landsberg%
}{%
{\protect \APACyear {2013}}%
}]{%
friedman2013hierarchical}
\APACinsertmetastar {%
friedman2013hierarchical}%
\begin{APACrefauthors}%
Friedman, E\BPBI J.%
\BCBT {}\ \BBA {} Landsberg, A\BPBI S.%
\end{APACrefauthors}%
\unskip\
\newblock
\APACrefYearMonthDay{2013}{}{}.
\newblock
{\BBOQ}\APACrefatitle {Hierarchical networks, power laws, and neuronal
  avalanches} {Hierarchical networks, power laws, and neuronal
  avalanches}.{\BBCQ}
\newblock
\APACjournalVolNumPages{Chaos: An Interdisciplinary Journal of Nonlinear
  Science}{23}{1}{013135}.
\PrintBackRefs{\CurrentBib}

\bibitem [\protect \citeauthoryear {%
N.~Friedman%
\ \protect \BOthers {.}}{%
N.~Friedman%
\ \protect \BOthers {.}}{%
{\protect \APACyear {2012}}%
}]{%
friedman2012universal}
\APACinsertmetastar {%
friedman2012universal}%
\begin{APACrefauthors}%
Friedman, N.%
, Ito, S.%
, Brinkman, B\BPBI A.%
, Shimono, M.%
, DeVille, R\BPBI L.%
, Dahmen, K\BPBI A.%
\BDBL {}Butler, T\BPBI C.%
\end{APACrefauthors}%
\unskip\
\newblock
\APACrefYearMonthDay{2012}{}{}.
\newblock
{\BBOQ}\APACrefatitle {Universal critical dynamics in high resolution neuronal
  avalanche data} {Universal critical dynamics in high resolution neuronal
  avalanche data}.{\BBCQ}
\newblock
\APACjournalVolNumPages{Physical review letters}{108}{20}{208102}.
\PrintBackRefs{\CurrentBib}

\bibitem [\protect \citeauthoryear {%
Fristedt%
\ \BBA {} Gray%
}{%
Fristedt%
\ \BBA {} Gray%
}{%
{\protect \APACyear {2013}}%
}]{%
fristedt2013modern}
\APACinsertmetastar {%
fristedt2013modern}%
\begin{APACrefauthors}%
Fristedt, B\BPBI E.%
\BCBT {}\ \BBA {} Gray, L\BPBI F.%
\end{APACrefauthors}%
\unskip\
\newblock
\APACrefYear{2013}.
\newblock
\APACrefbtitle {A modern approach to probability theory} {A modern approach to
  probability theory}.
\newblock
\APACaddressPublisher{}{Springer Science \& Business Media}.
\PrintBackRefs{\CurrentBib}

\bibitem [\protect \citeauthoryear {%
Friston%
}{%
Friston%
}{%
{\protect \APACyear {2009}}%
}]{%
friston2009free}
\APACinsertmetastar {%
friston2009free}%
\begin{APACrefauthors}%
Friston, K.%
\end{APACrefauthors}%
\unskip\
\newblock
\APACrefYearMonthDay{2009}{}{}.
\newblock
{\BBOQ}\APACrefatitle {The free-energy principle: a rough guide to the brain?}
  {The free-energy principle: a rough guide to the brain?}{\BBCQ}
\newblock
\APACjournalVolNumPages{Trends in cognitive sciences}{13}{7}{293--301}.
\PrintBackRefs{\CurrentBib}

\bibitem [\protect \citeauthoryear {%
Friston%
}{%
Friston%
}{%
{\protect \APACyear {2010}}%
}]{%
friston2010free}
\APACinsertmetastar {%
friston2010free}%
\begin{APACrefauthors}%
Friston, K.%
\end{APACrefauthors}%
\unskip\
\newblock
\APACrefYearMonthDay{2010}{}{}.
\newblock
{\BBOQ}\APACrefatitle {The free-energy principle: a unified brain theory?} {The
  free-energy principle: a unified brain theory?}{\BBCQ}
\newblock
\APACjournalVolNumPages{Nature reviews neuroscience}{11}{2}{127--138}.
\PrintBackRefs{\CurrentBib}

\bibitem [\protect \citeauthoryear {%
Galv{\'a}n%
}{%
Galv{\'a}n%
}{%
{\protect \APACyear {2010}}%
}]{%
galvan2010neural}
\APACinsertmetastar {%
galvan2010neural}%
\begin{APACrefauthors}%
Galv{\'a}n, A.%
\end{APACrefauthors}%
\unskip\
\newblock
\APACrefYearMonthDay{2010}{}{}.
\newblock
{\BBOQ}\APACrefatitle {Neural plasticity of development and learning} {Neural
  plasticity of development and learning}.{\BBCQ}
\newblock
\APACjournalVolNumPages{Human brain mapping}{31}{6}{879--890}.
\PrintBackRefs{\CurrentBib}

\bibitem [\protect \citeauthoryear {%
Ganmor%
, Segev%
\BCBL {}\ \BBA {} Schneidman%
}{%
Ganmor%
\ \protect \BOthers {.}}{%
{\protect \APACyear {2011}}%
}]{%
ganmor2011sparse}
\APACinsertmetastar {%
ganmor2011sparse}%
\begin{APACrefauthors}%
Ganmor, E.%
, Segev, R.%
\BCBL {}\ \BBA {} Schneidman, E.%
\end{APACrefauthors}%
\unskip\
\newblock
\APACrefYearMonthDay{2011}{}{}.
\newblock
{\BBOQ}\APACrefatitle {Sparse low-order interaction network underlies a highly
  correlated and learnable neural population code} {Sparse low-order
  interaction network underlies a highly correlated and learnable neural
  population code}.{\BBCQ}
\newblock
\APACjournalVolNumPages{Proceedings of the National Academy of
  sciences}{108}{23}{9679--9684}.
\PrintBackRefs{\CurrentBib}

\bibitem [\protect \citeauthoryear {%
Garc{\'\i}a-Pelayo%
, Salazar%
\BCBL {}\ \BBA {} Schieve%
}{%
Garc{\'\i}a-Pelayo%
\ \protect \BOthers {.}}{%
{\protect \APACyear {1993}}%
}]{%
garcia1993branching}
\APACinsertmetastar {%
garcia1993branching}%
\begin{APACrefauthors}%
Garc{\'\i}a-Pelayo, R.%
, Salazar, I.%
\BCBL {}\ \BBA {} Schieve, W\BPBI C.%
\end{APACrefauthors}%
\unskip\
\newblock
\APACrefYearMonthDay{1993}{}{}.
\newblock
{\BBOQ}\APACrefatitle {A branching process model for sand avalanches} {A
  branching process model for sand avalanches}.{\BBCQ}
\newblock
\APACjournalVolNumPages{Journal of statistical physics}{72}{1}{167--187}.
\PrintBackRefs{\CurrentBib}

\bibitem [\protect \citeauthoryear {%
Gautam%
, Hoang%
, McClanahan%
, Grady%
\BCBL {}\ \BBA {} Shew%
}{%
Gautam%
\ \protect \BOthers {.}}{%
{\protect \APACyear {2015}}%
}]{%
gautam2015maximizing}
\APACinsertmetastar {%
gautam2015maximizing}%
\begin{APACrefauthors}%
Gautam, S\BPBI H.%
, Hoang, T\BPBI T.%
, McClanahan, K.%
, Grady, S\BPBI K.%
\BCBL {}\ \BBA {} Shew, W\BPBI L.%
\end{APACrefauthors}%
\unskip\
\newblock
\APACrefYearMonthDay{2015}{}{}.
\newblock
{\BBOQ}\APACrefatitle {Maximizing sensory dynamic range by tuning the cortical
  state to criticality} {Maximizing sensory dynamic range by tuning the
  cortical state to criticality}.{\BBCQ}
\newblock
\APACjournalVolNumPages{PLoS computational biology}{11}{12}{e1004576}.
\PrintBackRefs{\CurrentBib}

\bibitem [\protect \citeauthoryear {%
Gerstner%
, Kistler%
, Naud%
\BCBL {}\ \BBA {} Paninski%
}{%
Gerstner%
\ \protect \BOthers {.}}{%
{\protect \APACyear {2014}}%
}]{%
gerstner2014neuronal}
\APACinsertmetastar {%
gerstner2014neuronal}%
\begin{APACrefauthors}%
Gerstner, W.%
, Kistler, W\BPBI M.%
, Naud, R.%
\BCBL {}\ \BBA {} Paninski, L.%
\end{APACrefauthors}%
\unskip\
\newblock
\APACrefYear{2014}.
\newblock
\APACrefbtitle {Neuronal dynamics: From single neurons to networks and models
  of cognition} {Neuronal dynamics: From single neurons to networks and models
  of cognition}.
\newblock
\APACaddressPublisher{}{Cambridge University Press}.
\PrintBackRefs{\CurrentBib}

\bibitem [\protect \citeauthoryear {%
Girardi-Schappo%
}{%
Girardi-Schappo%
}{%
{\protect \APACyear {2021}}%
}]{%
girardi2021brain}
\APACinsertmetastar {%
girardi2021brain}%
\begin{APACrefauthors}%
Girardi-Schappo, M.%
\end{APACrefauthors}%
\unskip\
\newblock
\APACrefYearMonthDay{2021}{}{}.
\newblock
{\BBOQ}\APACrefatitle {Brain criticality beyond avalanches: open problems and
  how to approach them} {Brain criticality beyond avalanches: open problems and
  how to approach them}.{\BBCQ}
\newblock
\APACjournalVolNumPages{Journal of Physics: Complexity}{}{}{}.
\PrintBackRefs{\CurrentBib}

\bibitem [\protect \citeauthoryear {%
Girardi-Schappo%
, Bortolotto%
, Gonsalves%
, Pinto%
\BCBL {}\ \BBA {} Tragtenberg%
}{%
Girardi-Schappo%
\ \protect \BOthers {.}}{%
{\protect \APACyear {2016}}%
}]{%
girardi2016griffiths}
\APACinsertmetastar {%
girardi2016griffiths}%
\begin{APACrefauthors}%
Girardi-Schappo, M.%
, Bortolotto, G\BPBI S.%
, Gonsalves, J\BPBI J.%
, Pinto, L\BPBI T.%
\BCBL {}\ \BBA {} Tragtenberg, M\BPBI H.%
\end{APACrefauthors}%
\unskip\
\newblock
\APACrefYearMonthDay{2016}{}{}.
\newblock
{\BBOQ}\APACrefatitle {Griffiths phase and long-range correlations in a
  biologically motivated visual cortex model} {Griffiths phase and long-range
  correlations in a biologically motivated visual cortex model}.{\BBCQ}
\newblock
\APACjournalVolNumPages{Scientific reports}{6}{1}{1--12}.
\PrintBackRefs{\CurrentBib}

\bibitem [\protect \citeauthoryear {%
Girardi-Schappo%
\ \protect \BOthers {.}}{%
Girardi-Schappo%
\ \protect \BOthers {.}}{%
{\protect \APACyear {2021}}%
}]{%
girardi2021unified}
\APACinsertmetastar {%
girardi2021unified}%
\begin{APACrefauthors}%
Girardi-Schappo, M.%
, Galera, E\BPBI F.%
, Carvalho, T\BPBI T.%
, Brochini, L.%
, Kamiji, N\BPBI L.%
, Roque, A\BPBI C.%
\BCBL {}\ \BBA {} Kinouchi, O.%
\end{APACrefauthors}%
\unskip\
\newblock
\APACrefYearMonthDay{2021}{}{}.
\newblock
{\BBOQ}\APACrefatitle {A unified theory of E/I synaptic balance, quasicritical
  neuronal avalanches and asynchronous irregular spiking} {A unified theory of
  e/i synaptic balance, quasicritical neuronal avalanches and asynchronous
  irregular spiking}.{\BBCQ}
\newblock
\APACjournalVolNumPages{Journal of Physics: Complexity}{2}{4}{045001}.
\PrintBackRefs{\CurrentBib}

\bibitem [\protect \citeauthoryear {%
Gireesh%
\ \BBA {} Plenz%
}{%
Gireesh%
\ \BBA {} Plenz%
}{%
{\protect \APACyear {2008}}%
}]{%
gireesh2008neuronal}
\APACinsertmetastar {%
gireesh2008neuronal}%
\begin{APACrefauthors}%
Gireesh, E\BPBI D.%
\BCBT {}\ \BBA {} Plenz, D.%
\end{APACrefauthors}%
\unskip\
\newblock
\APACrefYearMonthDay{2008}{}{}.
\newblock
{\BBOQ}\APACrefatitle {Neuronal avalanches organize as nested theta-and
  beta/gamma-oscillations during development of cortical layer 2/3} {Neuronal
  avalanches organize as nested theta-and beta/gamma-oscillations during
  development of cortical layer 2/3}.{\BBCQ}
\newblock
\APACjournalVolNumPages{Proceedings of the National Academy of
  Sciences}{105}{21}{7576--7581}.
\PrintBackRefs{\CurrentBib}

\bibitem [\protect \citeauthoryear {%
Gnesotto%
, Mura%
, Gladrow%
\BCBL {}\ \BBA {} Broedersz%
}{%
Gnesotto%
\ \protect \BOthers {.}}{%
{\protect \APACyear {2018}}%
}]{%
gnesotto2018broken}
\APACinsertmetastar {%
gnesotto2018broken}%
\begin{APACrefauthors}%
Gnesotto, F\BPBI S.%
, Mura, F.%
, Gladrow, J.%
\BCBL {}\ \BBA {} Broedersz, C\BPBI P.%
\end{APACrefauthors}%
\unskip\
\newblock
\APACrefYearMonthDay{2018}{}{}.
\newblock
{\BBOQ}\APACrefatitle {Broken detailed balance and non-equilibrium dynamics in
  living systems: a review} {Broken detailed balance and non-equilibrium
  dynamics in living systems: a review}.{\BBCQ}
\newblock
\APACjournalVolNumPages{Reports on Progress in Physics}{81}{6}{066601}.
\PrintBackRefs{\CurrentBib}

\bibitem [\protect \citeauthoryear {%
Gong%
\ \protect \BOthers {.}}{%
Gong%
\ \protect \BOthers {.}}{%
{\protect \APACyear {2009}}%
}]{%
gong2009mapping}
\APACinsertmetastar {%
gong2009mapping}%
\begin{APACrefauthors}%
Gong, G.%
, He, Y.%
, Concha, L.%
, Lebel, C.%
, Gross, D\BPBI W.%
, Evans, A\BPBI C.%
\BCBL {}\ \BBA {} Beaulieu, C.%
\end{APACrefauthors}%
\unskip\
\newblock
\APACrefYearMonthDay{2009}{}{}.
\newblock
{\BBOQ}\APACrefatitle {Mapping anatomical connectivity patterns of human
  cerebral cortex using in vivo diffusion tensor imaging tractography} {Mapping
  anatomical connectivity patterns of human cerebral cortex using in vivo
  diffusion tensor imaging tractography}.{\BBCQ}
\newblock
\APACjournalVolNumPages{Cerebral cortex}{19}{3}{524--536}.
\PrintBackRefs{\CurrentBib}

\bibitem [\protect \citeauthoryear {%
Gros%
}{%
Gros%
}{%
{\protect \APACyear {2010}}%
}]{%
gros2010complex}
\APACinsertmetastar {%
gros2010complex}%
\begin{APACrefauthors}%
Gros, C.%
\end{APACrefauthors}%
\unskip\
\newblock
\APACrefYear{2010}.
\newblock
\APACrefbtitle {Complex and adaptive dynamical systems} {Complex and adaptive
  dynamical systems}.
\newblock
\APACaddressPublisher{}{Springer}.
\PrintBackRefs{\CurrentBib}

\bibitem [\protect \citeauthoryear {%
Guevara%
}{%
Guevara%
}{%
{\protect \APACyear {2021}}%
}]{%
guevara2021synchronization}
\APACinsertmetastar {%
guevara2021synchronization}%
\begin{APACrefauthors}%
Guevara, R.%
\end{APACrefauthors}%
\unskip\
\newblock
\APACrefYearMonthDay{2021}{}{}.
\newblock
{\BBOQ}\APACrefatitle {Synchronization, free energy and the embryogenesis of
  the cortex} {Synchronization, free energy and the embryogenesis of the
  cortex}.{\BBCQ}
\newblock
\APACjournalVolNumPages{Physics of Life Reviews}{36}{}{5--6}.
\PrintBackRefs{\CurrentBib}

\bibitem [\protect \citeauthoryear {%
Guo%
\ \protect \BOthers {.}}{%
Guo%
\ \protect \BOthers {.}}{%
{\protect \APACyear {2021}}%
}]{%
guo2021percolation}
\APACinsertmetastar {%
guo2021percolation}%
\begin{APACrefauthors}%
Guo, S.%
, Chen, X.%
, Liu, Y.%
, Kang, R.%
, Liu, T.%
\BCBL {}\ \BBA {} Li, D.%
\end{APACrefauthors}%
\unskip\
\newblock
\APACrefYearMonthDay{2021}{}{}.
\newblock
{\BBOQ}\APACrefatitle {Percolation analysis of brain structural network}
  {Percolation analysis of brain structural network}.{\BBCQ}
\newblock
\APACjournalVolNumPages{Frontiers in Physics}{9}{}{345}.
\PrintBackRefs{\CurrentBib}

\bibitem [\protect \citeauthoryear {%
Hahn%
\ \protect \BOthers {.}}{%
Hahn%
\ \protect \BOthers {.}}{%
{\protect \APACyear {2017}}%
}]{%
hahn2017spontaneous}
\APACinsertmetastar {%
hahn2017spontaneous}%
\begin{APACrefauthors}%
Hahn, G.%
, Ponce-Alvarez, A.%
, Monier, C.%
, Benvenuti, G.%
, Kumar, A.%
, Chavane, F.%
\BDBL {}Fr{\'e}gnac, Y.%
\end{APACrefauthors}%
\unskip\
\newblock
\APACrefYearMonthDay{2017}{}{}.
\newblock
{\BBOQ}\APACrefatitle {Spontaneous cortical activity is transiently poised
  close to criticality} {Spontaneous cortical activity is transiently poised
  close to criticality}.{\BBCQ}
\newblock
\APACjournalVolNumPages{PLoS computational biology}{13}{5}{e1005543}.
\PrintBackRefs{\CurrentBib}

\bibitem [\protect \citeauthoryear {%
Haimovici%
, Tagliazucchi%
, Balenzuela%
\BCBL {}\ \BBA {} Chialvo%
}{%
Haimovici%
\ \protect \BOthers {.}}{%
{\protect \APACyear {2013}}%
}]{%
haimovici2013brain}
\APACinsertmetastar {%
haimovici2013brain}%
\begin{APACrefauthors}%
Haimovici, A.%
, Tagliazucchi, E.%
, Balenzuela, P.%
\BCBL {}\ \BBA {} Chialvo, D\BPBI R.%
\end{APACrefauthors}%
\unskip\
\newblock
\APACrefYearMonthDay{2013}{}{}.
\newblock
{\BBOQ}\APACrefatitle {Brain organization into resting state networks emerges
  at criticality on a model of the human connectome} {Brain organization into
  resting state networks emerges at criticality on a model of the human
  connectome}.{\BBCQ}
\newblock
\APACjournalVolNumPages{Physical review letters}{110}{17}{178101}.
\PrintBackRefs{\CurrentBib}

\bibitem [\protect \citeauthoryear {%
Haldeman%
\ \BBA {} Beggs%
}{%
Haldeman%
\ \BBA {} Beggs%
}{%
{\protect \APACyear {2005}}%
}]{%
haldeman2005critical}
\APACinsertmetastar {%
haldeman2005critical}%
\begin{APACrefauthors}%
Haldeman, C.%
\BCBT {}\ \BBA {} Beggs, J\BPBI M.%
\end{APACrefauthors}%
\unskip\
\newblock
\APACrefYearMonthDay{2005}{}{}.
\newblock
{\BBOQ}\APACrefatitle {Critical branching captures activity in living neural
  networks and maximizes the number of metastable states} {Critical branching
  captures activity in living neural networks and maximizes the number of
  metastable states}.{\BBCQ}
\newblock
\APACjournalVolNumPages{Physical review letters}{94}{5}{058101}.
\PrintBackRefs{\CurrentBib}

\bibitem [\protect \citeauthoryear {%
Hardstone%
, Mansvelder%
\BCBL {}\ \BBA {} Linkenkaer-Hansen%
}{%
Hardstone%
\ \protect \BOthers {.}}{%
{\protect \APACyear {2014}}%
}]{%
hardstone2014neuronal}
\APACinsertmetastar {%
hardstone2014neuronal}%
\begin{APACrefauthors}%
Hardstone, R.%
, Mansvelder, H\BPBI D.%
\BCBL {}\ \BBA {} Linkenkaer-Hansen, K.%
\end{APACrefauthors}%
\unskip\
\newblock
\APACrefYearMonthDay{2014}{}{}.
\newblock
{\BBOQ}\APACrefatitle {The Neuronal Network Oscillation as a Critical
  Phenomenon} {The neuronal network oscillation as a critical
  phenomenon}.{\BBCQ}
\newblock
\APACjournalVolNumPages{Criticality in Neural Systems. Wiley}{}{}{293--316}.
\PrintBackRefs{\CurrentBib}

\bibitem [\protect \citeauthoryear {%
Harris%
\ \BBA {} Edward%
}{%
Harris%
\ \BBA {} Edward%
}{%
{\protect \APACyear {1963}}%
}]{%
harris1963theory}
\APACinsertmetastar {%
harris1963theory}%
\begin{APACrefauthors}%
Harris%
\BCBT {}\ \BBA {} Edward, T.%
\end{APACrefauthors}%
\unskip\
\newblock
\APACrefYear{1963}.
\newblock
\APACrefbtitle {The theory of branching processes} {The theory of branching
  processes}\ (\BVOL~6).
\newblock
\APACaddressPublisher{}{Springer Berlin}.
\PrintBackRefs{\CurrentBib}

\bibitem [\protect \citeauthoryear {%
He%
}{%
He%
}{%
{\protect \APACyear {2014}}%
}]{%
he2014scale}
\APACinsertmetastar {%
he2014scale}%
\begin{APACrefauthors}%
He, B\BPBI J.%
\end{APACrefauthors}%
\unskip\
\newblock
\APACrefYearMonthDay{2014}{}{}.
\newblock
{\BBOQ}\APACrefatitle {Scale-free brain activity: past, present, and future}
  {Scale-free brain activity: past, present, and future}.{\BBCQ}
\newblock
\APACjournalVolNumPages{Trends in cognitive sciences}{18}{9}{480--487}.
\PrintBackRefs{\CurrentBib}

\bibitem [\protect \citeauthoryear {%
Henkel%
, Hinrichsen%
, L{\"u}beck%
\BCBL {}\ \BBA {} Pleimling%
}{%
Henkel%
\ \protect \BOthers {.}}{%
{\protect \APACyear {2008}}%
}]{%
henkel2008non}
\APACinsertmetastar {%
henkel2008non}%
\begin{APACrefauthors}%
Henkel, M.%
, Hinrichsen, H.%
, L{\"u}beck, S.%
\BCBL {}\ \BBA {} Pleimling, M.%
\end{APACrefauthors}%
\unskip\
\newblock
\APACrefYear{2008}.
\newblock
\APACrefbtitle {Non-equilibrium phase transitions} {Non-equilibrium phase
  transitions}\ (\BVOL~1).
\newblock
\APACaddressPublisher{}{Springer}.
\PrintBackRefs{\CurrentBib}

\bibitem [\protect \citeauthoryear {%
Hernandez-Urbina%
\ \BBA {} Herrmann%
}{%
Hernandez-Urbina%
\ \BBA {} Herrmann%
}{%
{\protect \APACyear {2017}}%
}]{%
hernandez2017self}
\APACinsertmetastar {%
hernandez2017self}%
\begin{APACrefauthors}%
Hernandez-Urbina, V.%
\BCBT {}\ \BBA {} Herrmann, J\BPBI M.%
\end{APACrefauthors}%
\unskip\
\newblock
\APACrefYearMonthDay{2017}{}{}.
\newblock
{\BBOQ}\APACrefatitle {Self-organized criticality via retro-synaptic signals}
  {Self-organized criticality via retro-synaptic signals}.{\BBCQ}
\newblock
\APACjournalVolNumPages{Frontiers in Physics}{4}{}{54}.
\PrintBackRefs{\CurrentBib}

\bibitem [\protect \citeauthoryear {%
Hesse%
\ \BBA {} Gross%
}{%
Hesse%
\ \BBA {} Gross%
}{%
{\protect \APACyear {2014}}%
}]{%
hesse2014self}
\APACinsertmetastar {%
hesse2014self}%
\begin{APACrefauthors}%
Hesse, J.%
\BCBT {}\ \BBA {} Gross, T.%
\end{APACrefauthors}%
\unskip\
\newblock
\APACrefYearMonthDay{2014}{}{}.
\newblock
{\BBOQ}\APACrefatitle {Self-organized criticality as a fundamental property of
  neural systems} {Self-organized criticality as a fundamental property of
  neural systems}.{\BBCQ}
\newblock
\APACjournalVolNumPages{Frontiers in systems neuroscience}{8}{}{166}.
\PrintBackRefs{\CurrentBib}

\bibitem [\protect \citeauthoryear {%
Hinrichsen%
}{%
Hinrichsen%
}{%
{\protect \APACyear {2000}}%
}]{%
hinrichsen2000non}
\APACinsertmetastar {%
hinrichsen2000non}%
\begin{APACrefauthors}%
Hinrichsen, H.%
\end{APACrefauthors}%
\unskip\
\newblock
\APACrefYearMonthDay{2000}{}{}.
\newblock
{\BBOQ}\APACrefatitle {Non-equilibrium critical phenomena and phase transitions
  into absorbing states} {Non-equilibrium critical phenomena and phase
  transitions into absorbing states}.{\BBCQ}
\newblock
\APACjournalVolNumPages{Advances in physics}{49}{7}{815--958}.
\PrintBackRefs{\CurrentBib}

\bibitem [\protect \citeauthoryear {%
Holcman%
\ \BBA {} Tsodyks%
}{%
Holcman%
\ \BBA {} Tsodyks%
}{%
{\protect \APACyear {2006}}%
}]{%
holcman2006emergence}
\APACinsertmetastar {%
holcman2006emergence}%
\begin{APACrefauthors}%
Holcman, D.%
\BCBT {}\ \BBA {} Tsodyks, M.%
\end{APACrefauthors}%
\unskip\
\newblock
\APACrefYearMonthDay{2006}{}{}.
\newblock
{\BBOQ}\APACrefatitle {The emergence of up and down states in cortical
  networks} {The emergence of up and down states in cortical networks}.{\BBCQ}
\newblock
\APACjournalVolNumPages{PLoS computational biology}{2}{3}{e23}.
\PrintBackRefs{\CurrentBib}

\bibitem [\protect \citeauthoryear {%
Hopfield%
}{%
Hopfield%
}{%
{\protect \APACyear {1982}}%
}]{%
hopfield1982neural}
\APACinsertmetastar {%
hopfield1982neural}%
\begin{APACrefauthors}%
Hopfield, J\BPBI J.%
\end{APACrefauthors}%
\unskip\
\newblock
\APACrefYearMonthDay{1982}{}{}.
\newblock
{\BBOQ}\APACrefatitle {Neural networks and physical systems with emergent
  collective computational abilities} {Neural networks and physical systems
  with emergent collective computational abilities}.{\BBCQ}
\newblock
\APACjournalVolNumPages{Proceedings of the national academy of
  sciences}{79}{8}{2554--2558}.
\PrintBackRefs{\CurrentBib}

\bibitem [\protect \citeauthoryear {%
Janowsky%
\ \BBA {} Laberge%
}{%
Janowsky%
\ \BBA {} Laberge%
}{%
{\protect \APACyear {1993}}%
}]{%
janowsky1993exact}
\APACinsertmetastar {%
janowsky1993exact}%
\begin{APACrefauthors}%
Janowsky, S\BPBI A.%
\BCBT {}\ \BBA {} Laberge, C\BPBI A.%
\end{APACrefauthors}%
\unskip\
\newblock
\APACrefYearMonthDay{1993}{}{}.
\newblock
{\BBOQ}\APACrefatitle {Exact solutions for a mean-field Abelian sandpile}
  {Exact solutions for a mean-field abelian sandpile}.{\BBCQ}
\newblock
\APACjournalVolNumPages{Journal of Physics A: Mathematical and
  General}{26}{19}{L973}.
\PrintBackRefs{\CurrentBib}

\bibitem [\protect \citeauthoryear {%
Jercog%
\ \protect \BOthers {.}}{%
Jercog%
\ \protect \BOthers {.}}{%
{\protect \APACyear {2017}}%
}]{%
jercog2017up}
\APACinsertmetastar {%
jercog2017up}%
\begin{APACrefauthors}%
Jercog, D.%
, Roxin, A.%
, Bartho, P.%
, Luczak, A.%
, Compte, A.%
\BCBL {}\ \BBA {} de~la Rocha, J.%
\end{APACrefauthors}%
\unskip\
\newblock
\APACrefYearMonthDay{2017}{}{}.
\newblock
{\BBOQ}\APACrefatitle {UP-DOWN cortical dynamics reflect state transitions in a
  bistable network} {Up-down cortical dynamics reflect state transitions in a
  bistable network}.{\BBCQ}
\newblock
\APACjournalVolNumPages{Elife}{6}{}{e22425}.
\PrintBackRefs{\CurrentBib}

\bibitem [\protect \citeauthoryear {%
Jung%
, Le%
, Lee%
\BCBL {}\ \BBA {} Lee%
}{%
Jung%
\ \protect \BOthers {.}}{%
{\protect \APACyear {2020}}%
}]{%
jung2020avalanche}
\APACinsertmetastar {%
jung2020avalanche}%
\begin{APACrefauthors}%
Jung, N.%
, Le, Q\BPBI A.%
, Lee, K\BHBI E.%
\BCBL {}\ \BBA {} Lee, J\BPBI W.%
\end{APACrefauthors}%
\unskip\
\newblock
\APACrefYearMonthDay{2020}{}{}.
\newblock
{\BBOQ}\APACrefatitle {Avalanche size distribution of an integrate-and-fire
  neural model on complex networks} {Avalanche size distribution of an
  integrate-and-fire neural model on complex networks}.{\BBCQ}
\newblock
\APACjournalVolNumPages{Chaos: An Interdisciplinary Journal of Nonlinear
  Science}{30}{6}{063118}.
\PrintBackRefs{\CurrentBib}

\bibitem [\protect \citeauthoryear {%
Kaiser%
\ \BBA {} Hilgetag%
}{%
Kaiser%
\ \BBA {} Hilgetag%
}{%
{\protect \APACyear {2006}}%
}]{%
kaiser2006nonoptimal}
\APACinsertmetastar {%
kaiser2006nonoptimal}%
\begin{APACrefauthors}%
Kaiser, M.%
\BCBT {}\ \BBA {} Hilgetag, C\BPBI C.%
\end{APACrefauthors}%
\unskip\
\newblock
\APACrefYearMonthDay{2006}{}{}.
\newblock
{\BBOQ}\APACrefatitle {Nonoptimal component placement, but short processing
  paths, due to long-distance projections in neural systems} {Nonoptimal
  component placement, but short processing paths, due to long-distance
  projections in neural systems}.{\BBCQ}
\newblock
\APACjournalVolNumPages{PLoS computational biology}{2}{7}{e95}.
\PrintBackRefs{\CurrentBib}

\bibitem [\protect \citeauthoryear {%
Kaiser%
\ \BBA {} Hilgetag%
}{%
Kaiser%
\ \BBA {} Hilgetag%
}{%
{\protect \APACyear {2010}}%
}]{%
kaiser2010optimal}
\APACinsertmetastar {%
kaiser2010optimal}%
\begin{APACrefauthors}%
Kaiser, M.%
\BCBT {}\ \BBA {} Hilgetag, C\BPBI C.%
\end{APACrefauthors}%
\unskip\
\newblock
\APACrefYearMonthDay{2010}{}{}.
\newblock
{\BBOQ}\APACrefatitle {Optimal hierarchical modular topologies for producing
  limited sustained activation of neural networks} {Optimal hierarchical
  modular topologies for producing limited sustained activation of neural
  networks}.{\BBCQ}
\newblock
\APACjournalVolNumPages{Frontiers in neuroinformatics}{4}{}{8}.
\PrintBackRefs{\CurrentBib}

\bibitem [\protect \citeauthoryear {%
Kanji%
}{%
Kanji%
}{%
{\protect \APACyear {2006}}%
}]{%
kanji2006100}
\APACinsertmetastar {%
kanji2006100}%
\begin{APACrefauthors}%
Kanji, G\BPBI K.%
\end{APACrefauthors}%
\unskip\
\newblock
\APACrefYear{2006}.
\newblock
\APACrefbtitle {100 statistical tests} {100 statistical tests}.
\newblock
\APACaddressPublisher{}{Sage}.
\PrintBackRefs{\CurrentBib}

\bibitem [\protect \citeauthoryear {%
Kara%
, Reinagel%
\BCBL {}\ \BBA {} Reid%
}{%
Kara%
\ \protect \BOthers {.}}{%
{\protect \APACyear {2000}}%
}]{%
kara2000low}
\APACinsertmetastar {%
kara2000low}%
\begin{APACrefauthors}%
Kara, P.%
, Reinagel, P.%
\BCBL {}\ \BBA {} Reid, R\BPBI C.%
\end{APACrefauthors}%
\unskip\
\newblock
\APACrefYearMonthDay{2000}{}{}.
\newblock
{\BBOQ}\APACrefatitle {Low response variability in simultaneously recorded
  retinal, thalamic, and cortical neurons} {Low response variability in
  simultaneously recorded retinal, thalamic, and cortical neurons}.{\BBCQ}
\newblock
\APACjournalVolNumPages{Neuron}{27}{3}{635--646}.
\PrintBackRefs{\CurrentBib}

\bibitem [\protect \citeauthoryear {%
Katsnelson%
, Vanchurin%
\BCBL {}\ \BBA {} Westerhout%
}{%
Katsnelson%
\ \protect \BOthers {.}}{%
{\protect \APACyear {2021}}%
}]{%
katsnelson2021self}
\APACinsertmetastar {%
katsnelson2021self}%
\begin{APACrefauthors}%
Katsnelson, M\BPBI I.%
, Vanchurin, V.%
\BCBL {}\ \BBA {} Westerhout, T.%
\end{APACrefauthors}%
\unskip\
\newblock
\APACrefYearMonthDay{2021}{}{}.
\newblock
{\BBOQ}\APACrefatitle {Self-organized criticality in Neural Networks}
  {Self-organized criticality in neural networks}.{\BBCQ}
\newblock
\APACjournalVolNumPages{arXiv preprint arXiv:2107.03402}{}{}{}.
\PrintBackRefs{\CurrentBib}

\bibitem [\protect \citeauthoryear {%
Keogh%
\ \BBA {} Pazzani%
}{%
Keogh%
\ \BBA {} Pazzani%
}{%
{\protect \APACyear {2001}}%
}]{%
keogh2001derivative}
\APACinsertmetastar {%
keogh2001derivative}%
\begin{APACrefauthors}%
Keogh, E\BPBI J.%
\BCBT {}\ \BBA {} Pazzani, M\BPBI J.%
\end{APACrefauthors}%
\unskip\
\newblock
\APACrefYearMonthDay{2001}{}{}.
\newblock
{\BBOQ}\APACrefatitle {Derivative dynamic time warping} {Derivative dynamic
  time warping}.{\BBCQ}
\newblock
\BIn{} \APACrefbtitle {Proceedings of the 2001 SIAM international conference on
  data mining} {Proceedings of the 2001 siam international conference on data
  mining}\ (\BPGS\ 1--11).
\PrintBackRefs{\CurrentBib}

\bibitem [\protect \citeauthoryear {%
Khambhati%
, Sizemore%
, Betzel%
\BCBL {}\ \BBA {} Bassett%
}{%
Khambhati%
\ \protect \BOthers {.}}{%
{\protect \APACyear {2018}}%
}]{%
khambhati2018modeling}
\APACinsertmetastar {%
khambhati2018modeling}%
\begin{APACrefauthors}%
Khambhati, A\BPBI N.%
, Sizemore, A\BPBI E.%
, Betzel, R\BPBI F.%
\BCBL {}\ \BBA {} Bassett, D\BPBI S.%
\end{APACrefauthors}%
\unskip\
\newblock
\APACrefYearMonthDay{2018}{}{}.
\newblock
{\BBOQ}\APACrefatitle {Modeling and interpreting mesoscale network dynamics}
  {Modeling and interpreting mesoscale network dynamics}.{\BBCQ}
\newblock
\APACjournalVolNumPages{NeuroImage}{180}{}{337--349}.
\PrintBackRefs{\CurrentBib}

\bibitem [\protect \citeauthoryear {%
Kinouchi%
\ \BBA {} Copelli%
}{%
Kinouchi%
\ \BBA {} Copelli%
}{%
{\protect \APACyear {2006}}%
}]{%
kinouchi2006optimal}
\APACinsertmetastar {%
kinouchi2006optimal}%
\begin{APACrefauthors}%
Kinouchi, O.%
\BCBT {}\ \BBA {} Copelli, M.%
\end{APACrefauthors}%
\unskip\
\newblock
\APACrefYearMonthDay{2006}{}{}.
\newblock
{\BBOQ}\APACrefatitle {Optimal dynamical range of excitable networks at
  criticality} {Optimal dynamical range of excitable networks at
  criticality}.{\BBCQ}
\newblock
\APACjournalVolNumPages{Nature physics}{2}{5}{348--351}.
\PrintBackRefs{\CurrentBib}

\bibitem [\protect \citeauthoryear {%
Krotov%
\ \BBA {} Hopfield%
}{%
Krotov%
\ \BBA {} Hopfield%
}{%
{\protect \APACyear {2020}}%
}]{%
krotov2020large}
\APACinsertmetastar {%
krotov2020large}%
\begin{APACrefauthors}%
Krotov, D.%
\BCBT {}\ \BBA {} Hopfield, J.%
\end{APACrefauthors}%
\unskip\
\newblock
\APACrefYearMonthDay{2020}{}{}.
\newblock
{\BBOQ}\APACrefatitle {Large associative memory problem in neurobiology and
  machine learning} {Large associative memory problem in neurobiology and
  machine learning}.{\BBCQ}
\newblock
\APACjournalVolNumPages{arXiv preprint arXiv:2008.06996}{}{}{}.
\PrintBackRefs{\CurrentBib}

\bibitem [\protect \citeauthoryear {%
Larremore%
, Carpenter%
, Ott%
\BCBL {}\ \BBA {} Restrepo%
}{%
Larremore%
\ \protect \BOthers {.}}{%
{\protect \APACyear {2012}}%
}]{%
larremore2012statistical}
\APACinsertmetastar {%
larremore2012statistical}%
\begin{APACrefauthors}%
Larremore, D\BPBI B.%
, Carpenter, M\BPBI Y.%
, Ott, E.%
\BCBL {}\ \BBA {} Restrepo, J\BPBI G.%
\end{APACrefauthors}%
\unskip\
\newblock
\APACrefYearMonthDay{2012}{}{}.
\newblock
{\BBOQ}\APACrefatitle {Statistical properties of avalanches in networks}
  {Statistical properties of avalanches in networks}.{\BBCQ}
\newblock
\APACjournalVolNumPages{Physical Review E}{85}{6}{066131}.
\PrintBackRefs{\CurrentBib}

\bibitem [\protect \citeauthoryear {%
Laurson%
\ \BBA {} Alava%
}{%
Laurson%
\ \BBA {} Alava%
}{%
{\protect \APACyear {2006}}%
}]{%
laurson20061}
\APACinsertmetastar {%
laurson20061}%
\begin{APACrefauthors}%
Laurson, L.%
\BCBT {}\ \BBA {} Alava, M\BPBI J.%
\end{APACrefauthors}%
\unskip\
\newblock
\APACrefYearMonthDay{2006}{}{}.
\newblock
{\BBOQ}\APACrefatitle {1/ f noise and avalanche scaling in plastic deformation}
  {1/ f noise and avalanche scaling in plastic deformation}.{\BBCQ}
\newblock
\APACjournalVolNumPages{Physical Review E}{74}{6}{066106}.
\PrintBackRefs{\CurrentBib}

\bibitem [\protect \citeauthoryear {%
Laurson%
, Illa%
\BCBL {}\ \BBA {} Alava%
}{%
Laurson%
\ \protect \BOthers {.}}{%
{\protect \APACyear {2009}}%
}]{%
laurson2009effect}
\APACinsertmetastar {%
laurson2009effect}%
\begin{APACrefauthors}%
Laurson, L.%
, Illa, X.%
\BCBL {}\ \BBA {} Alava, M\BPBI J.%
\end{APACrefauthors}%
\unskip\
\newblock
\APACrefYearMonthDay{2009}{}{}.
\newblock
{\BBOQ}\APACrefatitle {The effect of thresholding on temporal avalanche
  statistics} {The effect of thresholding on temporal avalanche
  statistics}.{\BBCQ}
\newblock
\APACjournalVolNumPages{Journal of Statistical Mechanics: Theory and
  Experiment}{2009}{01}{P01019}.
\PrintBackRefs{\CurrentBib}

\bibitem [\protect \citeauthoryear {%
Laurson%
\ \protect \BOthers {.}}{%
Laurson%
\ \protect \BOthers {.}}{%
{\protect \APACyear {2013}}%
}]{%
laurson2013evolution}
\APACinsertmetastar {%
laurson2013evolution}%
\begin{APACrefauthors}%
Laurson, L.%
, Illa, X.%
, Santucci, S.%
, Tore~Tallakstad, K.%
, M{\aa}l{\o}y, K\BPBI J.%
\BCBL {}\ \BBA {} Alava, M\BPBI J.%
\end{APACrefauthors}%
\unskip\
\newblock
\APACrefYearMonthDay{2013}{}{}.
\newblock
{\BBOQ}\APACrefatitle {Evolution of the average avalanche shape with the
  universality class} {Evolution of the average avalanche shape with the
  universality class}.{\BBCQ}
\newblock
\APACjournalVolNumPages{Nature communications}{4}{1}{1--6}.
\PrintBackRefs{\CurrentBib}

\bibitem [\protect \citeauthoryear {%
D\BPBI S.~Lee%
, Goh%
, Kahng%
\BCBL {}\ \BBA {} Kim%
}{%
D\BPBI S.~Lee%
\ \protect \BOthers {.}}{%
{\protect \APACyear {2004}}%
}]{%
lee2004branching}
\APACinsertmetastar {%
lee2004branching}%
\begin{APACrefauthors}%
Lee, D\BPBI S.%
, Goh, K\BPBI I.%
, Kahng, B.%
\BCBL {}\ \BBA {} Kim, D.%
\end{APACrefauthors}%
\unskip\
\newblock
\APACrefYearMonthDay{2004}{}{}.
\newblock
{\BBOQ}\APACrefatitle {Branching process approach to avalanche dynamics on
  complex networks} {Branching process approach to avalanche dynamics on
  complex networks}.{\BBCQ}
\newblock
\APACjournalVolNumPages{JOURNAL-KOREAN PHYSICAL SOCIETY}{44}{1}{633--637}.
\PrintBackRefs{\CurrentBib}

\bibitem [\protect \citeauthoryear {%
H.~Lee%
, Wang%
\BCBL {}\ \BBA {} Hudetz%
}{%
H.~Lee%
\ \protect \BOthers {.}}{%
{\protect \APACyear {2020}}%
}]{%
lee2020state}
\APACinsertmetastar {%
lee2020state}%
\begin{APACrefauthors}%
Lee, H.%
, Wang, S.%
\BCBL {}\ \BBA {} Hudetz, A\BPBI G.%
\end{APACrefauthors}%
\unskip\
\newblock
\APACrefYearMonthDay{2020}{}{}.
\newblock
{\BBOQ}\APACrefatitle {State-dependent cortical unit activity reflects dynamic
  brain state transitions in anesthesia} {State-dependent cortical unit
  activity reflects dynamic brain state transitions in anesthesia}.{\BBCQ}
\newblock
\APACjournalVolNumPages{Journal of Neuroscience}{40}{49}{9440--9454}.
\PrintBackRefs{\CurrentBib}

\bibitem [\protect \citeauthoryear {%
Levina%
, Herrmann%
\BCBL {}\ \BBA {} Geisel%
}{%
Levina%
\ \protect \BOthers {.}}{%
{\protect \APACyear {2007}}%
}]{%
levina2007dynamical}
\APACinsertmetastar {%
levina2007dynamical}%
\begin{APACrefauthors}%
Levina, A.%
, Herrmann, J\BPBI M.%
\BCBL {}\ \BBA {} Geisel, T.%
\end{APACrefauthors}%
\unskip\
\newblock
\APACrefYearMonthDay{2007}{}{}.
\newblock
{\BBOQ}\APACrefatitle {Dynamical synapses causing self-organized criticality in
  neural networks} {Dynamical synapses causing self-organized criticality in
  neural networks}.{\BBCQ}
\newblock
\APACjournalVolNumPages{Nature physics}{3}{12}{857--860}.
\PrintBackRefs{\CurrentBib}

\bibitem [\protect \citeauthoryear {%
Levina%
, Herrmann%
\BCBL {}\ \BBA {} Geisel%
}{%
Levina%
\ \protect \BOthers {.}}{%
{\protect \APACyear {2009}}%
}]{%
levina2009phase}
\APACinsertmetastar {%
levina2009phase}%
\begin{APACrefauthors}%
Levina, A.%
, Herrmann, J\BPBI M.%
\BCBL {}\ \BBA {} Geisel, T.%
\end{APACrefauthors}%
\unskip\
\newblock
\APACrefYearMonthDay{2009}{}{}.
\newblock
{\BBOQ}\APACrefatitle {Phase transitions towards criticality in a neural system
  with adaptive interactions} {Phase transitions towards criticality in a
  neural system with adaptive interactions}.{\BBCQ}
\newblock
\APACjournalVolNumPages{Physical review letters}{102}{11}{118110}.
\PrintBackRefs{\CurrentBib}

\bibitem [\protect \citeauthoryear {%
J.~Li%
\ \BBA {} Shew%
}{%
J.~Li%
\ \BBA {} Shew%
}{%
{\protect \APACyear {2020}}%
}]{%
li2020tuning}
\APACinsertmetastar {%
li2020tuning}%
\begin{APACrefauthors}%
Li, J.%
\BCBT {}\ \BBA {} Shew, W\BPBI L.%
\end{APACrefauthors}%
\unskip\
\newblock
\APACrefYearMonthDay{2020}{}{}.
\newblock
{\BBOQ}\APACrefatitle {Tuning network dynamics from criticality to an
  asynchronous state} {Tuning network dynamics from criticality to an
  asynchronous state}.{\BBCQ}
\newblock
\APACjournalVolNumPages{PLOS Computational Biology}{16}{9}{e1008268}.
\PrintBackRefs{\CurrentBib}

\bibitem [\protect \citeauthoryear {%
X.~Li%
\ \BBA {} Small%
}{%
X.~Li%
\ \BBA {} Small%
}{%
{\protect \APACyear {2012}}%
}]{%
li2012neuronal}
\APACinsertmetastar {%
li2012neuronal}%
\begin{APACrefauthors}%
Li, X.%
\BCBT {}\ \BBA {} Small, M.%
\end{APACrefauthors}%
\unskip\
\newblock
\APACrefYearMonthDay{2012}{}{}.
\newblock
{\BBOQ}\APACrefatitle {Neuronal avalanches of a self-organized neural network
  with active-neuron-dominant structure} {Neuronal avalanches of a
  self-organized neural network with active-neuron-dominant structure}.{\BBCQ}
\newblock
\APACjournalVolNumPages{Chaos: An Interdisciplinary Journal of Nonlinear
  Science}{22}{2}{023104}.
\PrintBackRefs{\CurrentBib}

\bibitem [\protect \citeauthoryear {%
Liggett%
}{%
Liggett%
}{%
{\protect \APACyear {2006}}%
}]{%
liggett2006interacting}
\APACinsertmetastar {%
liggett2006interacting}%
\begin{APACrefauthors}%
Liggett, T\BPBI M.%
\end{APACrefauthors}%
\unskip\
\newblock
\APACrefYear{2006}.
\newblock
\APACrefbtitle {Interacting Particle Systems} {Interacting particle systems}.
\newblock
\APACaddressPublisher{}{Springer Science \& Business Media}.
\PrintBackRefs{\CurrentBib}

\bibitem [\protect \citeauthoryear {%
Linkenkaer-Hansen%
, Nikouline%
, Palva%
\BCBL {}\ \BBA {} Ilmoniemi%
}{%
Linkenkaer-Hansen%
\ \protect \BOthers {.}}{%
{\protect \APACyear {2001}}%
}]{%
linkenkaer2001long}
\APACinsertmetastar {%
linkenkaer2001long}%
\begin{APACrefauthors}%
Linkenkaer-Hansen, K.%
, Nikouline, V\BPBI V.%
, Palva, J\BPBI M.%
\BCBL {}\ \BBA {} Ilmoniemi, R\BPBI J.%
\end{APACrefauthors}%
\unskip\
\newblock
\APACrefYearMonthDay{2001}{}{}.
\newblock
{\BBOQ}\APACrefatitle {Long-range temporal correlations and scaling behavior in
  human brain oscillations} {Long-range temporal correlations and scaling
  behavior in human brain oscillations}.{\BBCQ}
\newblock
\APACjournalVolNumPages{Journal of Neuroscience}{21}{4}{1370--1377}.
\PrintBackRefs{\CurrentBib}

\bibitem [\protect \citeauthoryear {%
Lombardi%
, Herrmann%
\BCBL {}\ \BBA {} de Arcangelis%
}{%
Lombardi%
\ \protect \BOthers {.}}{%
{\protect \APACyear {2017}}%
}]{%
lombardi2017balance}
\APACinsertmetastar {%
lombardi2017balance}%
\begin{APACrefauthors}%
Lombardi, F.%
, Herrmann, H\BPBI J.%
\BCBL {}\ \BBA {} de Arcangelis, L.%
\end{APACrefauthors}%
\unskip\
\newblock
\APACrefYearMonthDay{2017}{}{}.
\newblock
{\BBOQ}\APACrefatitle {Balance of excitation and inhibition determines 1/f
  power spectrum in neuronal networks} {Balance of excitation and inhibition
  determines 1/f power spectrum in neuronal networks}.{\BBCQ}
\newblock
\APACjournalVolNumPages{Chaos: An Interdisciplinary Journal of Nonlinear
  Science}{27}{4}{047402}.
\PrintBackRefs{\CurrentBib}

\bibitem [\protect \citeauthoryear {%
L{\"u}beck%
}{%
L{\"u}beck%
}{%
{\protect \APACyear {2004}}%
}]{%
lubeck2004universal}
\APACinsertmetastar {%
lubeck2004universal}%
\begin{APACrefauthors}%
L{\"u}beck, S.%
\end{APACrefauthors}%
\unskip\
\newblock
\APACrefYearMonthDay{2004}{}{}.
\newblock
{\BBOQ}\APACrefatitle {Universal scaling behavior of non-equilibrium phase
  transitions} {Universal scaling behavior of non-equilibrium phase
  transitions}.{\BBCQ}
\newblock
\APACjournalVolNumPages{International Journal of Modern Physics
  B}{18}{31n32}{3977--4118}.
\PrintBackRefs{\CurrentBib}

\bibitem [\protect \citeauthoryear {%
L{\"u}beck%
\ \BBA {} Heger%
}{%
L{\"u}beck%
\ \BBA {} Heger%
}{%
{\protect \APACyear {2003}}%
}]{%
lubeck2003universal}
\APACinsertmetastar {%
lubeck2003universal}%
\begin{APACrefauthors}%
L{\"u}beck, S.%
\BCBT {}\ \BBA {} Heger, P.%
\end{APACrefauthors}%
\unskip\
\newblock
\APACrefYearMonthDay{2003}{}{}.
\newblock
{\BBOQ}\APACrefatitle {Universal finite-size scaling behavior and universal
  dynamical scaling behavior of absorbing phase transitions with a conserved
  field} {Universal finite-size scaling behavior and universal dynamical
  scaling behavior of absorbing phase transitions with a conserved
  field}.{\BBCQ}
\newblock
\APACjournalVolNumPages{Physical Review E}{68}{5}{056102}.
\PrintBackRefs{\CurrentBib}

\bibitem [\protect \citeauthoryear {%
Lynn%
\ \BBA {} Bassett%
}{%
Lynn%
\ \BBA {} Bassett%
}{%
{\protect \APACyear {2019}}%
}]{%
lynn2019physics}
\APACinsertmetastar {%
lynn2019physics}%
\begin{APACrefauthors}%
Lynn, C\BPBI W.%
\BCBT {}\ \BBA {} Bassett, D\BPBI S.%
\end{APACrefauthors}%
\unskip\
\newblock
\APACrefYearMonthDay{2019}{}{}.
\newblock
{\BBOQ}\APACrefatitle {The physics of brain network structure, function and
  control} {The physics of brain network structure, function and
  control}.{\BBCQ}
\newblock
\APACjournalVolNumPages{Nature Reviews Physics}{1}{5}{318--332}.
\PrintBackRefs{\CurrentBib}

\bibitem [\protect \citeauthoryear {%
Lynn%
, Cornblath%
, Papadopoulos%
, Bertolero%
\BCBL {}\ \BBA {} Bassett%
}{%
Lynn%
\ \protect \BOthers {.}}{%
{\protect \APACyear {2021}}%
}]{%
lynn2021broken}
\APACinsertmetastar {%
lynn2021broken}%
\begin{APACrefauthors}%
Lynn, C\BPBI W.%
, Cornblath, E\BPBI J.%
, Papadopoulos, L.%
, Bertolero, M\BPBI A.%
\BCBL {}\ \BBA {} Bassett, D\BPBI S.%
\end{APACrefauthors}%
\unskip\
\newblock
\APACrefYearMonthDay{2021}{}{}.
\newblock
{\BBOQ}\APACrefatitle {Broken detailed balance and entropy production in the
  human brain} {Broken detailed balance and entropy production in the human
  brain}.{\BBCQ}
\newblock
\APACjournalVolNumPages{Proceedings of the National Academy of
  Sciences}{118}{47}{}.
\PrintBackRefs{\CurrentBib}

\bibitem [\protect \citeauthoryear {%
Ma%
, Turrigiano%
, Wessel%
\BCBL {}\ \BBA {} Hengen%
}{%
Ma%
\ \protect \BOthers {.}}{%
{\protect \APACyear {2019}}%
}]{%
ma2019cortical}
\APACinsertmetastar {%
ma2019cortical}%
\begin{APACrefauthors}%
Ma, Z.%
, Turrigiano, G\BPBI G.%
, Wessel, R.%
\BCBL {}\ \BBA {} Hengen, K\BPBI B.%
\end{APACrefauthors}%
\unskip\
\newblock
\APACrefYearMonthDay{2019}{}{}.
\newblock
{\BBOQ}\APACrefatitle {Cortical circuit dynamics are homeostatically tuned to
  criticality in vivo} {Cortical circuit dynamics are homeostatically tuned to
  criticality in vivo}.{\BBCQ}
\newblock
\APACjournalVolNumPages{Neuron}{104}{4}{655--664}.
\PrintBackRefs{\CurrentBib}

\bibitem [\protect \citeauthoryear {%
Malcai%
, Shilo%
\BCBL {}\ \BBA {} Biham%
}{%
Malcai%
\ \protect \BOthers {.}}{%
{\protect \APACyear {2006}}%
}]{%
malcai2006dissipative}
\APACinsertmetastar {%
malcai2006dissipative}%
\begin{APACrefauthors}%
Malcai, O.%
, Shilo, Y.%
\BCBL {}\ \BBA {} Biham, O.%
\end{APACrefauthors}%
\unskip\
\newblock
\APACrefYearMonthDay{2006}{}{}.
\newblock
{\BBOQ}\APACrefatitle {Dissipative sandpile models with universal exponents}
  {Dissipative sandpile models with universal exponents}.{\BBCQ}
\newblock
\APACjournalVolNumPages{Physical Review E}{73}{5}{056125}.
\PrintBackRefs{\CurrentBib}

\bibitem [\protect \citeauthoryear {%
Manna%
}{%
Manna%
}{%
{\protect \APACyear {1991}}%
}]{%
manna1991two}
\APACinsertmetastar {%
manna1991two}%
\begin{APACrefauthors}%
Manna, S\BPBI S.%
\end{APACrefauthors}%
\unskip\
\newblock
\APACrefYearMonthDay{1991}{}{}.
\newblock
{\BBOQ}\APACrefatitle {Two-state model of self-organized criticality}
  {Two-state model of self-organized criticality}.{\BBCQ}
\newblock
\APACjournalVolNumPages{Journal of Physics A: Mathematical and
  General}{24}{7}{L363}.
\PrintBackRefs{\CurrentBib}

\bibitem [\protect \citeauthoryear {%
Mariani%
\ \protect \BOthers {.}}{%
Mariani%
\ \protect \BOthers {.}}{%
{\protect \APACyear {2021}}%
}]{%
mariani2021beyond}
\APACinsertmetastar {%
mariani2021beyond}%
\begin{APACrefauthors}%
Mariani, B.%
, Nicoletti, G.%
, Bisio, M.%
, Maschietto, M.%
, Oboe, R.%
, Suweis, S.%
\BCBL {}\ \BBA {} Vassanelli, S.%
\end{APACrefauthors}%
\unskip\
\newblock
\APACrefYearMonthDay{2021}{}{}.
\newblock
{\BBOQ}\APACrefatitle {Beyond resting state neuronal avalanches in the
  somatosensory barrel cortex} {Beyond resting state neuronal avalanches in the
  somatosensory barrel cortex}.{\BBCQ}
\newblock
\APACjournalVolNumPages{bioRxiv}{}{}{}.
\PrintBackRefs{\CurrentBib}

\bibitem [\protect \citeauthoryear {%
Markovi{\'c}%
\ \BBA {} Gros%
}{%
Markovi{\'c}%
\ \BBA {} Gros%
}{%
{\protect \APACyear {2014}}%
}]{%
markovic2014power}
\APACinsertmetastar {%
markovic2014power}%
\begin{APACrefauthors}%
Markovi{\'c}, D.%
\BCBT {}\ \BBA {} Gros, C.%
\end{APACrefauthors}%
\unskip\
\newblock
\APACrefYearMonthDay{2014}{}{}.
\newblock
{\BBOQ}\APACrefatitle {Power laws and self-organized criticality in theory and
  nature} {Power laws and self-organized criticality in theory and
  nature}.{\BBCQ}
\newblock
\APACjournalVolNumPages{Physics Reports}{536}{2}{41--74}.
\PrintBackRefs{\CurrentBib}

\bibitem [\protect \citeauthoryear {%
Marshall%
\ \protect \BOthers {.}}{%
Marshall%
\ \protect \BOthers {.}}{%
{\protect \APACyear {2016}}%
}]{%
marshall2016analysis}
\APACinsertmetastar {%
marshall2016analysis}%
\begin{APACrefauthors}%
Marshall, N.%
, Timme, N\BPBI M.%
, Bennett, N.%
, Ripp, M.%
, Lautzenhiser, E.%
\BCBL {}\ \BBA {} Beggs, J\BPBI M.%
\end{APACrefauthors}%
\unskip\
\newblock
\APACrefYearMonthDay{2016}{}{}.
\newblock
{\BBOQ}\APACrefatitle {Analysis of power laws, shape collapses, and neural
  complexity: new techniques and matlab support via the ncc toolbox} {Analysis
  of power laws, shape collapses, and neural complexity: new techniques and
  matlab support via the ncc toolbox}.{\BBCQ}
\newblock
\APACjournalVolNumPages{Frontiers in physiology}{7}{}{250}.
\PrintBackRefs{\CurrentBib}

\bibitem [\protect \citeauthoryear {%
Martinello%
\ \protect \BOthers {.}}{%
Martinello%
\ \protect \BOthers {.}}{%
{\protect \APACyear {2017}}%
}]{%
martinello2017neutral}
\APACinsertmetastar {%
martinello2017neutral}%
\begin{APACrefauthors}%
Martinello, M.%
, Hidalgo, J.%
, Maritan, A.%
, Di~Santo, S.%
, Plenz, D.%
\BCBL {}\ \BBA {} Munoz, M\BPBI A.%
\end{APACrefauthors}%
\unskip\
\newblock
\APACrefYearMonthDay{2017}{}{}.
\newblock
{\BBOQ}\APACrefatitle {Neutral theory and scale-free neural dynamics} {Neutral
  theory and scale-free neural dynamics}.{\BBCQ}
\newblock
\APACjournalVolNumPages{Physical Review X}{7}{4}{041071}.
\PrintBackRefs{\CurrentBib}

\bibitem [\protect \citeauthoryear {%
Mehta%
, Mills%
, Dahmen%
\BCBL {}\ \BBA {} Sethna%
}{%
Mehta%
\ \protect \BOthers {.}}{%
{\protect \APACyear {2002}}%
}]{%
mehta2002universal}
\APACinsertmetastar {%
mehta2002universal}%
\begin{APACrefauthors}%
Mehta, A\BPBI P.%
, Mills, A\BPBI C.%
, Dahmen, K\BPBI A.%
\BCBL {}\ \BBA {} Sethna, J\BPBI P.%
\end{APACrefauthors}%
\unskip\
\newblock
\APACrefYearMonthDay{2002}{}{}.
\newblock
{\BBOQ}\APACrefatitle {Universal pulse shape scaling function and exponents:
  Critical test for avalanche models applied to Barkhausen noise} {Universal
  pulse shape scaling function and exponents: Critical test for avalanche
  models applied to barkhausen noise}.{\BBCQ}
\newblock
\APACjournalVolNumPages{Physical Review E}{65}{4}{046139}.
\PrintBackRefs{\CurrentBib}

\bibitem [\protect \citeauthoryear {%
Meisel%
\ \BBA {} Gross%
}{%
Meisel%
\ \BBA {} Gross%
}{%
{\protect \APACyear {2009}}%
}]{%
meisel2009adaptive}
\APACinsertmetastar {%
meisel2009adaptive}%
\begin{APACrefauthors}%
Meisel, C.%
\BCBT {}\ \BBA {} Gross, T.%
\end{APACrefauthors}%
\unskip\
\newblock
\APACrefYearMonthDay{2009}{}{}.
\newblock
{\BBOQ}\APACrefatitle {Adaptive self-organization in a realistic neural network
  model} {Adaptive self-organization in a realistic neural network
  model}.{\BBCQ}
\newblock
\APACjournalVolNumPages{Physical Review E}{80}{6}{061917}.
\PrintBackRefs{\CurrentBib}

\bibitem [\protect \citeauthoryear {%
Meisel%
, Olbrich%
, Shriki%
\BCBL {}\ \BBA {} Achermann%
}{%
Meisel%
\ \protect \BOthers {.}}{%
{\protect \APACyear {2013}}%
}]{%
meisel2013fading}
\APACinsertmetastar {%
meisel2013fading}%
\begin{APACrefauthors}%
Meisel, C.%
, Olbrich, E.%
, Shriki, O.%
\BCBL {}\ \BBA {} Achermann, P.%
\end{APACrefauthors}%
\unskip\
\newblock
\APACrefYearMonthDay{2013}{}{}.
\newblock
{\BBOQ}\APACrefatitle {Fading signatures of critical brain dynamics during
  sustained wakefulness in humans} {Fading signatures of critical brain
  dynamics during sustained wakefulness in humans}.{\BBCQ}
\newblock
\APACjournalVolNumPages{Journal of Neuroscience}{33}{44}{17363--17372}.
\PrintBackRefs{\CurrentBib}

\bibitem [\protect \citeauthoryear {%
Miller%
\ \BBA {} Wang%
}{%
Miller%
\ \BBA {} Wang%
}{%
{\protect \APACyear {2006}}%
}]{%
miller2006power}
\APACinsertmetastar {%
miller2006power}%
\begin{APACrefauthors}%
Miller, P.%
\BCBT {}\ \BBA {} Wang, X\BHBI J.%
\end{APACrefauthors}%
\unskip\
\newblock
\APACrefYearMonthDay{2006}{}{}.
\newblock
{\BBOQ}\APACrefatitle {Power-law neuronal fluctuations in a recurrent network
  model of parametric working memory} {Power-law neuronal fluctuations in a
  recurrent network model of parametric working memory}.{\BBCQ}
\newblock
\APACjournalVolNumPages{Journal of neurophysiology}{95}{2}{1099--1114}.
\PrintBackRefs{\CurrentBib}

\bibitem [\protect \citeauthoryear {%
Millman%
, Mihalas%
, Kirkwood%
\BCBL {}\ \BBA {} Niebur%
}{%
Millman%
\ \protect \BOthers {.}}{%
{\protect \APACyear {2010}}%
}]{%
millman2010self}
\APACinsertmetastar {%
millman2010self}%
\begin{APACrefauthors}%
Millman, D.%
, Mihalas, S.%
, Kirkwood, A.%
\BCBL {}\ \BBA {} Niebur, E.%
\end{APACrefauthors}%
\unskip\
\newblock
\APACrefYearMonthDay{2010}{}{}.
\newblock
{\BBOQ}\APACrefatitle {Self-organized criticality occurs in non-conservative
  neuronal networks during ‘up’states} {Self-organized criticality occurs
  in non-conservative neuronal networks during ‘up’states}.{\BBCQ}
\newblock
\APACjournalVolNumPages{Nature physics}{6}{10}{801--805}.
\PrintBackRefs{\CurrentBib}

\bibitem [\protect \citeauthoryear {%
Mitzenmacher%
}{%
Mitzenmacher%
}{%
{\protect \APACyear {2004}}%
}]{%
mitzenmacher2004brief}
\APACinsertmetastar {%
mitzenmacher2004brief}%
\begin{APACrefauthors}%
Mitzenmacher, M.%
\end{APACrefauthors}%
\unskip\
\newblock
\APACrefYearMonthDay{2004}{}{}.
\newblock
{\BBOQ}\APACrefatitle {A brief history of generative models for power law and
  lognormal distributions} {A brief history of generative models for power law
  and lognormal distributions}.{\BBCQ}
\newblock
\APACjournalVolNumPages{Internet mathematics}{1}{2}{226--251}.
\PrintBackRefs{\CurrentBib}

\bibitem [\protect \citeauthoryear {%
Molgedey%
, Schuchhardt%
\BCBL {}\ \BBA {} Schuster%
}{%
Molgedey%
\ \protect \BOthers {.}}{%
{\protect \APACyear {1992}}%
}]{%
molgedey1992suppressing}
\APACinsertmetastar {%
molgedey1992suppressing}%
\begin{APACrefauthors}%
Molgedey, L.%
, Schuchhardt, J.%
\BCBL {}\ \BBA {} Schuster, H\BPBI G.%
\end{APACrefauthors}%
\unskip\
\newblock
\APACrefYearMonthDay{1992}{}{}.
\newblock
{\BBOQ}\APACrefatitle {Suppressing chaos in neural networks by noise}
  {Suppressing chaos in neural networks by noise}.{\BBCQ}
\newblock
\APACjournalVolNumPages{Physical review letters}{69}{26}{3717}.
\PrintBackRefs{\CurrentBib}

\bibitem [\protect \citeauthoryear {%
Montague%
, Dayan%
\BCBL {}\ \BBA {} Sejnowski%
}{%
Montague%
\ \protect \BOthers {.}}{%
{\protect \APACyear {1996}}%
}]{%
montague1996framework}
\APACinsertmetastar {%
montague1996framework}%
\begin{APACrefauthors}%
Montague, P\BPBI R.%
, Dayan, P.%
\BCBL {}\ \BBA {} Sejnowski, T\BPBI J.%
\end{APACrefauthors}%
\unskip\
\newblock
\APACrefYearMonthDay{1996}{}{}.
\newblock
{\BBOQ}\APACrefatitle {A framework for mesencephalic dopamine systems based on
  predictive Hebbian learning} {A framework for mesencephalic dopamine systems
  based on predictive hebbian learning}.{\BBCQ}
\newblock
\APACjournalVolNumPages{Journal of neuroscience}{16}{5}{1936--1947}.
\PrintBackRefs{\CurrentBib}

\bibitem [\protect \citeauthoryear {%
Moretti%
\ \BBA {} Munoz%
}{%
Moretti%
\ \BBA {} Munoz%
}{%
{\protect \APACyear {2013}}%
}]{%
moretti2013griffiths}
\APACinsertmetastar {%
moretti2013griffiths}%
\begin{APACrefauthors}%
Moretti, P.%
\BCBT {}\ \BBA {} Munoz, M\BPBI A.%
\end{APACrefauthors}%
\unskip\
\newblock
\APACrefYearMonthDay{2013}{}{}.
\newblock
{\BBOQ}\APACrefatitle {Griffiths phases and the stretching of criticality in
  brain networks} {Griffiths phases and the stretching of criticality in brain
  networks}.{\BBCQ}
\newblock
\APACjournalVolNumPages{Nature communications}{4}{1}{1--10}.
\PrintBackRefs{\CurrentBib}

\bibitem [\protect \citeauthoryear {%
Munoz%
}{%
Munoz%
}{%
{\protect \APACyear {2018}}%
}]{%
munoz2018colloquium}
\APACinsertmetastar {%
munoz2018colloquium}%
\begin{APACrefauthors}%
Munoz, M\BPBI A.%
\end{APACrefauthors}%
\unskip\
\newblock
\APACrefYearMonthDay{2018}{}{}.
\newblock
{\BBOQ}\APACrefatitle {Colloquium: Criticality and dynamical scaling in living
  systems} {Colloquium: Criticality and dynamical scaling in living
  systems}.{\BBCQ}
\newblock
\APACjournalVolNumPages{Reviews of Modern Physics}{90}{3}{031001}.
\PrintBackRefs{\CurrentBib}

\bibitem [\protect \citeauthoryear {%
Narayan%
\ \BBA {} Middleton%
}{%
Narayan%
\ \BBA {} Middleton%
}{%
{\protect \APACyear {1994}}%
}]{%
narayan1994avalanches}
\APACinsertmetastar {%
narayan1994avalanches}%
\begin{APACrefauthors}%
Narayan, O.%
\BCBT {}\ \BBA {} Middleton, A\BPBI A.%
\end{APACrefauthors}%
\unskip\
\newblock
\APACrefYearMonthDay{1994}{}{}.
\newblock
{\BBOQ}\APACrefatitle {Avalanches and the renormalization group for pinned
  charge-density waves} {Avalanches and the renormalization group for pinned
  charge-density waves}.{\BBCQ}
\newblock
\APACjournalVolNumPages{Physical Review B}{49}{1}{244}.
\PrintBackRefs{\CurrentBib}

\bibitem [\protect \citeauthoryear {%
Otter%
}{%
Otter%
}{%
{\protect \APACyear {1949}}%
}]{%
otter1949multiplicative}
\APACinsertmetastar {%
otter1949multiplicative}%
\begin{APACrefauthors}%
Otter, R.%
\end{APACrefauthors}%
\unskip\
\newblock
\APACrefYearMonthDay{1949}{}{}.
\newblock
{\BBOQ}\APACrefatitle {The multiplicative process} {The multiplicative
  process}.{\BBCQ}
\newblock
\APACjournalVolNumPages{The Annals of Mathematical
  Statistics}{20}{2}{206--224}.
\PrintBackRefs{\CurrentBib}

\bibitem [\protect \citeauthoryear {%
Palva%
\ \protect \BOthers {.}}{%
Palva%
\ \protect \BOthers {.}}{%
{\protect \APACyear {2013}}%
}]{%
palva2013neuronal}
\APACinsertmetastar {%
palva2013neuronal}%
\begin{APACrefauthors}%
Palva, J\BPBI M.%
, Zhigalov, A.%
, Hirvonen, J.%
, Korhonen, O.%
, Linkenkaer-Hansen, K.%
\BCBL {}\ \BBA {} Palva, S.%
\end{APACrefauthors}%
\unskip\
\newblock
\APACrefYearMonthDay{2013}{}{}.
\newblock
{\BBOQ}\APACrefatitle {Neuronal long-range temporal correlations and avalanche
  dynamics are correlated with behavioral scaling laws} {Neuronal long-range
  temporal correlations and avalanche dynamics are correlated with behavioral
  scaling laws}.{\BBCQ}
\newblock
\APACjournalVolNumPages{Proceedings of the National Academy of
  Sciences}{110}{9}{3585--3590}.
\PrintBackRefs{\CurrentBib}

\bibitem [\protect \citeauthoryear {%
Papanikolaou%
\ \protect \BOthers {.}}{%
Papanikolaou%
\ \protect \BOthers {.}}{%
{\protect \APACyear {2011}}%
}]{%
papanikolaou2011universality}
\APACinsertmetastar {%
papanikolaou2011universality}%
\begin{APACrefauthors}%
Papanikolaou, S.%
, Bohn, F.%
, Sommer, R\BPBI L.%
, Durin, G.%
, Zapperi, S.%
\BCBL {}\ \BBA {} Sethna, J\BPBI P.%
\end{APACrefauthors}%
\unskip\
\newblock
\APACrefYearMonthDay{2011}{}{}.
\newblock
{\BBOQ}\APACrefatitle {Universality beyond power laws and the average avalanche
  shape} {Universality beyond power laws and the average avalanche
  shape}.{\BBCQ}
\newblock
\APACjournalVolNumPages{Nature Physics}{7}{4}{316--320}.
\PrintBackRefs{\CurrentBib}

\bibitem [\protect \citeauthoryear {%
Pausch%
, Garcia-Millan%
\BCBL {}\ \BBA {} Pruessner%
}{%
Pausch%
\ \protect \BOthers {.}}{%
{\protect \APACyear {2020}}%
}]{%
pausch2020time}
\APACinsertmetastar {%
pausch2020time}%
\begin{APACrefauthors}%
Pausch, J.%
, Garcia-Millan, R.%
\BCBL {}\ \BBA {} Pruessner, G.%
\end{APACrefauthors}%
\unskip\
\newblock
\APACrefYearMonthDay{2020}{}{}.
\newblock
{\BBOQ}\APACrefatitle {Time-dependent branching processes: a model of
  oscillating neuronal avalanches} {Time-dependent branching processes: a model
  of oscillating neuronal avalanches}.{\BBCQ}
\newblock
\APACjournalVolNumPages{Scientific Reports}{10}{1}{1--17}.
\PrintBackRefs{\CurrentBib}

\bibitem [\protect \citeauthoryear {%
Perl%
\ \protect \BOthers {.}}{%
Perl%
\ \protect \BOthers {.}}{%
{\protect \APACyear {2021}}%
}]{%
perl2021nonequilibrium}
\APACinsertmetastar {%
perl2021nonequilibrium}%
\begin{APACrefauthors}%
Perl, Y\BPBI S.%
, Bocaccio, H.%
, Pallavicini, C.%
, P{\'e}rez-Ipi{\~n}a, I.%
, Laureys, S.%
, Laufs, H.%
\BDBL {}Tagliazucchi, E.%
\end{APACrefauthors}%
\unskip\
\newblock
\APACrefYearMonthDay{2021}{}{}.
\newblock
{\BBOQ}\APACrefatitle {Nonequilibrium brain dynamics as a signature of
  consciousness} {Nonequilibrium brain dynamics as a signature of
  consciousness}.{\BBCQ}
\newblock
\APACjournalVolNumPages{Physical Review E}{104}{1}{014411}.
\PrintBackRefs{\CurrentBib}

\bibitem [\protect \citeauthoryear {%
Petermann%
\ \protect \BOthers {.}}{%
Petermann%
\ \protect \BOthers {.}}{%
{\protect \APACyear {2009}}%
}]{%
petermann2009spontaneous}
\APACinsertmetastar {%
petermann2009spontaneous}%
\begin{APACrefauthors}%
Petermann, T.%
, Thiagarajan, T\BPBI C.%
, Lebedev, M\BPBI A.%
, Nicolelis, M\BPBI A.%
, Chialvo, D\BPBI R.%
\BCBL {}\ \BBA {} Plenz, D.%
\end{APACrefauthors}%
\unskip\
\newblock
\APACrefYearMonthDay{2009}{}{}.
\newblock
{\BBOQ}\APACrefatitle {Spontaneous cortical activity in awake monkeys composed
  of neuronal avalanches} {Spontaneous cortical activity in awake monkeys
  composed of neuronal avalanches}.{\BBCQ}
\newblock
\APACjournalVolNumPages{Proceedings of the National Academy of
  Sciences}{106}{37}{15921--15926}.
\PrintBackRefs{\CurrentBib}

\bibitem [\protect \citeauthoryear {%
Poil%
, Hardstone%
, Mansvelder%
\BCBL {}\ \BBA {} Linkenkaer-Hansen%
}{%
Poil%
\ \protect \BOthers {.}}{%
{\protect \APACyear {2012}}%
}]{%
poil2012critical}
\APACinsertmetastar {%
poil2012critical}%
\begin{APACrefauthors}%
Poil, S\BHBI S.%
, Hardstone, R.%
, Mansvelder, H\BPBI D.%
\BCBL {}\ \BBA {} Linkenkaer-Hansen, K.%
\end{APACrefauthors}%
\unskip\
\newblock
\APACrefYearMonthDay{2012}{}{}.
\newblock
{\BBOQ}\APACrefatitle {Critical-state dynamics of avalanches and oscillations
  jointly emerge from balanced excitation/inhibition in neuronal networks}
  {Critical-state dynamics of avalanches and oscillations jointly emerge from
  balanced excitation/inhibition in neuronal networks}.{\BBCQ}
\newblock
\APACjournalVolNumPages{Journal of Neuroscience}{32}{29}{9817--9823}.
\PrintBackRefs{\CurrentBib}

\bibitem [\protect \citeauthoryear {%
Poil%
, van Ooyen%
\BCBL {}\ \BBA {} Linkenkaer-Hansen%
}{%
Poil%
\ \protect \BOthers {.}}{%
{\protect \APACyear {2008}}%
}]{%
poil2008avalanche}
\APACinsertmetastar {%
poil2008avalanche}%
\begin{APACrefauthors}%
Poil, S\BHBI S.%
, van Ooyen, A.%
\BCBL {}\ \BBA {} Linkenkaer-Hansen, K.%
\end{APACrefauthors}%
\unskip\
\newblock
\APACrefYearMonthDay{2008}{}{}.
\newblock
{\BBOQ}\APACrefatitle {Avalanche dynamics of human brain oscillations: relation
  to critical branching processes and temporal correlations} {Avalanche
  dynamics of human brain oscillations: relation to critical branching
  processes and temporal correlations}.{\BBCQ}
\newblock
\APACjournalVolNumPages{Human brain mapping}{29}{7}{770--777}.
\PrintBackRefs{\CurrentBib}

\bibitem [\protect \citeauthoryear {%
Ponce-Alvarez%
, Jouary%
, Privat%
, Deco%
\BCBL {}\ \BBA {} Sumbre%
}{%
Ponce-Alvarez%
\ \protect \BOthers {.}}{%
{\protect \APACyear {2018}}%
}]{%
ponce2018whole}
\APACinsertmetastar {%
ponce2018whole}%
\begin{APACrefauthors}%
Ponce-Alvarez, A.%
, Jouary, A.%
, Privat, M.%
, Deco, G.%
\BCBL {}\ \BBA {} Sumbre, G.%
\end{APACrefauthors}%
\unskip\
\newblock
\APACrefYearMonthDay{2018}{}{}.
\newblock
{\BBOQ}\APACrefatitle {Whole-brain neuronal activity displays crackling noise
  dynamics} {Whole-brain neuronal activity displays crackling noise
  dynamics}.{\BBCQ}
\newblock
\APACjournalVolNumPages{Neuron}{100}{6}{1446--1459}.
\PrintBackRefs{\CurrentBib}

\bibitem [\protect \citeauthoryear {%
Rao%
\ \BBA {} Swift%
}{%
Rao%
\ \BBA {} Swift%
}{%
{\protect \APACyear {2006}}%
}]{%
rao2006probability}
\APACinsertmetastar {%
rao2006probability}%
\begin{APACrefauthors}%
Rao, M\BPBI M.%
\BCBT {}\ \BBA {} Swift, R\BPBI J.%
\end{APACrefauthors}%
\unskip\
\newblock
\APACrefYear{2006}.
\newblock
\APACrefbtitle {Probability theory with applications} {Probability theory with
  applications}\ (\BVOL~582).
\newblock
\APACaddressPublisher{}{Springer Science \& Business Media}.
\PrintBackRefs{\CurrentBib}

\bibitem [\protect \citeauthoryear {%
Reed%
\ \BBA {} Hughes%
}{%
Reed%
\ \BBA {} Hughes%
}{%
{\protect \APACyear {2002}}%
}]{%
reed2002gene}
\APACinsertmetastar {%
reed2002gene}%
\begin{APACrefauthors}%
Reed, W\BPBI J.%
\BCBT {}\ \BBA {} Hughes, B\BPBI D.%
\end{APACrefauthors}%
\unskip\
\newblock
\APACrefYearMonthDay{2002}{}{}.
\newblock
{\BBOQ}\APACrefatitle {From gene families and genera to incomes and internet
  file sizes: Why power laws are so common in nature} {From gene families and
  genera to incomes and internet file sizes: Why power laws are so common in
  nature}.{\BBCQ}
\newblock
\APACjournalVolNumPages{Physical Review E}{66}{6}{067103}.
\PrintBackRefs{\CurrentBib}

\bibitem [\protect \citeauthoryear {%
Reimer%
\ \protect \BOthers {.}}{%
Reimer%
\ \protect \BOthers {.}}{%
{\protect \APACyear {2014}}%
}]{%
reimer2014pupil}
\APACinsertmetastar {%
reimer2014pupil}%
\begin{APACrefauthors}%
Reimer, J.%
, Froudarakis, E.%
, Cadwell, C\BPBI R.%
, Yatsenko, D.%
, Denfield, G\BPBI H.%
\BCBL {}\ \BBA {} Tolias, A\BPBI S.%
\end{APACrefauthors}%
\unskip\
\newblock
\APACrefYearMonthDay{2014}{}{}.
\newblock
{\BBOQ}\APACrefatitle {Pupil fluctuations track fast switching of cortical
  states during quiet wakefulness} {Pupil fluctuations track fast switching of
  cortical states during quiet wakefulness}.{\BBCQ}
\newblock
\APACjournalVolNumPages{neuron}{84}{2}{355--362}.
\PrintBackRefs{\CurrentBib}

\bibitem [\protect \citeauthoryear {%
Reiss%
\ \BBA {} Thomas%
}{%
Reiss%
\ \BBA {} Thomas%
}{%
{\protect \APACyear {2007}}%
}]{%
reiss2007statistical}
\APACinsertmetastar {%
reiss2007statistical}%
\begin{APACrefauthors}%
Reiss, R\BHBI D.%
\BCBT {}\ \BBA {} Thomas, M.%
\end{APACrefauthors}%
\unskip\
\newblock
\APACrefYear{2007}.
\newblock
\APACrefbtitle {Statistical Analysis of Extreme Values: With Applications to
  Insurance, Finance, Hydrology and Other Fields} {Statistical analysis of
  extreme values: With applications to insurance, finance, hydrology and other
  fields}.
\newblock
\APACaddressPublisher{}{Springer Science \& Business Media}.
\PrintBackRefs{\CurrentBib}

\bibitem [\protect \citeauthoryear {%
Ribeiro%
\ \protect \BOthers {.}}{%
Ribeiro%
\ \protect \BOthers {.}}{%
{\protect \APACyear {2010}}%
}]{%
ribeiro2010spike}
\APACinsertmetastar {%
ribeiro2010spike}%
\begin{APACrefauthors}%
Ribeiro, T\BPBI L.%
, Copelli, M.%
, Caixeta, F.%
, Belchior, H.%
, Chialvo, D\BPBI R.%
, Nicolelis, M\BPBI A.%
\BCBL {}\ \BBA {} Ribeiro, S.%
\end{APACrefauthors}%
\unskip\
\newblock
\APACrefYearMonthDay{2010}{}{}.
\newblock
{\BBOQ}\APACrefatitle {Spike avalanches exhibit universal dynamics across the
  sleep-wake cycle} {Spike avalanches exhibit universal dynamics across the
  sleep-wake cycle}.{\BBCQ}
\newblock
\APACjournalVolNumPages{PloS one}{5}{11}{e14129}.
\PrintBackRefs{\CurrentBib}

\bibitem [\protect \citeauthoryear {%
Robert%
\ \BBA {} Vignoud%
}{%
Robert%
\ \BBA {} Vignoud%
}{%
{\protect \APACyear {2021}}%
}]{%
robert2021stochastic}
\APACinsertmetastar {%
robert2021stochastic}%
\begin{APACrefauthors}%
Robert, P.%
\BCBT {}\ \BBA {} Vignoud, G.%
\end{APACrefauthors}%
\unskip\
\newblock
\APACrefYearMonthDay{2021}{}{}.
\newblock
{\BBOQ}\APACrefatitle {Stochastic Models of Neural Synaptic Plasticity: A
  Scaling Approach} {Stochastic models of neural synaptic plasticity: A scaling
  approach}.{\BBCQ}
\newblock
\APACjournalVolNumPages{SIAM Journal on Applied
  Mathematics}{81}{6}{2362--2386}.
\PrintBackRefs{\CurrentBib}

\bibitem [\protect \citeauthoryear {%
Robinson%
}{%
Robinson%
}{%
{\protect \APACyear {2021}}%
}]{%
robinson2021neural}
\APACinsertmetastar {%
robinson2021neural}%
\begin{APACrefauthors}%
Robinson, P.%
\end{APACrefauthors}%
\unskip\
\newblock
\APACrefYearMonthDay{2021}{}{}.
\newblock
{\BBOQ}\APACrefatitle {Neural field theory of neural avalanche exponents}
  {Neural field theory of neural avalanche exponents}.{\BBCQ}
\newblock
\APACjournalVolNumPages{Biological cybernetics}{115}{3}{237--243}.
\PrintBackRefs{\CurrentBib}

\bibitem [\protect \citeauthoryear {%
Rubinov%
, Sporns%
, Thivierge%
\BCBL {}\ \BBA {} Breakspear%
}{%
Rubinov%
\ \protect \BOthers {.}}{%
{\protect \APACyear {2011}}%
}]{%
rubinov2011neurobiologically}
\APACinsertmetastar {%
rubinov2011neurobiologically}%
\begin{APACrefauthors}%
Rubinov, M.%
, Sporns, O.%
, Thivierge, J\BHBI P.%
\BCBL {}\ \BBA {} Breakspear, M.%
\end{APACrefauthors}%
\unskip\
\newblock
\APACrefYearMonthDay{2011}{}{}.
\newblock
{\BBOQ}\APACrefatitle {Neurobiologically realistic determinants of
  self-organized criticality in networks of spiking neurons} {Neurobiologically
  realistic determinants of self-organized criticality in networks of spiking
  neurons}.{\BBCQ}
\newblock
\APACjournalVolNumPages{PLoS computational biology}{7}{6}{e1002038}.
\PrintBackRefs{\CurrentBib}

\bibitem [\protect \citeauthoryear {%
Sartori%
, Granger%
, Lee%
\BCBL {}\ \BBA {} Horowitz%
}{%
Sartori%
\ \protect \BOthers {.}}{%
{\protect \APACyear {2014}}%
}]{%
sartori2014thermodynamic}
\APACinsertmetastar {%
sartori2014thermodynamic}%
\begin{APACrefauthors}%
Sartori, P.%
, Granger, L.%
, Lee, C\BPBI F.%
\BCBL {}\ \BBA {} Horowitz, J\BPBI M.%
\end{APACrefauthors}%
\unskip\
\newblock
\APACrefYearMonthDay{2014}{}{}.
\newblock
{\BBOQ}\APACrefatitle {Thermodynamic costs of information processing in sensory
  adaptation} {Thermodynamic costs of information processing in sensory
  adaptation}.{\BBCQ}
\newblock
\APACjournalVolNumPages{PLoS computational biology}{10}{12}{e1003974}.
\PrintBackRefs{\CurrentBib}

\bibitem [\protect \citeauthoryear {%
Schaworonkow%
, Blythe%
, Kegeles%
, Curio%
\BCBL {}\ \BBA {} Nikulin%
}{%
Schaworonkow%
\ \protect \BOthers {.}}{%
{\protect \APACyear {2015}}%
}]{%
schaworonkow2015power}
\APACinsertmetastar {%
schaworonkow2015power}%
\begin{APACrefauthors}%
Schaworonkow, N.%
, Blythe, D\BPBI A.%
, Kegeles, J.%
, Curio, G.%
\BCBL {}\ \BBA {} Nikulin, V\BPBI V.%
\end{APACrefauthors}%
\unskip\
\newblock
\APACrefYearMonthDay{2015}{}{}.
\newblock
{\BBOQ}\APACrefatitle {Power-law dynamics in neuronal and behavioral data
  introduce spurious correlations} {Power-law dynamics in neuronal and
  behavioral data introduce spurious correlations}.{\BBCQ}
\newblock
\APACjournalVolNumPages{Human brain mapping}{36}{8}{2901--2914}.
\PrintBackRefs{\CurrentBib}

\bibitem [\protect \citeauthoryear {%
Schneidman%
, Berry%
, Segev%
\BCBL {}\ \BBA {} Bialek%
}{%
Schneidman%
\ \protect \BOthers {.}}{%
{\protect \APACyear {2006}}%
}]{%
schneidman2006weak}
\APACinsertmetastar {%
schneidman2006weak}%
\begin{APACrefauthors}%
Schneidman, E.%
, Berry, M\BPBI J.%
, Segev, R.%
\BCBL {}\ \BBA {} Bialek, W.%
\end{APACrefauthors}%
\unskip\
\newblock
\APACrefYearMonthDay{2006}{}{}.
\newblock
{\BBOQ}\APACrefatitle {Weak pairwise correlations imply strongly correlated
  network states in a neural population} {Weak pairwise correlations imply
  strongly correlated network states in a neural population}.{\BBCQ}
\newblock
\APACjournalVolNumPages{Nature}{440}{7087}{1007--1012}.
\PrintBackRefs{\CurrentBib}

\bibitem [\protect \citeauthoryear {%
Schwarz%
}{%
Schwarz%
}{%
{\protect \APACyear {1978}}%
}]{%
schwarz1978estimating}
\APACinsertmetastar {%
schwarz1978estimating}%
\begin{APACrefauthors}%
Schwarz, G.%
\end{APACrefauthors}%
\unskip\
\newblock
\APACrefYearMonthDay{1978}{}{}.
\newblock
{\BBOQ}\APACrefatitle {Estimating the dimension of a model} {Estimating the
  dimension of a model}.{\BBCQ}
\newblock
\APACjournalVolNumPages{The annals of statistics}{}{}{461--464}.
\PrintBackRefs{\CurrentBib}

\bibitem [\protect \citeauthoryear {%
Scott%
\ \BBA {} Alwyn%
}{%
Scott%
\ \BBA {} Alwyn%
}{%
{\protect \APACyear {1977}}%
}]{%
scott1977neurophysics}
\APACinsertmetastar {%
scott1977neurophysics}%
\begin{APACrefauthors}%
Scott%
\BCBT {}\ \BBA {} Alwyn.%
\end{APACrefauthors}%
\unskip\
\newblock
\APACrefYear{1977}.
\newblock
\APACrefbtitle {Neurophysics} {Neurophysics}.
\newblock
\APACaddressPublisher{}{John Wiley \& Sons Incorporated}.
\PrintBackRefs{\CurrentBib}

\bibitem [\protect \citeauthoryear {%
G.~Scott%
\ \protect \BOthers {.}}{%
G.~Scott%
\ \protect \BOthers {.}}{%
{\protect \APACyear {2014}}%
}]{%
scott2014voltage}
\APACinsertmetastar {%
scott2014voltage}%
\begin{APACrefauthors}%
Scott, G.%
, Fagerholm, E\BPBI D.%
, Mutoh, H.%
, Leech, R.%
, Sharp, D\BPBI J.%
, Shew, W\BPBI L.%
\BCBL {}\ \BBA {} Kn{\"o}pfel, T.%
\end{APACrefauthors}%
\unskip\
\newblock
\APACrefYearMonthDay{2014}{}{}.
\newblock
{\BBOQ}\APACrefatitle {Voltage imaging of waking mouse cortex reveals emergence
  of critical neuronal dynamics} {Voltage imaging of waking mouse cortex
  reveals emergence of critical neuronal dynamics}.{\BBCQ}
\newblock
\APACjournalVolNumPages{Journal of Neuroscience}{34}{50}{16611--16620}.
\PrintBackRefs{\CurrentBib}

\bibitem [\protect \citeauthoryear {%
Senzai%
, Fernandez-Ruiz%
\BCBL {}\ \BBA {} Buzs{\'a}ki%
}{%
Senzai%
\ \protect \BOthers {.}}{%
{\protect \APACyear {2019}}%
}]{%
senzai2019layer}
\APACinsertmetastar {%
senzai2019layer}%
\begin{APACrefauthors}%
Senzai, Y.%
, Fernandez-Ruiz, A.%
\BCBL {}\ \BBA {} Buzs{\'a}ki, G.%
\end{APACrefauthors}%
\unskip\
\newblock
\APACrefYearMonthDay{2019}{}{}.
\newblock
{\BBOQ}\APACrefatitle {Layer-specific physiological features and interlaminar
  interactions in the primary visual cortex of the mouse} {Layer-specific
  physiological features and interlaminar interactions in the primary visual
  cortex of the mouse}.{\BBCQ}
\newblock
\APACjournalVolNumPages{Neuron}{101}{3}{500--513}.
\PrintBackRefs{\CurrentBib}

\bibitem [\protect \citeauthoryear {%
Sethna%
, Dahmen%
\BCBL {}\ \BBA {} Myers%
}{%
Sethna%
\ \protect \BOthers {.}}{%
{\protect \APACyear {2001}}%
}]{%
sethna2001crackling}
\APACinsertmetastar {%
sethna2001crackling}%
\begin{APACrefauthors}%
Sethna, J\BPBI P.%
, Dahmen, K\BPBI A.%
\BCBL {}\ \BBA {} Myers, C\BPBI R.%
\end{APACrefauthors}%
\unskip\
\newblock
\APACrefYearMonthDay{2001}{}{}.
\newblock
{\BBOQ}\APACrefatitle {Crackling noise} {Crackling noise}.{\BBCQ}
\newblock
\APACjournalVolNumPages{Nature}{410}{6825}{242--250}.
\PrintBackRefs{\CurrentBib}

\bibitem [\protect \citeauthoryear {%
Shaukat%
\ \BBA {} Thivierge%
}{%
Shaukat%
\ \BBA {} Thivierge%
}{%
{\protect \APACyear {2016}}%
}]{%
shaukat2016statistical}
\APACinsertmetastar {%
shaukat2016statistical}%
\begin{APACrefauthors}%
Shaukat, A.%
\BCBT {}\ \BBA {} Thivierge, J\BHBI P.%
\end{APACrefauthors}%
\unskip\
\newblock
\APACrefYearMonthDay{2016}{}{}.
\newblock
{\BBOQ}\APACrefatitle {Statistical evaluation of waveform collapse reveals
  scale-free properties of neuronal avalanches} {Statistical evaluation of
  waveform collapse reveals scale-free properties of neuronal
  avalanches}.{\BBCQ}
\newblock
\APACjournalVolNumPages{Frontiers in computational neuroscience}{10}{}{29}.
\PrintBackRefs{\CurrentBib}

\bibitem [\protect \citeauthoryear {%
Shew%
\ \protect \BOthers {.}}{%
Shew%
\ \protect \BOthers {.}}{%
{\protect \APACyear {2015}}%
}]{%
shew2015adaptation}
\APACinsertmetastar {%
shew2015adaptation}%
\begin{APACrefauthors}%
Shew, W\BPBI L.%
, Clawson, W\BPBI P.%
, Pobst, J.%
, Karimipanah, Y.%
, Wright, N\BPBI C.%
\BCBL {}\ \BBA {} Wessel, R.%
\end{APACrefauthors}%
\unskip\
\newblock
\APACrefYearMonthDay{2015}{}{}.
\newblock
{\BBOQ}\APACrefatitle {Adaptation to sensory input tunes visual cortex to
  criticality} {Adaptation to sensory input tunes visual cortex to
  criticality}.{\BBCQ}
\newblock
\APACjournalVolNumPages{Nature Physics}{11}{8}{659--663}.
\PrintBackRefs{\CurrentBib}

\bibitem [\protect \citeauthoryear {%
Shew%
\ \BBA {} Plenz%
}{%
Shew%
\ \BBA {} Plenz%
}{%
{\protect \APACyear {2013}}%
}]{%
shew2013functional}
\APACinsertmetastar {%
shew2013functional}%
\begin{APACrefauthors}%
Shew, W\BPBI L.%
\BCBT {}\ \BBA {} Plenz, D.%
\end{APACrefauthors}%
\unskip\
\newblock
\APACrefYearMonthDay{2013}{}{}.
\newblock
{\BBOQ}\APACrefatitle {The functional benefits of criticality in the cortex}
  {The functional benefits of criticality in the cortex}.{\BBCQ}
\newblock
\APACjournalVolNumPages{The neuroscientist}{19}{1}{88--100}.
\PrintBackRefs{\CurrentBib}

\bibitem [\protect \citeauthoryear {%
Shew%
, Yang%
, Petermann%
, Roy%
\BCBL {}\ \BBA {} Plenz%
}{%
Shew%
\ \protect \BOthers {.}}{%
{\protect \APACyear {2009}}%
}]{%
shew2009neuronal}
\APACinsertmetastar {%
shew2009neuronal}%
\begin{APACrefauthors}%
Shew, W\BPBI L.%
, Yang, H.%
, Petermann, T.%
, Roy, R.%
\BCBL {}\ \BBA {} Plenz, D.%
\end{APACrefauthors}%
\unskip\
\newblock
\APACrefYearMonthDay{2009}{}{}.
\newblock
{\BBOQ}\APACrefatitle {Neuronal avalanches imply maximum dynamic range in
  cortical networks at criticality} {Neuronal avalanches imply maximum dynamic
  range in cortical networks at criticality}.{\BBCQ}
\newblock
\APACjournalVolNumPages{Journal of neuroscience}{29}{49}{15595--15600}.
\PrintBackRefs{\CurrentBib}

\bibitem [\protect \citeauthoryear {%
Shew%
, Yang%
, Yu%
, Roy%
\BCBL {}\ \BBA {} Plenz%
}{%
Shew%
\ \protect \BOthers {.}}{%
{\protect \APACyear {2011}}%
}]{%
shew2011information}
\APACinsertmetastar {%
shew2011information}%
\begin{APACrefauthors}%
Shew, W\BPBI L.%
, Yang, H.%
, Yu, S.%
, Roy, R.%
\BCBL {}\ \BBA {} Plenz, D.%
\end{APACrefauthors}%
\unskip\
\newblock
\APACrefYearMonthDay{2011}{}{}.
\newblock
{\BBOQ}\APACrefatitle {Information capacity and transmission are maximized in
  balanced cortical networks with neuronal avalanches} {Information capacity
  and transmission are maximized in balanced cortical networks with neuronal
  avalanches}.{\BBCQ}
\newblock
\APACjournalVolNumPages{Journal of neuroscience}{31}{1}{55--63}.
\PrintBackRefs{\CurrentBib}

\bibitem [\protect \citeauthoryear {%
Shin%
\ \BBA {} Kim%
}{%
Shin%
\ \BBA {} Kim%
}{%
{\protect \APACyear {2006}}%
}]{%
shin2006self}
\APACinsertmetastar {%
shin2006self}%
\begin{APACrefauthors}%
Shin, C\BHBI W.%
\BCBT {}\ \BBA {} Kim, S.%
\end{APACrefauthors}%
\unskip\
\newblock
\APACrefYearMonthDay{2006}{}{}.
\newblock
{\BBOQ}\APACrefatitle {Self-organized criticality and scale-free properties in
  emergent functional neural networks} {Self-organized criticality and
  scale-free properties in emergent functional neural networks}.{\BBCQ}
\newblock
\APACjournalVolNumPages{Physical Review E}{74}{4}{045101}.
\PrintBackRefs{\CurrentBib}

\bibitem [\protect \citeauthoryear {%
Shriki%
\ \protect \BOthers {.}}{%
Shriki%
\ \protect \BOthers {.}}{%
{\protect \APACyear {2013}}%
}]{%
shriki2013neuronal}
\APACinsertmetastar {%
shriki2013neuronal}%
\begin{APACrefauthors}%
Shriki, O.%
, Alstott, J.%
, Carver, F.%
, Holroyd, T.%
, Henson, R\BPBI N.%
, Smith, M\BPBI L.%
\BDBL {}Plenz, D.%
\end{APACrefauthors}%
\unskip\
\newblock
\APACrefYearMonthDay{2013}{}{}.
\newblock
{\BBOQ}\APACrefatitle {Neuronal avalanches in the resting MEG of the human
  brain} {Neuronal avalanches in the resting meg of the human brain}.{\BBCQ}
\newblock
\APACjournalVolNumPages{Journal of Neuroscience}{33}{16}{7079--7090}.
\PrintBackRefs{\CurrentBib}

\bibitem [\protect \citeauthoryear {%
Smit%
\ \protect \BOthers {.}}{%
Smit%
\ \protect \BOthers {.}}{%
{\protect \APACyear {2011}}%
}]{%
smit2011scale}
\APACinsertmetastar {%
smit2011scale}%
\begin{APACrefauthors}%
Smit, D\BPBI J.%
, de Geus, E\BPBI J.%
, van~de Nieuwenhuijzen, M\BPBI E.%
, van Beijsterveldt, C\BPBI E.%
, van Baal, G\BPBI C\BPBI M.%
, Mansvelder, H\BPBI D.%
\BDBL {}Linkenkaer-Hansen, K.%
\end{APACrefauthors}%
\unskip\
\newblock
\APACrefYearMonthDay{2011}{}{}.
\newblock
{\BBOQ}\APACrefatitle {Scale-free modulation of resting-state neuronal
  oscillations reflects prolonged brain maturation in humans} {Scale-free
  modulation of resting-state neuronal oscillations reflects prolonged brain
  maturation in humans}.{\BBCQ}
\newblock
\APACjournalVolNumPages{Journal of Neuroscience}{31}{37}{13128--13136}.
\PrintBackRefs{\CurrentBib}

\bibitem [\protect \citeauthoryear {%
Softky%
\ \BBA {} Koch%
}{%
Softky%
\ \BBA {} Koch%
}{%
{\protect \APACyear {1993}}%
}]{%
softky1993highly}
\APACinsertmetastar {%
softky1993highly}%
\begin{APACrefauthors}%
Softky, W\BPBI R.%
\BCBT {}\ \BBA {} Koch, C.%
\end{APACrefauthors}%
\unskip\
\newblock
\APACrefYearMonthDay{1993}{}{}.
\newblock
{\BBOQ}\APACrefatitle {The highly irregular firing of cortical cells is
  inconsistent with temporal integration of random EPSPs} {The highly irregular
  firing of cortical cells is inconsistent with temporal integration of random
  epsps}.{\BBCQ}
\newblock
\APACjournalVolNumPages{Journal of neuroscience}{13}{1}{334--350}.
\PrintBackRefs{\CurrentBib}

\bibitem [\protect \citeauthoryear {%
Song%
, Miller%
\BCBL {}\ \BBA {} Abbott%
}{%
Song%
\ \protect \BOthers {.}}{%
{\protect \APACyear {2000}}%
}]{%
song2000competitive}
\APACinsertmetastar {%
song2000competitive}%
\begin{APACrefauthors}%
Song, S.%
, Miller, K\BPBI D.%
\BCBL {}\ \BBA {} Abbott, L\BPBI F.%
\end{APACrefauthors}%
\unskip\
\newblock
\APACrefYearMonthDay{2000}{}{}.
\newblock
{\BBOQ}\APACrefatitle {Competitive Hebbian learning through
  spike-timing-dependent synaptic plasticity} {Competitive hebbian learning
  through spike-timing-dependent synaptic plasticity}.{\BBCQ}
\newblock
\APACjournalVolNumPages{Nature neuroscience}{3}{9}{919--926}.
\PrintBackRefs{\CurrentBib}

\bibitem [\protect \citeauthoryear {%
Sornette%
, Johansen%
\BCBL {}\ \BBA {} Dornic%
}{%
Sornette%
\ \protect \BOthers {.}}{%
{\protect \APACyear {1995}}%
}]{%
sornette1995mapping}
\APACinsertmetastar {%
sornette1995mapping}%
\begin{APACrefauthors}%
Sornette, D.%
, Johansen, A.%
\BCBL {}\ \BBA {} Dornic, I.%
\end{APACrefauthors}%
\unskip\
\newblock
\APACrefYearMonthDay{1995}{}{}.
\newblock
{\BBOQ}\APACrefatitle {Mapping self-organized criticality onto criticality}
  {Mapping self-organized criticality onto criticality}.{\BBCQ}
\newblock
\APACjournalVolNumPages{Journal de Physique I}{5}{3}{325--335}.
\PrintBackRefs{\CurrentBib}

\bibitem [\protect \citeauthoryear {%
Squire%
\ \protect \BOthers {.}}{%
Squire%
\ \protect \BOthers {.}}{%
{\protect \APACyear {2012}}%
}]{%
squire2012fundamental}
\APACinsertmetastar {%
squire2012fundamental}%
\begin{APACrefauthors}%
Squire, L.%
, Berg, D.%
, Bloom, F\BPBI E.%
, Du~Lac, S.%
, Ghosh, A.%
\BCBL {}\ \BBA {} Spitzer, N\BPBI C.%
\end{APACrefauthors}%
\unskip\
\newblock
\APACrefYear{2012}.
\newblock
\APACrefbtitle {Fundamental neuroscience} {Fundamental neuroscience}.
\newblock
\APACaddressPublisher{}{Academic press}.
\PrintBackRefs{\CurrentBib}

\bibitem [\protect \citeauthoryear {%
Steif%
}{%
Steif%
}{%
{\protect \APACyear {2009}}%
}]{%
steif2009survey}
\APACinsertmetastar {%
steif2009survey}%
\begin{APACrefauthors}%
Steif, J\BPBI E.%
\end{APACrefauthors}%
\unskip\
\newblock
\APACrefYearMonthDay{2009}{}{}.
\newblock
{\BBOQ}\APACrefatitle {A survey of dynamical percolation} {A survey of
  dynamical percolation}.{\BBCQ}
\newblock
\BIn{} \APACrefbtitle {Fractal geometry and stochastics IV} {Fractal geometry
  and stochastics iv}\ (\BPGS\ 145--174).
\newblock
\APACaddressPublisher{}{Springer}.
\PrintBackRefs{\CurrentBib}

\bibitem [\protect \citeauthoryear {%
Stein%
, Gossen%
\BCBL {}\ \BBA {} Jones%
}{%
Stein%
\ \protect \BOthers {.}}{%
{\protect \APACyear {2005}}%
}]{%
stein2005neuronal}
\APACinsertmetastar {%
stein2005neuronal}%
\begin{APACrefauthors}%
Stein, R\BPBI B.%
, Gossen, E\BPBI R.%
\BCBL {}\ \BBA {} Jones, K\BPBI E.%
\end{APACrefauthors}%
\unskip\
\newblock
\APACrefYearMonthDay{2005}{}{}.
\newblock
{\BBOQ}\APACrefatitle {Neuronal variability: noise or part of the signal?}
  {Neuronal variability: noise or part of the signal?}{\BBCQ}
\newblock
\APACjournalVolNumPages{Nature Reviews Neuroscience}{6}{5}{389--397}.
\PrintBackRefs{\CurrentBib}

\bibitem [\protect \citeauthoryear {%
Stepp%
, Plenz%
\BCBL {}\ \BBA {} Srinivasa%
}{%
Stepp%
\ \protect \BOthers {.}}{%
{\protect \APACyear {2015}}%
}]{%
stepp2015synaptic}
\APACinsertmetastar {%
stepp2015synaptic}%
\begin{APACrefauthors}%
Stepp, N.%
, Plenz, D.%
\BCBL {}\ \BBA {} Srinivasa, N.%
\end{APACrefauthors}%
\unskip\
\newblock
\APACrefYearMonthDay{2015}{}{}.
\newblock
{\BBOQ}\APACrefatitle {Synaptic plasticity enables adaptive self-tuning
  critical networks} {Synaptic plasticity enables adaptive self-tuning critical
  networks}.{\BBCQ}
\newblock
\APACjournalVolNumPages{PLoS computational biology}{11}{1}{e1004043}.
\PrintBackRefs{\CurrentBib}

\bibitem [\protect \citeauthoryear {%
Stewart%
\ \BBA {} Plenz%
}{%
Stewart%
\ \BBA {} Plenz%
}{%
{\protect \APACyear {2006}}%
}]{%
stewart2006inverted}
\APACinsertmetastar {%
stewart2006inverted}%
\begin{APACrefauthors}%
Stewart, C\BPBI V.%
\BCBT {}\ \BBA {} Plenz, D.%
\end{APACrefauthors}%
\unskip\
\newblock
\APACrefYearMonthDay{2006}{}{}.
\newblock
{\BBOQ}\APACrefatitle {Inverted-U profile of dopamine--NMDA-mediated
  spontaneous avalanche recurrence in superficial layers of rat prefrontal
  cortex} {Inverted-u profile of dopamine--nmda-mediated spontaneous avalanche
  recurrence in superficial layers of rat prefrontal cortex}.{\BBCQ}
\newblock
\APACjournalVolNumPages{Journal of neuroscience}{26}{31}{8148--8159}.
\PrintBackRefs{\CurrentBib}

\bibitem [\protect \citeauthoryear {%
Stewart%
\ \BBA {} Plenz%
}{%
Stewart%
\ \BBA {} Plenz%
}{%
{\protect \APACyear {2008}}%
}]{%
stewart2008homeostasis}
\APACinsertmetastar {%
stewart2008homeostasis}%
\begin{APACrefauthors}%
Stewart, C\BPBI V.%
\BCBT {}\ \BBA {} Plenz, D.%
\end{APACrefauthors}%
\unskip\
\newblock
\APACrefYearMonthDay{2008}{}{}.
\newblock
{\BBOQ}\APACrefatitle {Homeostasis of neuronal avalanches during postnatal
  cortex development in vitro} {Homeostasis of neuronal avalanches during
  postnatal cortex development in vitro}.{\BBCQ}
\newblock
\APACjournalVolNumPages{Journal of neuroscience methods}{169}{2}{405--416}.
\PrintBackRefs{\CurrentBib}

\bibitem [\protect \citeauthoryear {%
Tagliazucchi%
, Balenzuela%
, Fraiman%
\BCBL {}\ \BBA {} Chialvo%
}{%
Tagliazucchi%
\ \protect \BOthers {.}}{%
{\protect \APACyear {2012}}%
}]{%
tagliazucchi2012criticality}
\APACinsertmetastar {%
tagliazucchi2012criticality}%
\begin{APACrefauthors}%
Tagliazucchi, E.%
, Balenzuela, P.%
, Fraiman, D.%
\BCBL {}\ \BBA {} Chialvo, D\BPBI R.%
\end{APACrefauthors}%
\unskip\
\newblock
\APACrefYearMonthDay{2012}{}{}.
\newblock
{\BBOQ}\APACrefatitle {Criticality in large-scale brain fMRI dynamics unveiled
  by a novel point process analysis} {Criticality in large-scale brain fmri
  dynamics unveiled by a novel point process analysis}.{\BBCQ}
\newblock
\APACjournalVolNumPages{Frontiers in physiology}{3}{}{15}.
\PrintBackRefs{\CurrentBib}

\bibitem [\protect \citeauthoryear {%
Tian%
, Li%
\BCBL {}\ \BBA {} Sun%
}{%
Tian%
\ \protect \BOthers {.}}{%
{\protect \APACyear {2021}}%
}]{%
tian2021bridging}
\APACinsertmetastar {%
tian2021bridging}%
\begin{APACrefauthors}%
Tian, Y.%
, Li, G.%
\BCBL {}\ \BBA {} Sun, P.%
\end{APACrefauthors}%
\unskip\
\newblock
\APACrefYearMonthDay{2021}{}{}.
\newblock
{\BBOQ}\APACrefatitle {Bridging the information and dynamics attributes of
  neural activities} {Bridging the information and dynamics attributes of
  neural activities}.{\BBCQ}
\newblock
\APACjournalVolNumPages{Physical Review Research}{3}{4}{043085}.
\PrintBackRefs{\CurrentBib}

\bibitem [\protect \citeauthoryear {%
Tian%
\ \BBA {} Sun%
}{%
Tian%
\ \BBA {} Sun%
}{%
{\protect \APACyear {2021}}%
}]{%
tian2021characteristics}
\APACinsertmetastar {%
tian2021characteristics}%
\begin{APACrefauthors}%
Tian, Y.%
\BCBT {}\ \BBA {} Sun, P.%
\end{APACrefauthors}%
\unskip\
\newblock
\APACrefYearMonthDay{2021}{}{}.
\newblock
{\BBOQ}\APACrefatitle {Characteristics of the neural coding of causality}
  {Characteristics of the neural coding of causality}.{\BBCQ}
\newblock
\APACjournalVolNumPages{Physical Review E}{103}{1}{012406}.
\PrintBackRefs{\CurrentBib}

\bibitem [\protect \citeauthoryear {%
Tian%
\ \BBA {} Sun%
}{%
Tian%
\ \BBA {} Sun%
}{%
{\protect \APACyear {2022}}%
}]{%
tian2022information}
\APACinsertmetastar {%
tian2022information}%
\begin{APACrefauthors}%
Tian, Y.%
\BCBT {}\ \BBA {} Sun, P.%
\end{APACrefauthors}%
\unskip\
\newblock
\APACrefYearMonthDay{2022}{}{}.
\newblock
{\BBOQ}\APACrefatitle {Information thermodynamics of encoding and encoders}
  {Information thermodynamics of encoding and encoders}.{\BBCQ}
\newblock
\APACjournalVolNumPages{Chaos: An Interdisciplinary Journal of Nonlinear
  Science}{32}{6}{063109}.
\PrintBackRefs{\CurrentBib}

\bibitem [\protect \citeauthoryear {%
Tka{\v{c}}ik%
\ \protect \BOthers {.}}{%
Tka{\v{c}}ik%
\ \protect \BOthers {.}}{%
{\protect \APACyear {2015}}%
}]{%
tkavcik2015thermodynamics}
\APACinsertmetastar {%
tkavcik2015thermodynamics}%
\begin{APACrefauthors}%
Tka{\v{c}}ik, G.%
, Mora, T.%
, Marre, O.%
, Amodei, D.%
, Palmer, S\BPBI E.%
, Berry, M\BPBI J.%
\BCBL {}\ \BBA {} Bialek, W.%
\end{APACrefauthors}%
\unskip\
\newblock
\APACrefYearMonthDay{2015}{}{}.
\newblock
{\BBOQ}\APACrefatitle {Thermodynamics and signatures of criticality in a
  network of neurons} {Thermodynamics and signatures of criticality in a
  network of neurons}.{\BBCQ}
\newblock
\APACjournalVolNumPages{Proceedings of the National Academy of
  Sciences}{112}{37}{11508--11513}.
\PrintBackRefs{\CurrentBib}

\bibitem [\protect \citeauthoryear {%
Touboul%
\ \BBA {} Destexhe%
}{%
Touboul%
\ \BBA {} Destexhe%
}{%
{\protect \APACyear {2010}}%
}]{%
touboul2010can}
\APACinsertmetastar {%
touboul2010can}%
\begin{APACrefauthors}%
Touboul, J.%
\BCBT {}\ \BBA {} Destexhe, A.%
\end{APACrefauthors}%
\unskip\
\newblock
\APACrefYearMonthDay{2010}{}{}.
\newblock
{\BBOQ}\APACrefatitle {Can power-law scaling and neuronal avalanches arise from
  stochastic dynamics?} {Can power-law scaling and neuronal avalanches arise
  from stochastic dynamics?}{\BBCQ}
\newblock
\APACjournalVolNumPages{PloS one}{5}{2}{e8982}.
\PrintBackRefs{\CurrentBib}

\bibitem [\protect \citeauthoryear {%
Touboul%
\ \BBA {} Destexhe%
}{%
Touboul%
\ \BBA {} Destexhe%
}{%
{\protect \APACyear {2017}}%
}]{%
touboul2017power}
\APACinsertmetastar {%
touboul2017power}%
\begin{APACrefauthors}%
Touboul, J.%
\BCBT {}\ \BBA {} Destexhe, A.%
\end{APACrefauthors}%
\unskip\
\newblock
\APACrefYearMonthDay{2017}{}{}.
\newblock
{\BBOQ}\APACrefatitle {Power-law statistics and universal scaling in the
  absence of criticality} {Power-law statistics and universal scaling in the
  absence of criticality}.{\BBCQ}
\newblock
\APACjournalVolNumPages{Physical Review E}{95}{1}{012413}.
\PrintBackRefs{\CurrentBib}

\bibitem [\protect \citeauthoryear {%
Touboul%
, Wendling%
, Chauvel%
\BCBL {}\ \BBA {} Faugeras%
}{%
Touboul%
\ \protect \BOthers {.}}{%
{\protect \APACyear {2011}}%
}]{%
touboul2011neural}
\APACinsertmetastar {%
touboul2011neural}%
\begin{APACrefauthors}%
Touboul, J.%
, Wendling, F.%
, Chauvel, P.%
\BCBL {}\ \BBA {} Faugeras, O.%
\end{APACrefauthors}%
\unskip\
\newblock
\APACrefYearMonthDay{2011}{}{}.
\newblock
{\BBOQ}\APACrefatitle {Neural mass activity, bifurcations, and epilepsy}
  {Neural mass activity, bifurcations, and epilepsy}.{\BBCQ}
\newblock
\APACjournalVolNumPages{Neural computation}{23}{12}{3232--3286}.
\PrintBackRefs{\CurrentBib}

\bibitem [\protect \citeauthoryear {%
Tyulmankov%
, Fang%
, Vadaparty%
\BCBL {}\ \BBA {} Yang%
}{%
Tyulmankov%
\ \protect \BOthers {.}}{%
{\protect \APACyear {2021}}%
}]{%
tyulmankov2021biological}
\APACinsertmetastar {%
tyulmankov2021biological}%
\begin{APACrefauthors}%
Tyulmankov, D.%
, Fang, C.%
, Vadaparty, A.%
\BCBL {}\ \BBA {} Yang, G\BPBI R.%
\end{APACrefauthors}%
\unskip\
\newblock
\APACrefYearMonthDay{2021}{}{}.
\newblock
{\BBOQ}\APACrefatitle {Biological key-value memory networks} {Biological
  key-value memory networks}.{\BBCQ}
\newblock
\APACjournalVolNumPages{Advances in Neural Information Processing
  Systems}{34}{}{}.
\PrintBackRefs{\CurrentBib}

\bibitem [\protect \citeauthoryear {%
Van~Hateren%
\ \BBA {} van~der Schaaf%
}{%
Van~Hateren%
\ \BBA {} van~der Schaaf%
}{%
{\protect \APACyear {1998}}%
}]{%
van1998independent}
\APACinsertmetastar {%
van1998independent}%
\begin{APACrefauthors}%
Van~Hateren, J\BPBI H.%
\BCBT {}\ \BBA {} van~der Schaaf, A.%
\end{APACrefauthors}%
\unskip\
\newblock
\APACrefYearMonthDay{1998}{}{}.
\newblock
{\BBOQ}\APACrefatitle {Independent component filters of natural images compared
  with simple cells in primary visual cortex} {Independent component filters of
  natural images compared with simple cells in primary visual cortex}.{\BBCQ}
\newblock
\APACjournalVolNumPages{Proceedings of the Royal Society of London. Series B:
  Biological Sciences}{265}{1394}{359--366}.
\PrintBackRefs{\CurrentBib}

\bibitem [\protect \citeauthoryear {%
Van~Vreeswijk%
\ \BBA {} Sompolinsky%
}{%
Van~Vreeswijk%
\ \BBA {} Sompolinsky%
}{%
{\protect \APACyear {1996}}%
}]{%
van1996chaos}
\APACinsertmetastar {%
van1996chaos}%
\begin{APACrefauthors}%
Van~Vreeswijk, C.%
\BCBT {}\ \BBA {} Sompolinsky, H.%
\end{APACrefauthors}%
\unskip\
\newblock
\APACrefYearMonthDay{1996}{}{}.
\newblock
{\BBOQ}\APACrefatitle {Chaos in neuronal networks with balanced excitatory and
  inhibitory activity} {Chaos in neuronal networks with balanced excitatory and
  inhibitory activity}.{\BBCQ}
\newblock
\APACjournalVolNumPages{Science}{274}{5293}{1724--1726}.
\PrintBackRefs{\CurrentBib}

\bibitem [\protect \citeauthoryear {%
Varley%
, Sporns%
, Puce%
\BCBL {}\ \BBA {} Beggs%
}{%
Varley%
\ \protect \BOthers {.}}{%
{\protect \APACyear {2020}}%
}]{%
varley2020differential}
\APACinsertmetastar {%
varley2020differential}%
\begin{APACrefauthors}%
Varley, T\BPBI F.%
, Sporns, O.%
, Puce, A.%
\BCBL {}\ \BBA {} Beggs, J.%
\end{APACrefauthors}%
\unskip\
\newblock
\APACrefYearMonthDay{2020}{}{}.
\newblock
{\BBOQ}\APACrefatitle {Differential effects of propofol and ketamine on
  critical brain dynamics} {Differential effects of propofol and ketamine on
  critical brain dynamics}.{\BBCQ}
\newblock
\APACjournalVolNumPages{PLoS computational biology}{16}{12}{e1008418}.
\PrintBackRefs{\CurrentBib}

\bibitem [\protect \citeauthoryear {%
Villegas%
, di Santo%
, Burioni%
\BCBL {}\ \BBA {} Mu{\~n}oz%
}{%
Villegas%
\ \protect \BOthers {.}}{%
{\protect \APACyear {2019}}%
}]{%
villegas2019time}
\APACinsertmetastar {%
villegas2019time}%
\begin{APACrefauthors}%
Villegas, P.%
, di Santo, S.%
, Burioni, R.%
\BCBL {}\ \BBA {} Mu{\~n}oz, M\BPBI A.%
\end{APACrefauthors}%
\unskip\
\newblock
\APACrefYearMonthDay{2019}{}{}.
\newblock
{\BBOQ}\APACrefatitle {Time-series thresholding and the definition of avalanche
  size} {Time-series thresholding and the definition of avalanche size}.{\BBCQ}
\newblock
\APACjournalVolNumPages{Physical Review E}{100}{1}{012133}.
\PrintBackRefs{\CurrentBib}

\bibitem [\protect \citeauthoryear {%
Villegas%
, Moretti%
\BCBL {}\ \BBA {} Munoz%
}{%
Villegas%
\ \protect \BOthers {.}}{%
{\protect \APACyear {2014}}%
}]{%
villegas2014frustrated}
\APACinsertmetastar {%
villegas2014frustrated}%
\begin{APACrefauthors}%
Villegas, P.%
, Moretti, P.%
\BCBL {}\ \BBA {} Munoz, M\BPBI A.%
\end{APACrefauthors}%
\unskip\
\newblock
\APACrefYearMonthDay{2014}{}{}.
\newblock
{\BBOQ}\APACrefatitle {Frustrated hierarchical synchronization and emergent
  complexity in the human connectome network} {Frustrated hierarchical
  synchronization and emergent complexity in the human connectome
  network}.{\BBCQ}
\newblock
\APACjournalVolNumPages{Scientific reports}{4}{1}{1--7}.
\PrintBackRefs{\CurrentBib}

\bibitem [\protect \citeauthoryear {%
Virkar%
\ \BBA {} Clauset%
}{%
Virkar%
\ \BBA {} Clauset%
}{%
{\protect \APACyear {2014}}%
}]{%
virkar2014power}
\APACinsertmetastar {%
virkar2014power}%
\begin{APACrefauthors}%
Virkar, Y.%
\BCBT {}\ \BBA {} Clauset, A.%
\end{APACrefauthors}%
\unskip\
\newblock
\APACrefYearMonthDay{2014}{}{}.
\newblock
{\BBOQ}\APACrefatitle {Power-law distributions in binned empirical data}
  {Power-law distributions in binned empirical data}.{\BBCQ}
\newblock
\APACjournalVolNumPages{The Annals of Applied Statistics}{8}{1}{89--119}.
\PrintBackRefs{\CurrentBib}

\bibitem [\protect \citeauthoryear {%
Vuong%
}{%
Vuong%
}{%
{\protect \APACyear {1989}}%
}]{%
vuong1989likelihood}
\APACinsertmetastar {%
vuong1989likelihood}%
\begin{APACrefauthors}%
Vuong, Q\BPBI H.%
\end{APACrefauthors}%
\unskip\
\newblock
\APACrefYearMonthDay{1989}{}{}.
\newblock
{\BBOQ}\APACrefatitle {Likelihood ratio tests for model selection and
  non-nested hypotheses} {Likelihood ratio tests for model selection and
  non-nested hypotheses}.{\BBCQ}
\newblock
\APACjournalVolNumPages{Econometrica: Journal of the Econometric
  Society}{}{}{307--333}.
\PrintBackRefs{\CurrentBib}

\bibitem [\protect \citeauthoryear {%
S.~Wang%
\ \BBA {} Zhou%
}{%
S.~Wang%
\ \BBA {} Zhou%
}{%
{\protect \APACyear {2012}}%
}]{%
wang2012hierarchical}
\APACinsertmetastar {%
wang2012hierarchical}%
\begin{APACrefauthors}%
Wang, S.%
\BCBT {}\ \BBA {} Zhou, C.%
\end{APACrefauthors}%
\unskip\
\newblock
\APACrefYearMonthDay{2012}{}{}.
\newblock
{\BBOQ}\APACrefatitle {Hierarchical modular structure enhances the robustness
  of self-organized criticality in neural networks} {Hierarchical modular
  structure enhances the robustness of self-organized criticality in neural
  networks}.{\BBCQ}
\newblock
\APACjournalVolNumPages{New Journal of Physics}{14}{2}{023005}.
\PrintBackRefs{\CurrentBib}

\bibitem [\protect \citeauthoryear {%
X\BPBI R.~Wang%
, Lizier%
\BCBL {}\ \BBA {} Prokopenko%
}{%
X\BPBI R.~Wang%
\ \protect \BOthers {.}}{%
{\protect \APACyear {2011}}%
}]{%
wang2011fisher}
\APACinsertmetastar {%
wang2011fisher}%
\begin{APACrefauthors}%
Wang, X\BPBI R.%
, Lizier, J\BPBI T.%
\BCBL {}\ \BBA {} Prokopenko, M.%
\end{APACrefauthors}%
\unskip\
\newblock
\APACrefYearMonthDay{2011}{}{}.
\newblock
{\BBOQ}\APACrefatitle {Fisher information at the edge of chaos in random
  Boolean networks} {Fisher information at the edge of chaos in random boolean
  networks}.{\BBCQ}
\newblock
\APACjournalVolNumPages{Artificial life}{17}{4}{315--329}.
\PrintBackRefs{\CurrentBib}

\bibitem [\protect \citeauthoryear {%
Williams-Garc{\'\i}a%
, Moore%
, Beggs%
\BCBL {}\ \BBA {} Ortiz%
}{%
Williams-Garc{\'\i}a%
\ \protect \BOthers {.}}{%
{\protect \APACyear {2014}}%
}]{%
williams2014quasicritical}
\APACinsertmetastar {%
williams2014quasicritical}%
\begin{APACrefauthors}%
Williams-Garc{\'\i}a, R\BPBI V.%
, Moore, M.%
, Beggs, J\BPBI M.%
\BCBL {}\ \BBA {} Ortiz, G.%
\end{APACrefauthors}%
\unskip\
\newblock
\APACrefYearMonthDay{2014}{}{}.
\newblock
{\BBOQ}\APACrefatitle {Quasicritical brain dynamics on a nonequilibrium Widom
  line} {Quasicritical brain dynamics on a nonequilibrium widom line}.{\BBCQ}
\newblock
\APACjournalVolNumPages{Physical Review E}{90}{6}{062714}.
\PrintBackRefs{\CurrentBib}

\bibitem [\protect \citeauthoryear {%
Wilting%
\ \BBA {} Priesemann%
}{%
Wilting%
\ \BBA {} Priesemann%
}{%
{\protect \APACyear {2019}}%
}]{%
wilting2019between}
\APACinsertmetastar {%
wilting2019between}%
\begin{APACrefauthors}%
Wilting, J.%
\BCBT {}\ \BBA {} Priesemann, V.%
\end{APACrefauthors}%
\unskip\
\newblock
\APACrefYearMonthDay{2019}{}{}.
\newblock
{\BBOQ}\APACrefatitle {Between perfectly critical and fully irregular: A
  reverberating model captures and predicts cortical spike propagation}
  {Between perfectly critical and fully irregular: A reverberating model
  captures and predicts cortical spike propagation}.{\BBCQ}
\newblock
\APACjournalVolNumPages{Cerebral Cortex}{29}{6}{2759--2770}.
\PrintBackRefs{\CurrentBib}

\bibitem [\protect \citeauthoryear {%
Wolf%
}{%
Wolf%
}{%
{\protect \APACyear {2005}}%
}]{%
wolf2005symmetry}
\APACinsertmetastar {%
wolf2005symmetry}%
\begin{APACrefauthors}%
Wolf, F.%
\end{APACrefauthors}%
\unskip\
\newblock
\APACrefYearMonthDay{2005}{}{}.
\newblock
{\BBOQ}\APACrefatitle {Symmetry, multistability, and long-range interactions in
  brain development} {Symmetry, multistability, and long-range interactions in
  brain development}.{\BBCQ}
\newblock
\APACjournalVolNumPages{Physical review letters}{95}{20}{208701}.
\PrintBackRefs{\CurrentBib}

\bibitem [\protect \citeauthoryear {%
Yaghoubi%
\ \protect \BOthers {.}}{%
Yaghoubi%
\ \protect \BOthers {.}}{%
{\protect \APACyear {2018}}%
}]{%
yaghoubi2018neuronal}
\APACinsertmetastar {%
yaghoubi2018neuronal}%
\begin{APACrefauthors}%
Yaghoubi, M.%
, de Graaf, T.%
, Orlandi, J\BPBI G.%
, Girotto, F.%
, Colicos, M\BPBI A.%
\BCBL {}\ \BBA {} Davidsen, J.%
\end{APACrefauthors}%
\unskip\
\newblock
\APACrefYearMonthDay{2018}{}{}.
\newblock
{\BBOQ}\APACrefatitle {Neuronal avalanche dynamics indicates different
  universality classes in neuronal cultures} {Neuronal avalanche dynamics
  indicates different universality classes in neuronal cultures}.{\BBCQ}
\newblock
\APACjournalVolNumPages{Scientific reports}{8}{1}{1--11}.
\PrintBackRefs{\CurrentBib}

\bibitem [\protect \citeauthoryear {%
Yang%
\ \BBA {} Schoenholz%
}{%
Yang%
\ \BBA {} Schoenholz%
}{%
{\protect \APACyear {2017}}%
}]{%
yang2017mean}
\APACinsertmetastar {%
yang2017mean}%
\begin{APACrefauthors}%
Yang, G.%
\BCBT {}\ \BBA {} Schoenholz, S.%
\end{APACrefauthors}%
\unskip\
\newblock
\APACrefYearMonthDay{2017}{}{}.
\newblock
{\BBOQ}\APACrefatitle {Mean field residual networks: On the edge of chaos}
  {Mean field residual networks: On the edge of chaos}.{\BBCQ}
\newblock
\APACjournalVolNumPages{Advances in neural information processing
  systems}{30}{}{}.
\PrintBackRefs{\CurrentBib}

\bibitem [\protect \citeauthoryear {%
Yu%
, Klaus%
, Yang%
\BCBL {}\ \BBA {} Plenz%
}{%
Yu%
\ \protect \BOthers {.}}{%
{\protect \APACyear {2014}}%
}]{%
yu2014scale}
\APACinsertmetastar {%
yu2014scale}%
\begin{APACrefauthors}%
Yu, S.%
, Klaus, A.%
, Yang, H.%
\BCBL {}\ \BBA {} Plenz, D.%
\end{APACrefauthors}%
\unskip\
\newblock
\APACrefYearMonthDay{2014}{}{}.
\newblock
{\BBOQ}\APACrefatitle {Scale-invariant neuronal avalanche dynamics and the
  cut-off in size distributions} {Scale-invariant neuronal avalanche dynamics
  and the cut-off in size distributions}.{\BBCQ}
\newblock
\APACjournalVolNumPages{PloS one}{9}{6}{e99761}.
\PrintBackRefs{\CurrentBib}

\bibitem [\protect \citeauthoryear {%
Zhigalov%
, Arnulfo%
, Nobili%
, Palva%
\BCBL {}\ \BBA {} Palva%
}{%
Zhigalov%
\ \protect \BOthers {.}}{%
{\protect \APACyear {2015}}%
}]{%
zhigalov2015relationship}
\APACinsertmetastar {%
zhigalov2015relationship}%
\begin{APACrefauthors}%
Zhigalov, A.%
, Arnulfo, G.%
, Nobili, L.%
, Palva, S.%
\BCBL {}\ \BBA {} Palva, J\BPBI M.%
\end{APACrefauthors}%
\unskip\
\newblock
\APACrefYearMonthDay{2015}{}{}.
\newblock
{\BBOQ}\APACrefatitle {Relationship of fast-and slow-timescale neuronal
  dynamics in human MEG and SEEG} {Relationship of fast-and slow-timescale
  neuronal dynamics in human meg and seeg}.{\BBCQ}
\newblock
\APACjournalVolNumPages{Journal of Neuroscience}{35}{13}{5385--5396}.
\PrintBackRefs{\CurrentBib}

\end{thebibliography}


\end{document}